\documentclass[11pt]{article}
\usepackage{epsfig,color,natbib,authblk}
\usepackage{graphicx}

\def\lesssim{\mathrel{\hbox{\rlap{\hbox{\lower4pt\hbox{$\sim$}}}\hbox{$<$}}}}
\def\gtrsim{\mathrel{\hbox{\rlap{\hbox{\lower4pt\hbox{$\sim$}}}\hbox{$>$}}}}

\begin{document}

\title{Evolution of active galactic nuclei} 
\author[,1]{A. Merloni\footnote{{\tt am@mpe.mpg.de}}} 
\author[,2]{S. Heinz\footnote{{\tt heinzs@astro.wisc.edu}}}
\affil[1]{Max-Planck-Institut f\"ur Extraterrestrische Physik,
Giessenbachstr., D-85741, Garching, Germany}
\affil[2]{Astronomy
Department, University of Wisconsin-Madison, Madison, WI 53706}
\date{\today}

\maketitle

\begin{abstract}
  Supermassive black holes (SMBH) lurk in the nuclei of most massive
  galaxies, perhaps in all of them. The tight observed scaling
  relations between SMBH masses and structural properties of their
  host spheroids likely indicate that the processes fostering the
  growth of both components are physically linked, despite the many
  orders of magnitude difference in their physical size.  This chapter
  discusses how we constrain the evolution of SMBH, probed by their
  actively growing phases, when they shine as active galactic nuclei
  (AGN) with luminosities often in excess of that of the entire
  stellar population of their host galaxies.  Following loosely the
  chronological developments of the field, we begin by discussing
  early evolutionary studies, when AGN observed at various
  wavelengths represented beacons of light probing the most distant
  reaches of the universe and were used as tracers of the large scale
  structure ("Cosmography").  This early study turned into a more
  mundane enterprise of AGN ``Demography'', once it was realized that
  the strong evolution (in luminosity, number density) of the
  AGN population hindered any attempt to derive cosmological
  parameters from AGN observations directly.  Following a discussion
  of the state of the art in the study of AGN luminosity functions, we
  move on to discuss the ``modern'' view of AGN evolution, one in
  which a bigger emphasis is given to the physical relationships
  between the population of growing black holes and their environment
  (``Cosmology'').  This includes observational and theoretical efforts
  aimed at constraining and understanding the evolution of scaling
  relations, as well as the resulting limits on the evolution of the
  SMBH mass function. Physical models of AGN feedback and the ongoing
  efforts to isolate them observationally are discussed next. Finally,
  we touch upon the problem of when and how the first black holes
  formed and the role of black holes in the high-redshift universe.
\end{abstract}

\section*{Index Terms}
Active Galactic Nuclei (AGN), Quasars, Radio Galaxies,
Supermassive Black Holes, Eddington Luminosity, Accretion, Jets,
Scaling relations, AGN Number Counts, AGN Luminosity Functions, AGN
Clustering, Cosmic X-ray Background, Black Hole Mass Function, AGN
Feedback, Galaxy Evolution, Black Hole Formation

\section*{Keywords}

\section*{List of Abbreviations}
\noindent AGN: Active Galactic Nucleus\\
\noindent AU: Astronomical Unit\\
\noindent BAL: Broad Absorption Line\\
\noindent BHAR: Black Hole Accretion Rate\\
\noindent CSS: Compact Steep Spectrum\\
\noindent CXRB: Cosmic X-ray Background\\
\noindent DM: Dark Matter\\
\noindent FR I/II: Fanaroff-Riley class I/II\\
\noindent GPS: Gigahertz Peak Spectrum\\
\noindent ICM: Intra-Cluster Medium\\
\noindent IGM: Intra-Group Medium\\
\noindent LDDE: Luminosity Dependent Density Evolution\\
\noindent LADE: Luminosity And Density Evolution\\
\noindent LF: Luminosity Function\\
\noindent LLAGN: Low Luminosity Active Galactic Nuclei\\
\noindent PLE: Pure Luminosity Evolution\\
\noindent PDE: Pure Density Evolution\\
\noindent QSO: Quasi-Stellar Object\\
\noindent SED: Spectral Energy Distribution\\
\noindent SFR: Star Formation Rate\\
\noindent SMBH: Super-Massive Black Hole\\

\section{A historical perspective on AGN research}
\label{sec:intro}
The study of astrophysical black holes, as it has developed over the
last five decades, is driven by three main rationales and goals.
Because the mere existence of black holes is the most far-reaching
implication of the theory of General Relativity (together with the Big
Bang cosmological theory), they can first of all be used to test
theories of gravitation in the strong field regime.  Secondly,
astrophysical black holes are revealed to us through emission
processes taking place in accretion flows and relativistic jets, both
originating in the black hole's deep potential well, and they offer a
unique opportunity of studying interesting and complex
astrophysical problems, involving extreme physical conditions,
relativistic magnetohydrodynamics, and radiative effects.  Thirdly,
black hole formation and evolution might play an important role in a
broader cosmological context, affecting the formation and the
evolution of the structures they live in, such as galaxies, groups and
clusters.

During the first golden age\footnote{This definition was introduced by
  the Caltech graduate Bill Press \citep{thorne:94} to identify the
  years between the early '60s and the early '70s.} of black hole
astrophysics, efforts were focused on finding proof of the existence
of black holes and to define their basic interactions with the
environment (accretion and relativistic jet theory).  Such goals only
touched on the first two of the rationales listed above.  The history
of the development of black hole physics (both theoretical and
observational) in these years has been beautifully laid out by Kip
Thorne in his book ``Black Holes and time warps: Einstein's outrageous
legacy'' \citep{thorne:94}, where the reader can find a more complete
set of references and biographical notes, together with the historical
accounts presented elsewhere in this volume (see E. Perlman: ``Active galactic
nuclei'').

Beginning at about the turn of the 21st century, black hole
astrophysicists have acknowledged the relevance of their subject of
study for a broader community of cosmologists and extragalactic
astronomers, thanks to the multiple lines of evidence pointing towards
a fundamental role played by black holes in galaxy evolution.

\begin{figure*}[t!]
\centering
\psfig{figure=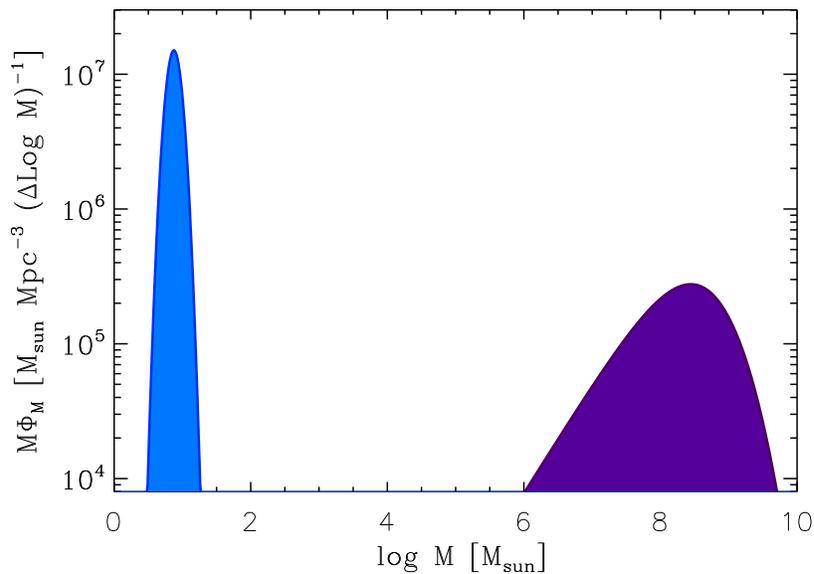,height=8.cm}
\caption{The local black hole mass function, plotted as $M \times
  \phi_M$ in order to highlight the location and height of the two
  main peaks in the distribution.  The stellar mass black hole peak
  has been drawn assuming a log-normal distribution with mean mass
  equal to 5 solar masses, width of 0.1 dex and a normalization
  yielding a density of about $1.1 \times 10^{7}$ $M_{\odot}$
  Mpc$^{-3}$ \citep{fukugita:04}, which is about $7 \times 10^{-5}$
  times the critical density of the universe.  The supermassive black
  hole peak, instead, contributes to an overall density of about $4.2
  \times 10^{5}$ $M_{\odot}$ Mpc$^{-3}$, or a fraction only $2.7 \times
  10^{-6}$ of the critical density (see
  section~\ref{sec:global_growth} for details).}
\label{fig:mf}
\end{figure*}

In fact, black holes in the local universe come in two main families
according to their size, as recognized by the strongly bi-modal
distribution of the local black hole mass function (see
Fig.~\ref{fig:mf}). While the height, width and exact mass scale of
the {\em stellar} mass peak should be understood as a by-product of
stellar (and binary) evolution, and of the physical processes that
make supernovae and gamma-ray bursts explode, the {\em supermassive}
black hole peak in this distribution is the outcome of the
cosmological growth of structures and of the evolution of accretion in
the nuclei of galaxies, likely modulated by the mergers these nuclear
black holes will experience as a result of the hierarchical
galaxy-galaxy coalescences.

\begin{figure}
  \centering \psfig{figure=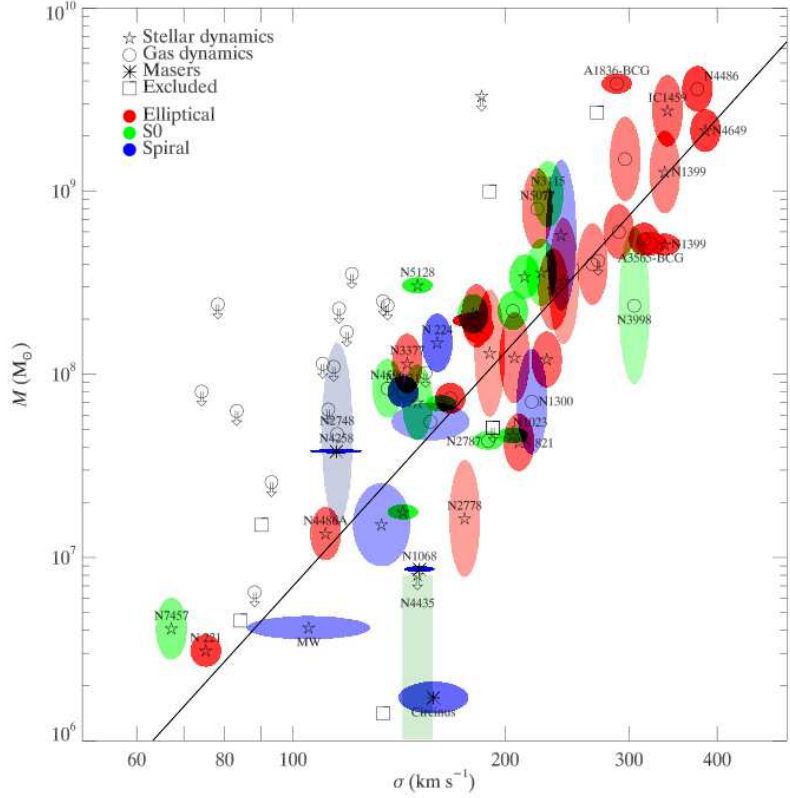,height=11.cm}
  \caption{The $M_{\rm BH}-\sigma_{*}$ relation for galaxies with
    dynamical BH mass measurements.  The symbols indicate the method
    of BH mass measurement: stellar dynamical (\emph{pentagrams}), gas
    dynamical (\emph{circles)}, masers (\emph{asterisks}).  Arrows
    indicate 3$\sigma$ confidence upper limits to the BH mass.  The
    color of the error ellipse indicates the Hubble type of the host
    galaxy: elliptical (\emph{red}), S0 (\emph{green}), and spiral
    (\emph{blue}).  The saturation of the colors in the error ellipses
    or boxes is inversely proportional to the area of the ellipse or
    box.  Squares are galaxies not included in the fit. This is shown
    as a solid line for the best fit relation to the full sample:
    $M_{\mathrm{BH}} =
    10^{8.12}~M_{\odot}(\sigma_{*}/200~{~\mathrm{km~s^{-1}}})^{4.24}$
    (adopted from \citealt{gultekin:09}).}
  \label{fig:m-sigma}
\end{figure}

This picture of the local demographics of black holes has been made
possible by the discovery of tight scaling relations between the
central black hole mass and various properties of their host spheroids
(velocity dispersion, $\sigma_{*}$, stellar mass, $M_{*}$, luminosity,
etc.) that characterize the structure of nearby {\em inactive}
galaxies
(\citealt{magorrian:98,gebhardt:00,ferrarese:00,haering:04,gultekin:09},
see Figure~\ref{fig:m-sigma}).

These correlations are a result of the search for local QSO relics via
the study of their dynamical influence on the surrounding stars and
gas made possible by the launch of the {\it Hubble Space Telescope}.
They have revolutionized the way we understand the physical link
between the evolution of galaxies and active galactic nuclei
(AGN\footnote{In this chapter, we will use both the term AGN and
  QSO/quasar to indicate actively growing supermassive black holes,
  implying no real physical distinction between the two, apart from
  one based on the total emitted luminosity: while AGN can be used for
  any objects, QSO/quasar usually identify those with bolometric
  luminosity $\log L_{\rm bol} > 46$ in cgs units.}).

In addition, it is now understood that supermassive black hole (SMBH)
growth is due mainly to radiatively efficient accretion over
cosmological times, taking place during active phases (see
\S~\ref{sec:global_growth} below).  This, together with the
understanding of a near universal presence of black holes in galactic
centers has led to the suggestion that most, if not all, galaxies went
through a phase of nuclear activity in the past, during which a strong
physical coupling (generally termed ``feedback'') might have
established a long-lasting link between host and black hole
properties.

Such a {\em renewed} interest for AGN in a cosmological
context requires a good understanding of the evolutionary properties
of this class of objects.  The fact that AGN and quasars were a
strongly evolving class of astronomical sources became evident very
soon after their discovery, as we will discuss at length in the
following sections.  Nonetheless, the appreciation that such an
evolution could not only mirror, but also influence that of galaxies,
groups and clusters, only became commonplace after the discovery of
the above mentioned scaling relations.

In this chapter we will focus on the current knowledge of AGN
evolution.  Following loosely the chronological developments of the
field, we will begin by discussing the ``first generation'' of AGN
evolutionary studies (\S~\ref{sec:cosmography}), during which AGN
observed at various wavelengths represented beacons of light probing
the most distant reaches of the universe and were used as tracers of
the structures themselves.

This short-lived epoch of AGN ``Cosmography'' quickly gave way to a
more mundane enterprise of AGN ``Demography'', once it was realized
that the strong evolution (in luminosity, number density, etc.) of the
AGN population hindered any serious attempt to derive cosmological
parameters from AGN observations directly.  The attention then moved
to the study of the {\em evolution} of active galactic nuclei by means
of determinations of their luminosity functions.  An update on the
most recent works on the luminosity functions of AGN selected in
different ways from different electromagnetic bands will also be given
in section~\ref{sec:cosmography}, which will be closed by a brief
discussion of AGN clustering as a natural complementary cosmographic
tool (\S~\ref{sec:clustering}).

We will then move to discuss the ``modern'' view of AGN evolution, one
in which a bigger emphasis is given to the physical relationships
between the population of growing black holes and their environments.
We call this the ``Cosmology'' phase of AGN studies, to highlight the
close link between these subject areas that has been established in
recent years.  We will first discuss observational and theoretical
efforts aimed at constraining and understanding the evolution of the
scaling relations, as well as the resulting limits on the evolution of
the SMBH mass function (\S~\ref{sec:cosmology}), and we will then
present physical models of AGN feedback and the ongoing efforts to
isolate them observationally in section~\ref{sec:feedback}.

Finally, in the last section of this chapter (\S~\ref{sec:cosmogony})
we will touch upon the problem of when and how the first black holes
form and the role of black holes in the high redshift universe.

Before all this, however, a small diversion is in place.  To be
consistent, the very notion of an evolutionary study of any particular
class of objects, not only in astrophysics, requires a definition of
the non-evolving {\it substratus} that allows us to first identify an
object as a member of the class, the evolution of which one wishes to
study.  To make just a simple example drawn from astronomical
research, the evolution of galaxies is very much complicated by the
never-ending morphological and photometric transformation of the
different populations, so that a non-trivial element of any such study
is the identification of progenitors and offspring along the Hubble
sequence (see the Chapter R. Buta: ``Galaxy Morphology'' in this volume).

\subsection{Redshift evolution in AGN Spectral Energy Distributions}
\label{sec:sed_vs_z}
The overall spectral energy distribution (SED) of AGN extends over
many decades in frequency, and is the result of a number of different
emission processes acting at different physical scales. We refer the
reader to Chapter "Active Galactic Nuclei" by E. Perlman in this book for a
thorough discussion of these processes and of the main characteristic
of AGN SED.

The observational appearance of an active galactic nucleus is
determined not only by its intrinsic emission properties, but also by
the nature, amount, dynamical and kinematic state of any intervening
material along the line of sight.  AGN obscuration is a crucial factor
for our general understanding of the AGN phenomenon. For example,
in the traditional unification-by-orientation schemes different
classes of AGN are explained on the basis of the line-of-sight
orientation with respect to the axis of rotational symmetry of the
system (see e.g.~\citealt{antonucci:93,urry:95}, and references
therein).

At odds with such simple schemes, evidence for a variation of the
fraction of obscured AGN as a function of {\it luminosity} has been
mounting recently \citep{ueda:03,steffen:03,simpson:05,hasinger:08}.
The fraction of absorbed AGN, defined in different and often
independent ways, appears to be lower at higher nuclear luminosities.
This might be considered a signature of AGN feedback (in the
``quasar'' mode, see \S~\ref{sec:qso_feedback} below), in that
powerful sources are able to clean up their immediate gaseous environments,
responsible for the nuclear obscuration, more efficiently.

After accounting for such a clear luminosity-dependence, it is
currently unclear whether the overall incidence of obscuration and
extinction in the nuclear regions of a galaxy evolve with redshift.
This would be expected if, for example,  nuclear obscuration were
causally linked to the overall amount of gas within galaxies, a
quantity that increases obviously with redshift.

What we are interested in here, however, is any possible evidence of
redshift evolution (or lack thereof) of the {\it intrinsic} AGN
spectral properties, i.e. those characterizing the emission processes
associated with the major mode of radiative energy release.
   
\begin{figure*}
\centering
\begin{tabular}{cc}
\psfig{figure=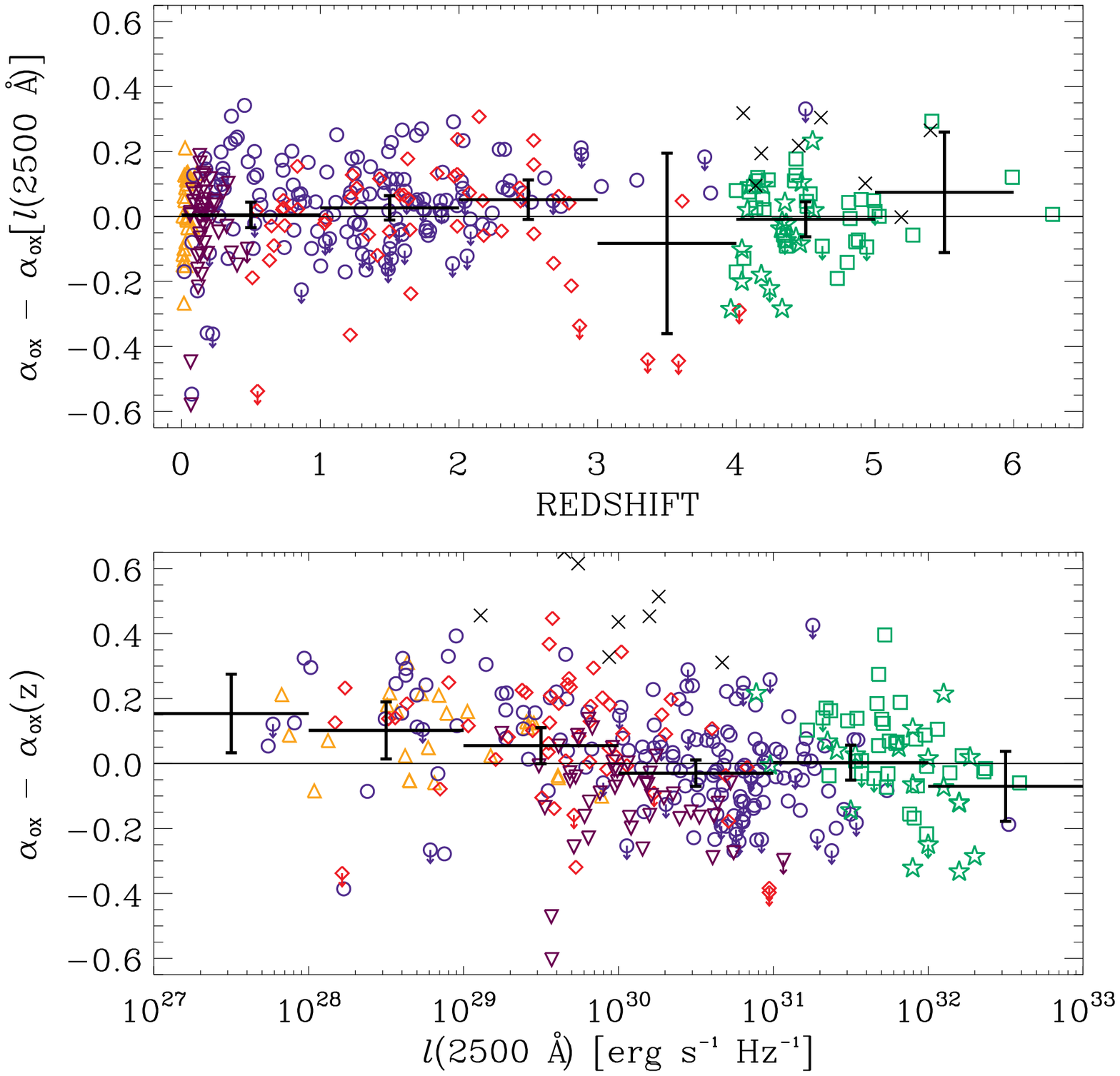,height=6.2cm}&
\psfig{figure=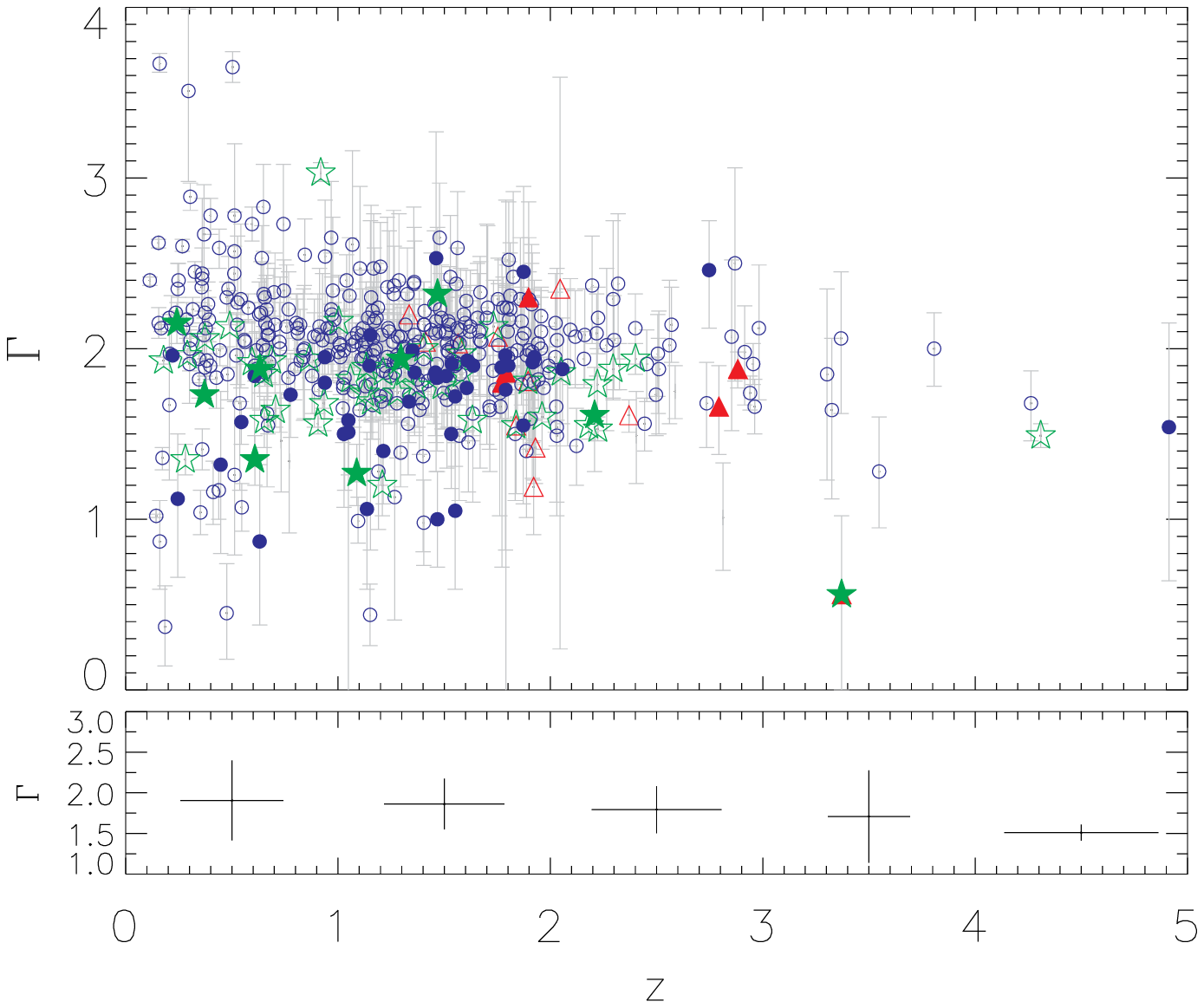,height=6cm}\\
\end{tabular}
\caption{{\it Left}: $\alpha_{\rm ox}$ residuals as a function of
  redshift (top panel) and luminosity density at 2500$\AA$ (bottom
  panel). The overlaid error bars denote the mean and the 3$\sigma$
  standard deviation of the mean of the residuals.  Limits are denoted
  with arrows. The systematic residuals in the lower plot indicate
  that $\alpha_{\rm ox}$ cannot be dependent on redshift alone
  (adopted from \citealt{steffen:06}); {\it
    Right}: X-ray photon index ($\Gamma$) vs.~redshift $z$. Blue
  circles represent radio quiet, non-BAL (Broad
  Absorption Line) quasars, green stars
  represent radio loud quasars, and red triangles represent BAL
  quasars.  The bottom plot shows the weighted mean
  $\Gamma$ values for bins of width $\Delta z=1$. No clear sign of
  evolution in the average X-ray spectral slope of AGN is detected
  over more than 90\% of the age of the universe (from
  \citealt{young:10}).}
\label{fig:young10}
\end{figure*}

X-ray emission is ubiquitous in AGN, and is a very effective way for
selecting accreting black holes due both to the minimal contamination
of star-forming processes and due to the decreasing importance of
obscuration at increasing X-ray energies.  Unfortunately, the exact
mechanism responsible for AGN X-ray emission and its physical location
are not fully understood yet (see Chapter ``Active Galactic
Nuclei'' by E. Perlman in this book).  Still, as a very general diagnostic, the
``X-ray loudness'', usually characterized by the $\alpha_{\rm ox}$
parameter, i.e. the slope of the spectrum between 2500 \AA$\,=5$ eV
and 2 keV: $\alpha_{\rm ox}= 0.3838 \log(F_{\rm 2keV}/F_{\rm 2500})$
can be used to characterize the fraction of bolometric light carried
away by high-energy X-ray photons.  Recent studies of large samples of
both X-ray and optical selected AGN have clearly demonstrated that
$\alpha_{\rm ox}$ is itself a function of UV luminosity (see
e.g.~\citealt{steffen:06,young:10}).  However, no redshift evolution
can be discerned in the data, as shown in the left panel of
Figure~\ref{fig:young10}.

Moreover, large collecting-area X-ray telescopes allow a more precise
determination of the X-ray spectra of AGN, which are usually
characterized by a power-law, upon which emission lines and
absorption features are superimposed.  Up to the highest redshift
where reliable spectral analysis of AGN can be performed, no clear
sign of evolution in the X-ray spectral slope $\Gamma$ has been
detected (see the right panel of Figure~\ref{fig:young10}).

\begin{figure*}
  \centering \psfig{figure=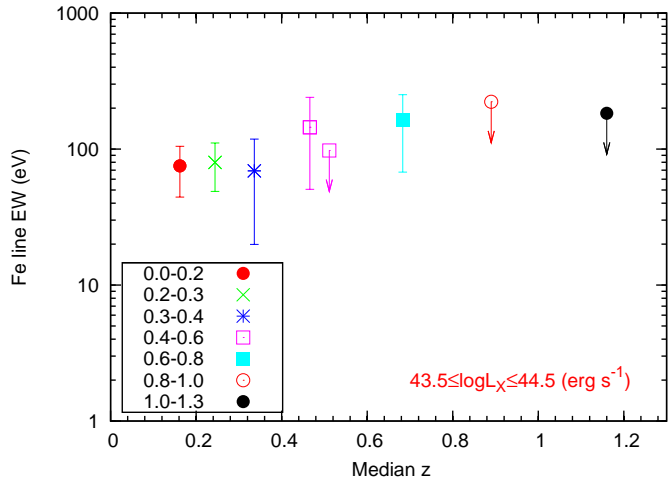,height=6.5cm}
  \caption{Rest frame equivalent width of the narrow Iron K$\alpha$
    emission line observed in the average spectrum of AGN as a
    function of redshift.  Only objects in a fixed (2-10 keV)
    luminosity range $10^{43.5}<L_{\rm 2-10}<10^{44.5}$ ergs/s were
    considered.  From \citealt{chaudhary:10}.}
\label{fig:pooja10}
\end{figure*}

Similarly, while the narrow iron K$\alpha$ emission line, the most
prominent feature in AGN X-ray spectra, is clearly dependent on
luminosity (the so-called Iwasawa-Taniguchi effect \citealt{iwasawa:93}), 
it shows {\em no}
sign of evolution in its equivalent width with redshift, at least up
to $z\simeq 1.2$ (Figure~\ref{fig:pooja10}; \citealt{chaudhary:10},
and references therein).

Even more surprising is the lack of evolution in the {\em optical}
emission line properties of QSOs.  The metallicities implied by the
relative strength of broad emission lines do not show any
significant redshift evolution: They are solar or super-solar, even in
the highest redshift QSOs known (see
e.g.~\citealt{hamann:92}), in contrast with the
strong evolution of the metallicity in star forming galaxies.

\begin{figure*}
\centering
\begin{tabular}{cc}
\psfig{figure=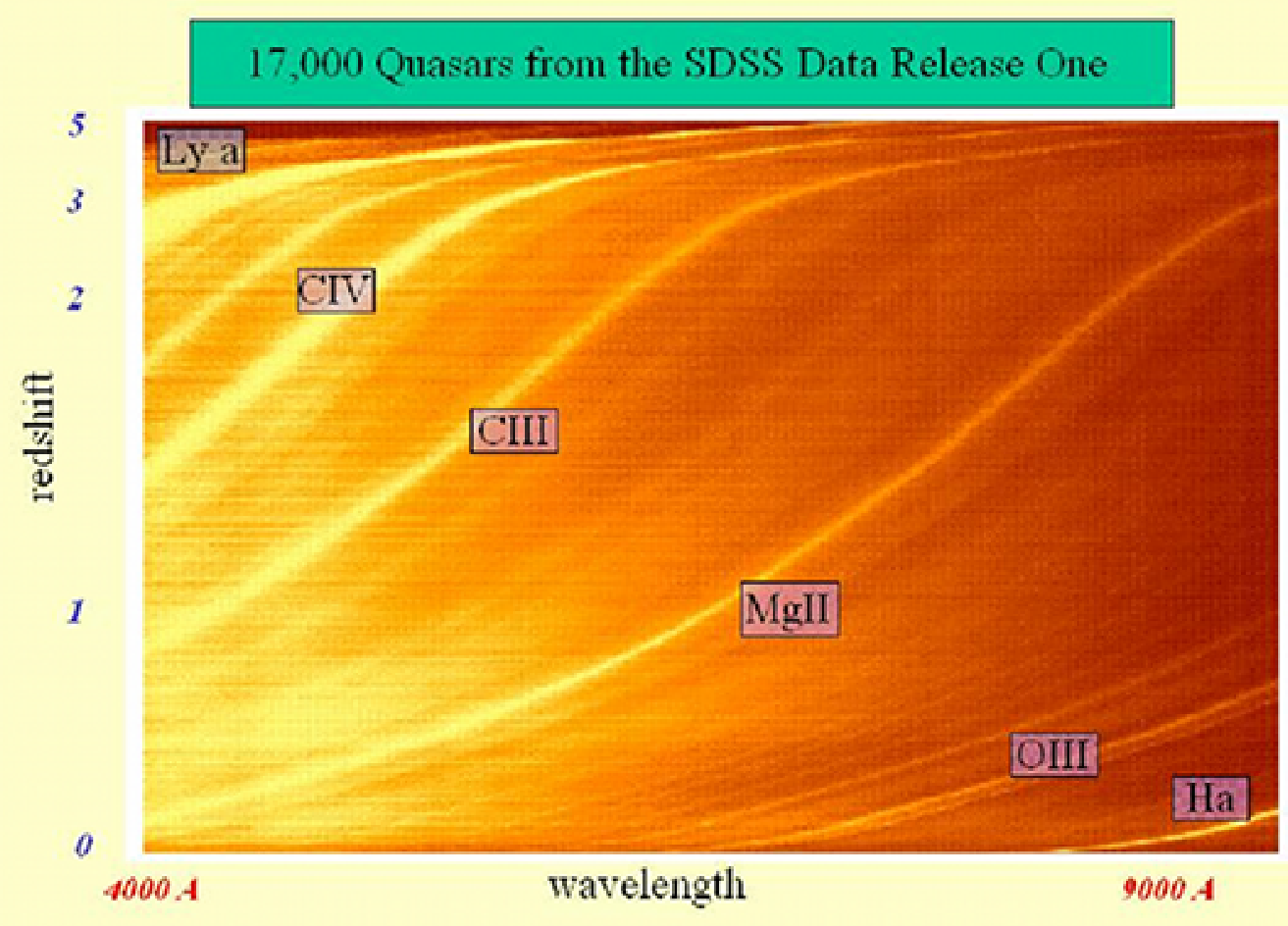,height=5.5cm}&
\psfig{figure=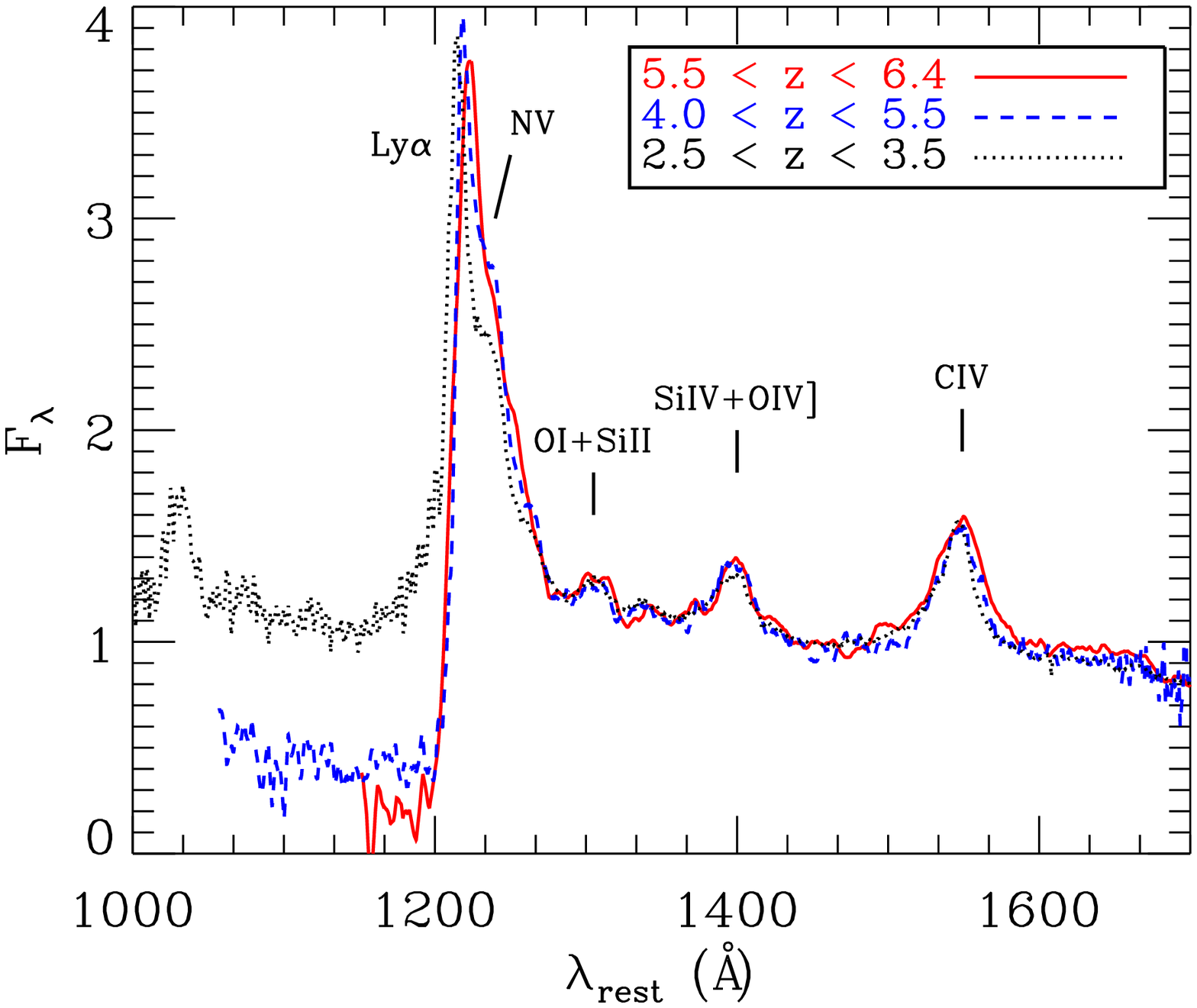,height=5.5cm}\\
\end{tabular}
\caption{{\it Left}: Spectra of 17,000 QSOs from SDSS. Notice the
  large degree of uniformity in the relative intensity of the main
  emission features (courtesy of X. Fan); {\it Right} Stacked spectra
  of quasars in different redshift bins.  Note that the relative
  intensity of the metal lines (and in particular the
  (SiIV+OIV])$/$CIV ratio) remains constant over the wide redshift
  interval 2.5<z<6.4, indicating that the metallicity in the observed
  quasars does not evolve with redshift.  From \citet{juarez:09}.}
\label{fig:fan08}
\end{figure*}

A pictorial view of this surprising uniformity is shown in the left
panel of figure~\ref{fig:fan08}, where the raw spectra from $\sim$
17,000 QSOs extracted from the Sloan Digital Sky Survey (SDSS) are
plotted next to each other in a sequence of increasing redshift from
bottom to top.  The right panel of Fig.~\ref{fig:fan08} shows a direct
comparison of stacked QSO spectra in three redshift intervals
\citep{juarez:09}, where it is clear that the flux ratios among the
most prominent lines stay almost constant up to the highest redshift
probed.

Summarizing, there exists a remarkably uniform set of spectral
characteristics that defines active nuclei at all epochs in the
history if the universe, at least if we consider objects of a fixed
total (bolometric) luminosity.  The simplest
explanation is that the {\it emission} properties from AGN, i.e.,
those which are (in most cases, at least) set by physical processes
taking place within the gravitational sphere of influence of the
central black hole, are essentially dictated by the gas and plasma
dynamics there, where the central object's gravity dominates.  We
should then expect them to be relatively insensitive to the
cosmological epoch, which instead greatly
affects the properties of matter (density, temperature, ionization
state, etc.) at the generic outer boundary, i.e., right outside the SMBH
gravitational sphere of influence.

\section{Cosmography and Demography}
\label{sec:cosmography}

Accreting supermassive black holes have long been the lighthouses of
our observable universe, holding the record of the most distant object
known for more than four decades.  As such, they have played a key
role in the early phases of cosmological investigations.

Already in 1955 the second Cambridge catalog (2C) of unresolved radio
sources (the so-called `radio stars') observed at 81 MHz (3.7 meters)
had shown both a remarkable {\em uniformity} in the distribution of
objects in the sky and an {\em increase} in the cumulative number
counts (see below) that allowed \citet{ryle:55} to unambiguously
demonstrate not only their extragalactic origin, but also that the
bulk of the source should lie at distances larger than a few tens of
Mpc, i.e. well beyond the edge of the optically observable universe at
the time.

\begin{figure*}
\centering
\psfig{figure=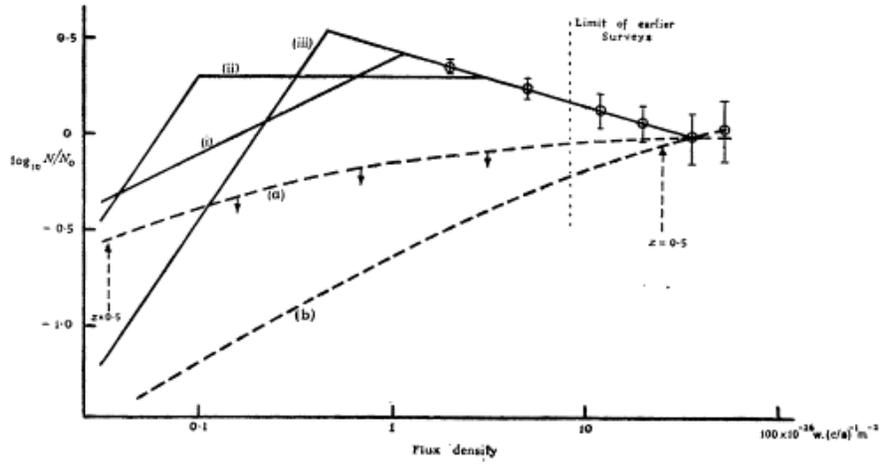,height=6.5cm}
\caption{Observed normalized radio number counts from the original 3C
  catalog at 178 MHz (from \citealt{ryle:61}). The observational
  points (open circles) are extrapolated at low fluxes with three
  empirical models (i,ii, and iii) made such as not to violate the
  total low-frequency radio background available at the time. Dashed
  lines marked with (a) and (b) denote the counts predicted by the
  steady state cosmological 
  model assuming two possible luminosity functions, given
  the observed sources. A clear discrepancy emerged between
  observations and non-evolving universe models. A more detailed
  discussion of current constraints on radio sources number counts is
  given in section~\ref{sec:radio_lf}}
\label{fig:ryle61}
\end{figure*}

The dispute over the exact shape of the radio source number count
distribution that ensued soon afterwards became a key part of the
debate between ``steady state'' and ``evolutionary'' models of the
universe, lending strong support against stationary universe models
(see Figure~\ref{fig:ryle61}).  Just two years after the discovery of
quasars \citep{schmidt:63} with their exceedingly large redshifts,
A.~Sandage wrote: {\em ``The objects would seem to be of major
  importance in the solution of the cosmological problem. They can be
  found at great distances because of their high luminosity. Studies
  of the [number counts] curves using [quasars] should eventually
  provide a crucial test of various cosmological models''}
\citep{sandage:65}.  Similar hopes were expressed by Longair the following year
\citep{longair:66}.

However, as we will discuss in more detail in the following sections,
quasars and radio galaxy source counts demonstrated clearly that the
populations being studied {\em did} evolve strongly with cosmic epoch:
the number of quasars per unit comoving volume was clearly larger in
the past, so that the information about the geometry of the universe
and the cosmological parameters is
buried underneath that about the evolution of the AGN themselves.

Progress in characterizing the intrinsic evolution of the QSO
population effectively quenched the hope to use black holes
as ideal tracers of the structure of the universe, but opened up the
study of the evolution of growing supermassive black holes, that we
outline below.  

\subsection{From Number Counts to Luminosity Functions}

By {\it number counts} one typically means the surface density in the
sky of a given class of sources as a function of the limiting flux of
the observations.  In astronomy, this is the simplest observational
tool that can be used to study the evolution of a sample of objects
(and to test cosmological models).

The space density of sources of different intrinsic luminosities, $L$,
is described by the {\it luminosity function} (LF), $\phi(L)$, so that
$dN=\phi(L) dL$ is the number of sources per unit volume with
luminosity in the range $L$ to $L+dL$.  Let us, for simplicity,
consider the local or nearby (Euclidean) universe uniformly filled
with sources with LF $\phi(L)$. If $S$ is the limiting flux that we
can detect, sources with luminosity $L$ can be observed out to a
distance $r=(L/4\pi S)^{1/2}$. The number of sources over the solid
angle $\Omega$, observable down to the flux $S$ are:

\begin{equation}
\label{eq:euclidean_counts}
N(>S)=\int \frac{\Omega}{3}r^3
\phi(L)dL=\frac{\Omega}{3(4\pi)^{3/2}}S^{-3/2}\int L^{3/2}\phi(L)dL \, .
\end{equation}

Thus, independent of the exact {\em shape} of the luminosity function
entering in the determination of a normalization constant, the {\em
  slope} of the cumulative number counts of any non-evolving class of
sources in a uniform, Euclidean universe should always be equal to
$d\log N(>S)/d\log S=-3/2$ (if we use magnitudes, $m$, instead of
luminosities, then $d\log N(>m)/dm=0.6$).

In general, the correct relativistic expression for number counts
differs from the Euclidean one because (a) the observed flux density
depends upon the spectrum of the source, as the radiation emitted at
frequency $\nu_1$ is observed at the redshifted frequency
$\nu_0=\nu_1/(1+z)$, and (b) curvature effects modify the volume
element per unit redshift, making it smaller with increasing $z$.
Overall, for typical source spectra which are not too strongly
``inverted'' (i.e. with flux density increasing with
frequency), the combination of these effects makes it more and more
difficult to detect sources at progressively higher redshift and
causes number counts to have slopes always shallower than the
Euclidean one (see e.g.~\citealt{longair:08}, chapter 17).  
As we will see below, strong evolutionary effects (i.e. luminosity
functions changing rapidly with time) can counteract such a behavior.

Before proceeding, a brief introduction of common terminology widely
adopted in the study of luminosity function evolution is necessary.
The simplest general approach describes an evolving
luminosity function with the aid of two functions, $f_{\rm l}(z)$ and
$f_{\rm d}(z)$, that take into account the evolution of the luminosity
and number density of the sources, respectively:
\begin{equation}
\phi(L,z) = f_{\rm d}(z)\phi(L/f_{\rm l}(z),z=0).
\end{equation}

In the {\it pure luminosity evolution (PLE)} case ($f_{\rm
  d}=$const.), the co-moving number density of sources is constant,
but luminosity varies with cosmic epoch; in the {\it pure density
  evolution (PDE)} case ($f_{\rm l}=$const.), but the co-moving
density of sources of any luminosity varies.

In the following sections we will discuss the observational state of
the art as far as AGN number counts and luminosity functions are
concerned, in the radio, X-rays and optical/IR bands.  More
comprehensive and specialized reviews have, of course, been published.
In particular, we refer to the recent work by \citet{dezotti:10} for a
discussion of observations at radio wavelengths, \citet{croom:09} for
optical QSOs and to \citet{brandt:05} for X-ray studies.

\subsubsection{The evolution of radio AGN}
\label{sec:radio_lf}

\begin{figure*}
\centering
\psfig{figure=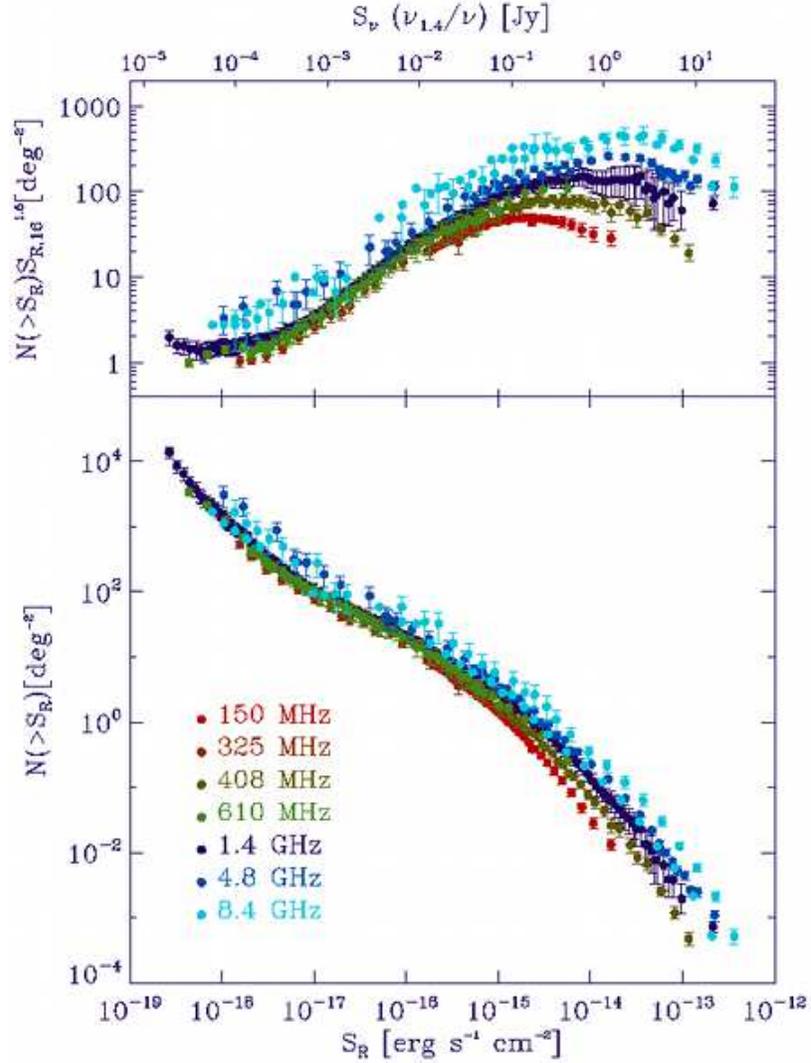,height=15.cm}
\caption{A compilation of cumulative radio source counts (number of
  objects brighter than a given flux per square degree) in various
  bands. The observational data are taken from (\citealt{massardi:10}
  see also references therein).  On the bottom horizontal axis the
  total radio flux $S_{\rm R}=\nu S_{\nu}$ in CGS units is shown,
  while the top horizontal axis shows the corresponding flux density
  in Jansky, where $\nu_{1.4}$ is the frequency of 1.4 GHz. The bottom
  panel shows the observed counts, while the top panel shows the
  counts after the Euclidean slope has been factored out.}
\label{fig:radio_counts}
\end{figure*}

Figure~\ref{fig:radio_counts} shows a compilation of cumulative source
number counts from a large number of surveys in different radio bands
(data points from \citealt{massardi:10}, see references therein). On
the bottom x-axis, the total radio flux is expressed as $S_{\rm
  R}\equiv \nu S_{\nu}$ (where $S_{\nu}$ is the observed radio flux
density at any given radio frequency $\nu$) in cgs units, while the
top axis shows the corresponding radio flux density at 1.4 GHz.
Overall, the shape of the radio counts is similar in all bands,
indicating the relative lack of spectral complexity of radio AGN. This
is best seen when normalizing the observed counts to the Euclidean
slope, as shown in the top panel.

At bright fluxes, counts rise more steeply than $S^{-3/2}$.  This was
already discovered by the first radio surveys at meter wavelengths
\citep{ryle:55}, as we have discussed above, lending strong support
for evolutionary cosmological models, as opposed to theories of a
steady state universe (see Fig.~\ref{fig:ryle61}).

At fluxes fainter than about a Jansky\footnote{A Jansky (named after
  Karl Jansky, who first discovered the existence of radio waves from
  space) is a flux measure, corresponding to $10^{-23}$ ergs cm$^{-2}$
  Hz$^{-1}$} (or $\approx 10^{-14}$ ergs s$^{-1}$ cm$^{-2}$ at 1 GHz)
the counts increase less steeply than $S^{-3/2}$, being
dominated by sources at high redshift, thus probing a substantial
volume of the observable universe.

At flux densities above a mJy the population of radio sources is
largely composed by AGN. For these sources, the observed
radio emission includes the classical extended jet and double lobe
radio sources as well as compact radio components more directly
associated with the energy generation and collimation near the central
engine.

The deepest radio surveys, however, (see e.g.~\citealt{padovani:09}
and references therein), probing well into the sub-mJy regime, clearly
show a further steepening of the counts. The nature of this change is
not completely understood yet, but in general it is attributed to the
emergence of a new class of radio sources, most likely that of
star-forming galaxies and/or radio quiet AGN.  Unambiguous solutions
of the population constituents at those faint flux levels requires not
only identification of the (optical/IR) counterparts of such faint
radio sources, but also a robust understanding of the physical
mechanisms responsible for the observed emission both at radio and
optical/IR wavelengths.

Thus, the complex shape of the observed number counts provides clues
about the evolution of radio AGN, as well as on their physical nature,
even before undertaking the daunting task of identifying substantial
fractions of the observed sources, determining their distances, and
translating the observed density of sources in the redshift-luminosity
plane into a (evolving) luminosity function.  Pioneering work from
\citet{longair:66} already demonstrated that, in order to reproduce
the narrowness of the observed 'bump' in the normalized counts around
$1$ Jy seen in Fig.~\ref{fig:radio_counts}, 
only the most luminous sources could evolve strongly (in density
and/or luminosity) with
redshift.  This was probably the first direct hint of the intimate
nature of the {\it differential} evolution AGN undergo over cosmological
times.

\begin{figure*}
\begin{center}
\resizebox{!}{0.48\textwidth}{\includegraphics{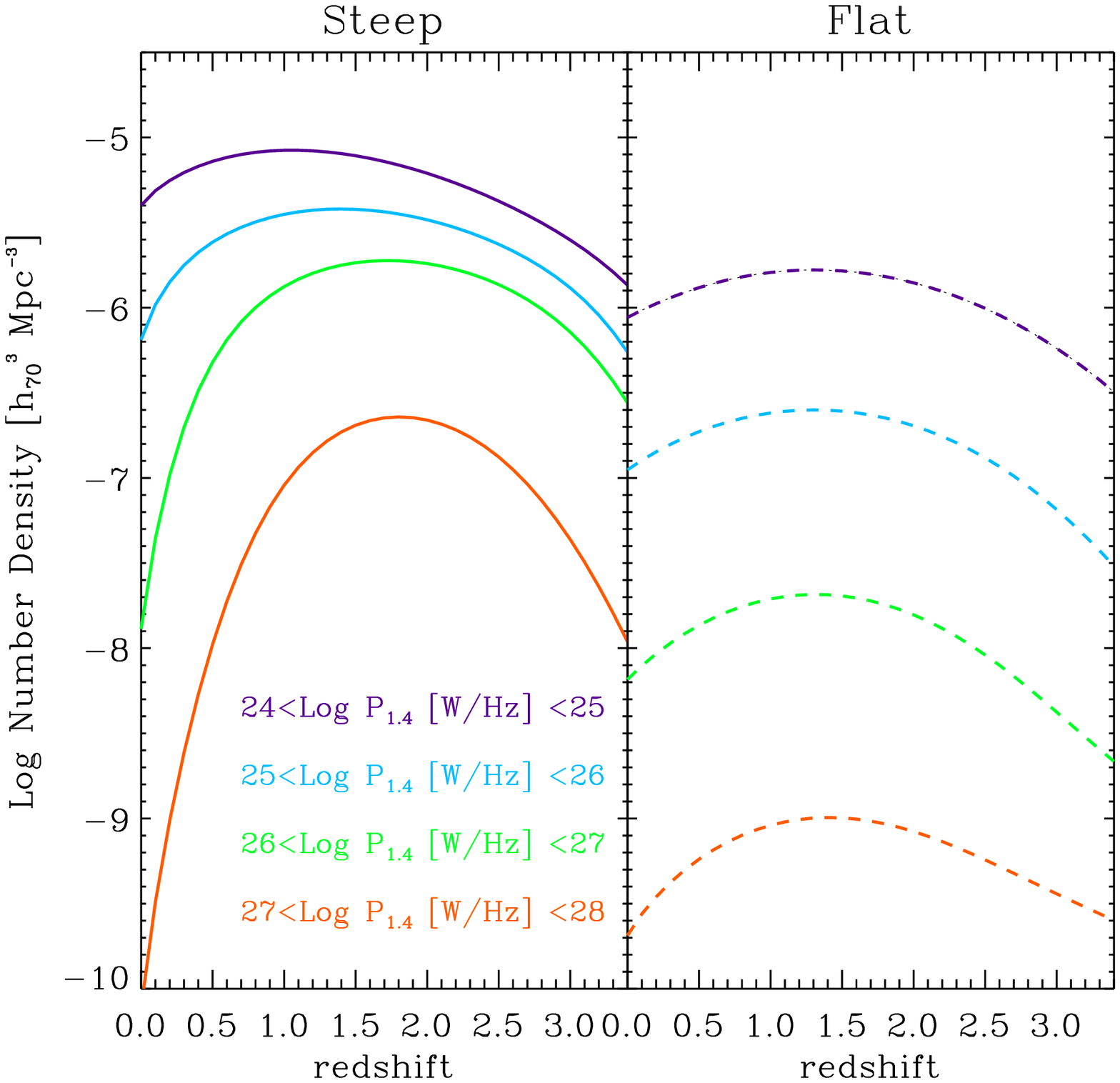}}
\resizebox{!}{0.48\textwidth}{\includegraphics{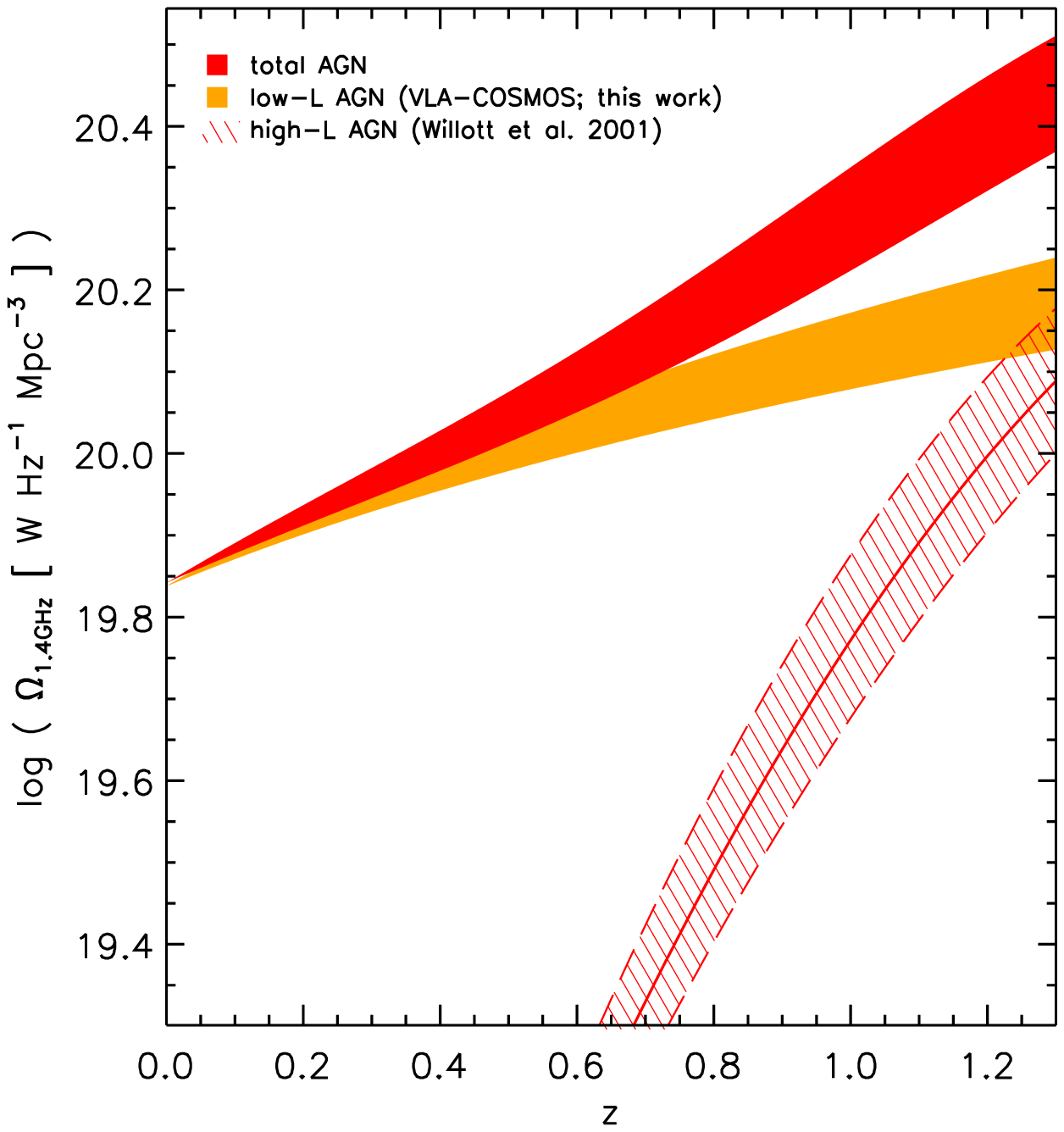}}
\end{center}
\caption{The radio view of AGN downsizing. {\it Left} Best fit number
  density evolution of radio sources of different power, taken from
  the models of \citet{massardi:10}, for steep and flat spectrum
  sources in the left and right panels, respectively. {\it Right:}
  Evolution of the comoving 20 cm integrated luminosity density for
  VLA-COSMOS AGN (orange curve) galaxies for $z < 1.3$.  Also shown is
  the evolution of the high-luminosity radio AGN, adopted from
  (\citealt{willott:01}, hatched region; the thick and dashed lines
  correspond to the mean, maximum and minimum results, respectively).
  The evolution for the total AGN population, obtained by co-adding
  the VLA-COSMOS and high luminosity AGN energy densities, is shown as
  the red-shaded curve (adopted from \citealt{smolcic:09}). }
\label{fig:smolcic}
\end{figure*}

Indeed, many early investigations of high redshift radio luminosity functions
(see, e.g., \citealt{danese:87}) demonstrated that neither PLE nor PDE
models could explain the observed evolution of radio sources, with
more powerful sources (often of FRII morphology) displaying a far more dramatic rise in
their number densities with increasing redshift (see also
\citealt{willott:01}).

Trying to assess the nature of radio AGN evolution across larger
redshift ranges requires a careful evaluation of radio spectral
properties of AGN.  Steeper synchrotron spectra are produced in the
extended lobes of radio jets, while flat spectra are usually
associated with compact cores.  For objects at distances such that no
radio morphological information is available, the combination of
observing frequency, K-corrections, intrinsic source variability and
orientation of the jet with respect to the line of sight may all
contribute to severe biases in the determination of the co-moving
number densities of sources, especially at high redshift
\citep{wall:05}.

In a very extensive and equally influential work \citet{dunlop:90}
studied the evolution of the luminosity functions of steep and flat
spectrum sources separately.  They showed that the overall redshift
evolution of the two classes of sources were similar, with steep
spectrum sources outnumbering flat ones by almost a factor of ten.
Uncertainties remained regarding the possibility of a high-redshift
decline of radio AGN number densities. The issue is still under
discussion, with the most clear evidence for such a decline observed
for flat-spectrum radio QSO at $z>3$ \citep{wall:05}, consistent with
the most recent findings of optical and X-ray surveys (see also
\S~\ref{sec:cosmogony} below).

Under the simplifying assumption that the overall radio AGN population
can be sub-divided into steep and flat spectrum sources, characterized
by a power-law synchrotron spectrum $S_{\nu} \propto \nu^{-\alpha}$,
with slope $\alpha_{\rm flat}=0.1$ and $\alpha_{\rm steep}=0.8$,
respectively, a redshift dependent luminosity function can be derived
for the two populations separately, by fitting simple models to a very
large and comprehensive set of data on multi-frequency source counts
and redshift distributions obtained by radio surveys at $\nu<5$ GHz
\citep{massardi:10}.  The comoving number densities in bins of
increasing radio power (at 1.4 GHz) from the resulting best fit
luminosity function models are shown in the left panel of
Figure~\ref{fig:smolcic}.

Radio AGN, both with steep and flat spectrum, show the distinctive
feature of a differential density evolution, with the most powerful
objects evolving more strongly towards higher redshift, a
phenomenological trend that, in the current cosmologist jargon, is
called ``downsizing''.

Recent radio observational campaigns of large multi-wavelength sky
surveys have also corroborated this view, by providing a much more
detailed picture of low luminosity radio AGN. For example, the work of
\citet{smolcic:09} on the COSMOS field showed that radio galaxies with
$L_{\rm 1.4 GHz} < few \times 10^{25}\,{\rm W Hz^{-1}}$ evolve up to
$z \simeq 1$, but much more mildly than their more luminous
counterparts, as shown in the right panel Figure~\ref{fig:smolcic}.

\subsubsection{X-ray surveys and the resolution of the X-ray
  background}
\label{sec:xray_rveys}

As already mentioned, active galactic nuclei are powerful X-ray
emitters.  The discovery of the intense cosmic X-ray background
radiation \citep[CXRB;][]{giacconi:62} in the early 1960's opened up a
privileged window for the study of the energetic phenomena associated
with accretion onto black holes.

Due to the relative weakness of X-ray emission from stars and stellar
remnants (magnetically active stars, cataclysmic variables and, more
importantly, X-ray binaries are the main stellar X-ray sources), the
X-ray sky is almost completely dominated by the evolving SMBH
population, at least down to the faintest fluxes probed by current
X-ray focusing telescopes.  The goal of reaching a complete census of
evolving AGN, and thus of the accretion power released by SMBH in the
history of the universe has therefore been intertwined with that of
fully resolving the CXRB into individual sources.  Accurate
determinations of the CXRB intensity and spectral shape, coupled with
the resolution of this radiation into individual sources, allow very
sensitive tests of how the AGN luminosity and obscuration evolve with
redshift.

\begin{figure*}
\centering
\psfig{figure=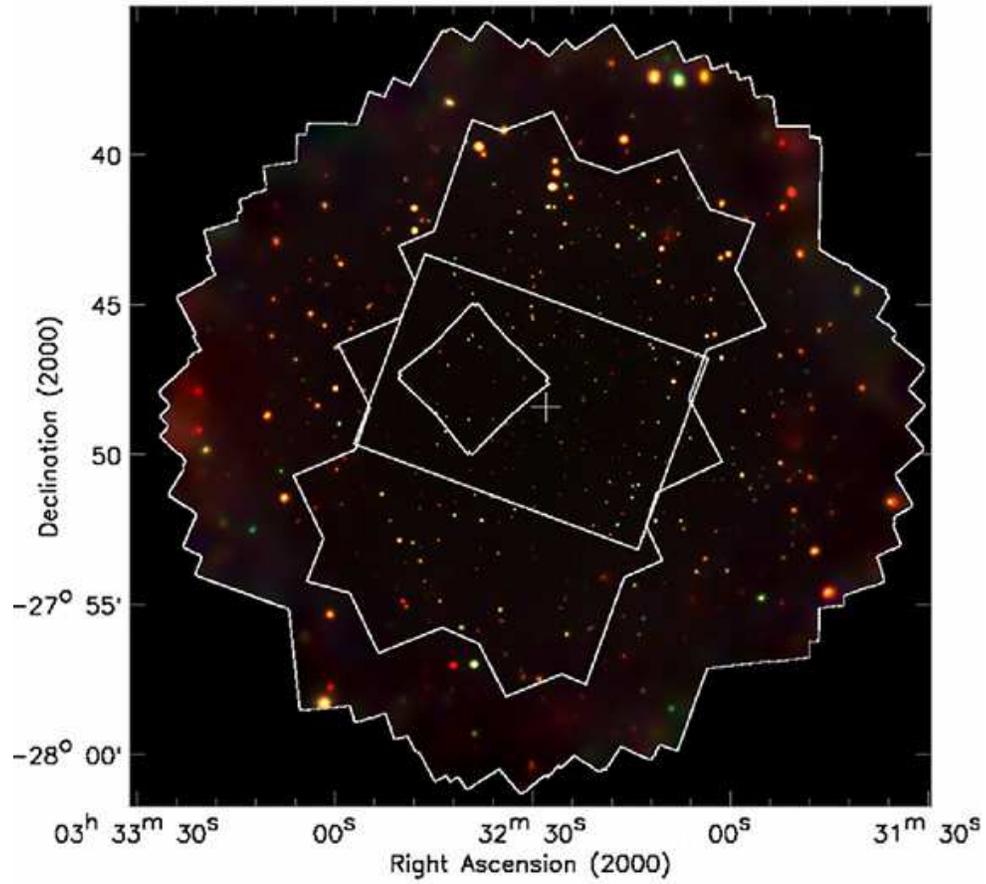,height=12.cm}
\caption{A smoothed, false color image of the deepest X-ray exposure
  to date, the $4\times 10^{6}$ second {\em Chandra} observation of
  the Chandra Deep Field South (CDFS). These observations resolve
  almost the entire CXRB radiation below $\sim$ 5 keV into individual
  sources, the vast majority of which are accreting supermassive black
  holes.  From \citet{xue:11}. }
\label{fig:cdfs_4ms}
\end{figure*}

\begin{figure*}
  \centering \psfig{figure=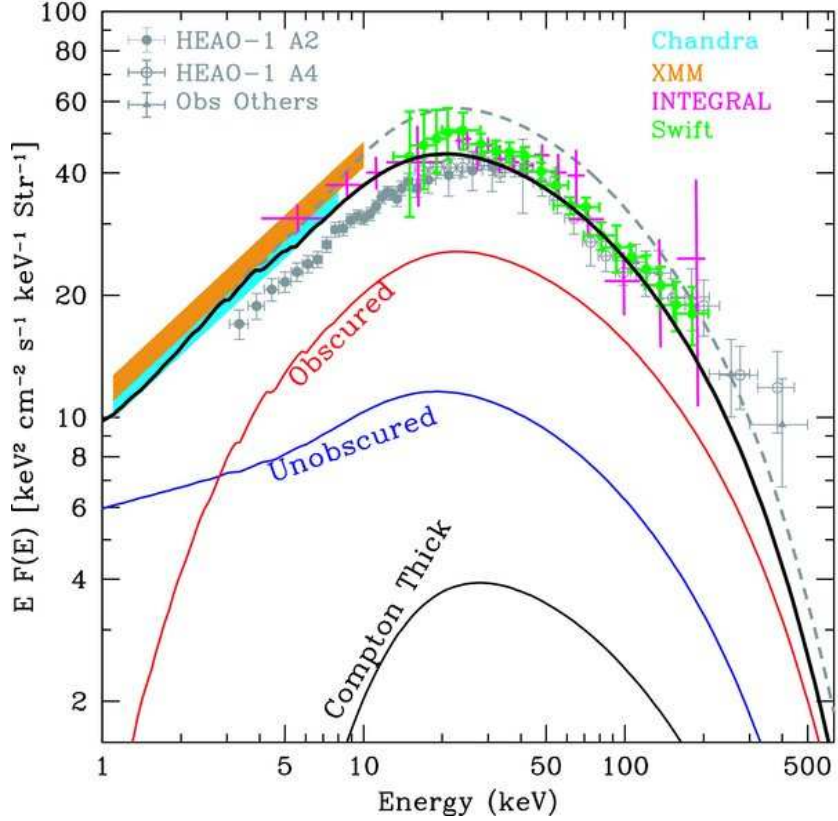,height=11.cm}
  \caption{Observed spectrum of the extragalactic CXRB from {\em HEAO
      1}, {\em Chandra}, {\em XMM-Newton}, {\em INTEGRAL}, and {\em
      Swift} data. The solid (red, blue, and black) thin lines show
    the contribution to this model from unobscured, obscured
    Compton-thin, and Compton Thick AGNs, respectively. The thick
    black solid and dashed grey lines are two different CXRB spectral
    models proposed in the literature depending on the assumed
    normalization of the {\em HEAO 1}; the main difference is the
    number of Compton Thick AGNs, which is reduced by a factor of 4 if
    the black solid line is assumed instead of the dashed one.
    Adopted from \citet{treister:09}, where a complete list of
    references of the data points shown is presented. }
\label{fig:cxrb}
\end{figure*}

New generations of synthesis models of the CXRB
\citep{gilli:07,treister:09} have quickly followed the publication of
increasingly larger and deeper surveys (for the current deepest view
of the X-ray sky, see Figure~\ref{fig:cdfs_4ms}).  
Figure~\ref{fig:cxrb} shows a recent compilation of the CXRB
measurements together with one incarnation of a synthesis model
\citep{treister:09} of AGN evolution that explains those data. The
hard slope of the background spectrum (well described by a power-law
with photon index $\Gamma_{\rm CXRB}\simeq 1.4$ at $E<10$ keV) and the
prominent peak at about 30 keV are accounted for by assuming that the
majority of active galactic nuclei are in fact obscured.

These new models
have progressively reduced the uncertainties in the absorbing column
density distribution.  When combined with the observed X-ray
luminosity functions, they provide an almost complete census of the
Compton thin AGN (i.e., those obscured by columns $N_{\rm
  H}<\sigma_{\rm T}^{-1} \simeq 1.5 \times 10^{24}$ cm$^{-2}$, where
$\sigma_{\rm T}$ is the Thomson cross section).  This class of objects
dominates the counts in the lower energy X-ray energy band, where
almost the entire CXRB radiation has been resolved into individual
sources \citep{worsley:05}.   It
should be noted, however, that at the peak energy of the observed CXRB
radiation only a small fraction (less than 5\%) of
the emission has so far been resolved into individual objects.

CXRB synthesis models, like the one shown in Figure~\ref{fig:cxrb}
ascribe a substantial fraction of this unresolved emission to heavily
obscured (Compton Thick) AGN. However, because of their faintness even
at hard X-ray energies, their redshift and luminosity distribution is
essentially unknown, and even their absolute contribution to the
overall CXRB sensitively depends on the quite uncertain normalization
of the unresolved emission at hard X-ray energies.  The quest for the
physical characterization of this ``missing'' AGN population, most
likely dominated by Compton thick AGN, represents one of the last
current frontiers of the study of AGN evolution at X-ray wavelengths.

\begin{figure*}
\centering
\psfig{figure=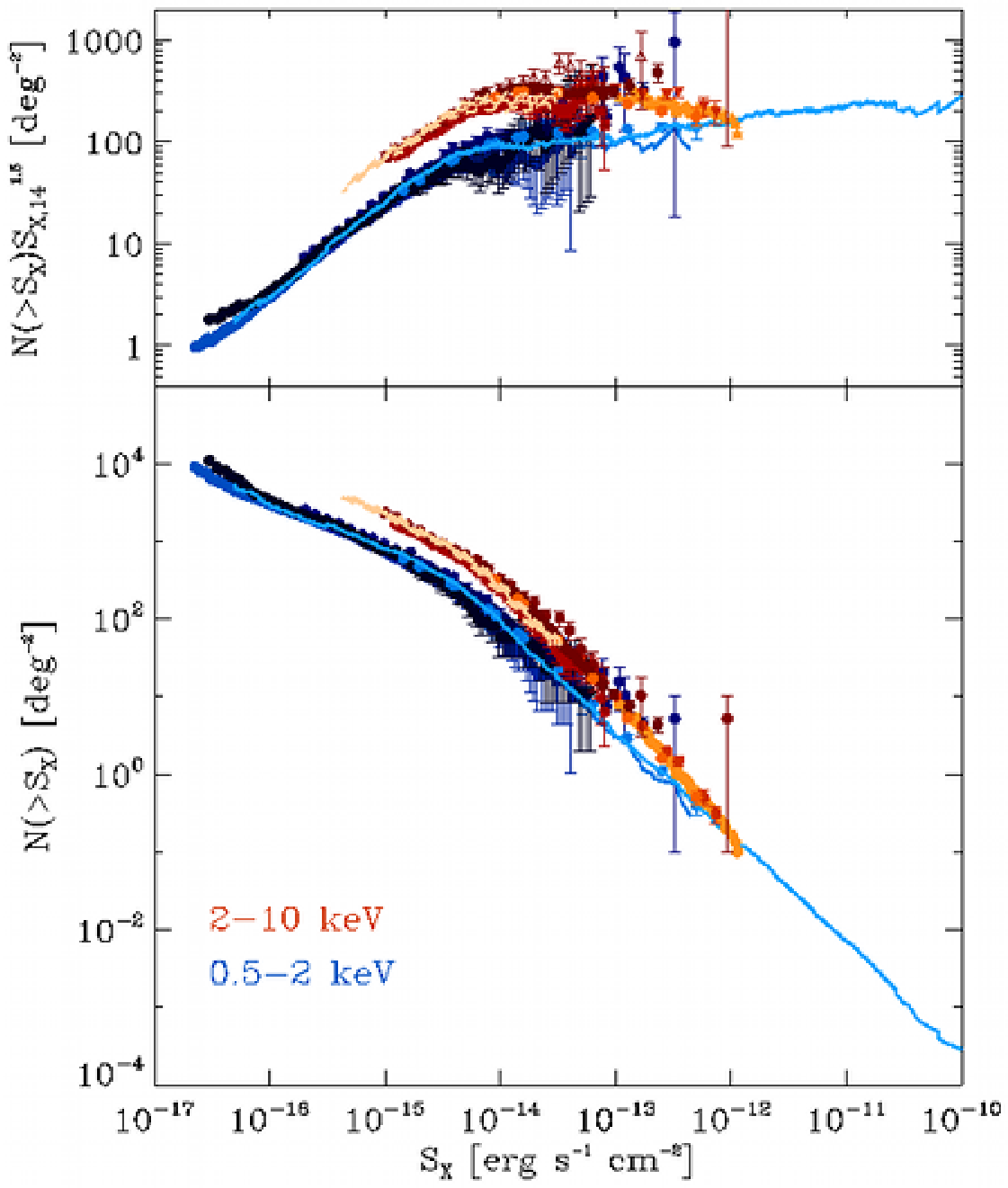,height=15.cm}
\caption{A compilation of cumulative X-ray source counts (number of
  objects brighter than a given flux per square degree) in the soft
  (0.5-2 keV, cyan-blue colors) and hard (2-10 keV, red-orange colors)
  bands. The observational data are taken from \citet{cappelluti:09}
  (see references therein) and \citet{mateos:08}.  The bottom panel
  shows the observed counts, while the top panel shows the counts
  after the Euclidean slope has been factored out.}
\label{fig:xray_counts}
\end{figure*}

Putting together the observational data from a large suite of
complementary surveys, Figure~\ref{fig:xray_counts} shows a
compilation of X-ray number counts, for both soft (0.5-2 keV) and hard
(2-10 keV) selected samples.

Given the steep frequency dependence of photo-electric absorption cross
sections, the harder the energy band probed the less affected by
obscuration the objects under study are.  Current technologies provide
the best compromise between telescope effective area and energy range
in the 2-10 keV band. Indeed, the density of AGN detected in this band
is higher than that of 0.5-2 keV selected ones, by at least a factor
of 3.

As for the radio $\log N$-$\log S$ of Fig.~\ref{fig:radio_counts}, the
counts become shallower than the Euclidean slope at intermediate
fluxes (about $10^{-14}$ ergs s$^{-1}$ cm$^{-2}$), where the largest
relative fraction of the CXRB is produced. Tentative evidence of a
steepening at the lowest fluxes might indicate the emergence of a
different, non-AGN, population of star-forming galaxies whose X-ray
emission is primarily due to stars and stellar remnants.

\begin{figure*}
\centering
\psfig{figure=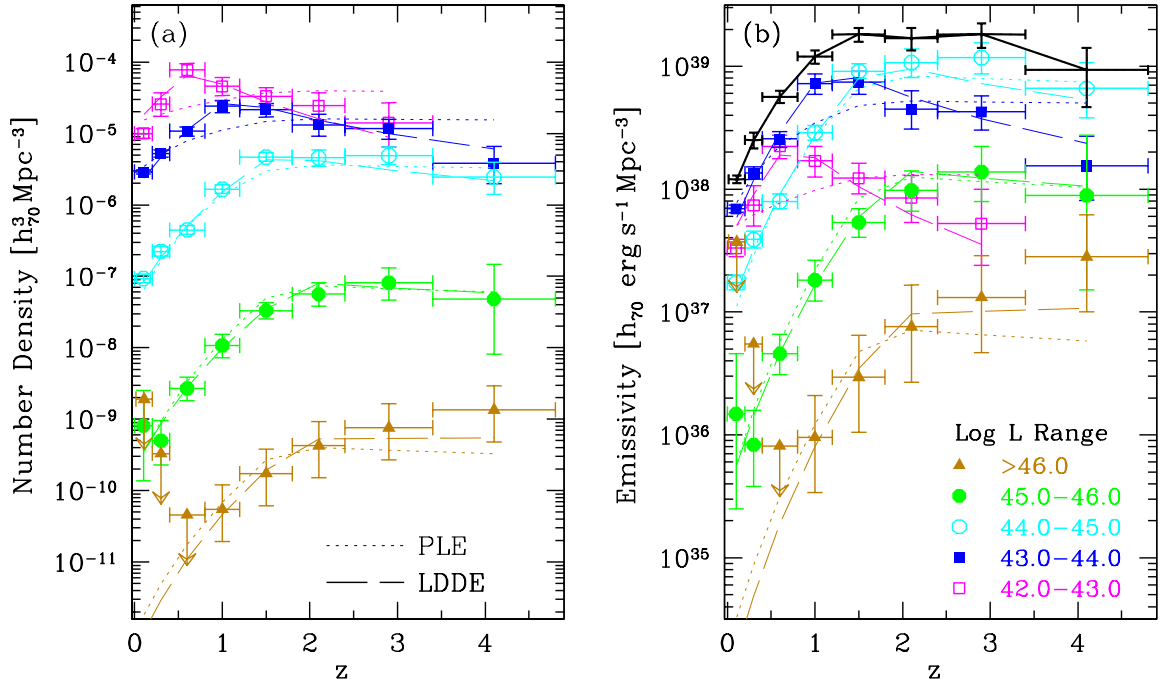,height=9.cm}
\caption{The soft X-ray view of AGN downsizing: (a) The space density
  of AGNs as a function of redshift in different luminosity classes
  and the sum over all luminosities with ${\rm log}\;L_{\rm x}\gtrsim
  42$. Densities from the PLE and LDDE models are over-plotted with
  solid lines. (b) The same as (a), except that the soft X-ray
  emissivities are plotted instead of number densities. The uppermost
  curve (black) shows the sum of emissivities in all luminosity
  classes in the plot.  From \citet{hasinger:05}.}
\label{fig:downsizing_xray}
\end{figure*}

The deepest surveys so far carried out in the soft X-ray energy range
(0.5-2 keV), supplemented by the painstaking work of optical
identification and redshift determination of the detected sources have
provided the most accurate description of the overall evolution of the
AGN luminosity function.  Neither PLE nor PDE provide a satisfactory
description of the X-ray LF evolution, with a good fit to the data achieved
with a ``Luminosity Dependent Density Evolution'' (LDDE) model, or
variations thereof.  In their influential work, \citet{hasinger:05}
unambiguously demonstrated that in the observed soft X-ray energy band
more luminous AGN peaked at higher redshift than lower luminosity ones
(see Fig.~\ref{fig:downsizing_xray}).

A good enough sampling of the luminosity redshift plane
necessary for accurate LF studies requires more extensive
observational efforts in the hard X-ray band, as obscured AGN are more
difficult to identify (and to obtain redshifts for) in the optical
band.  Nonetheless, the general ``downsizing'' trend illustrated by
the soft-X-ray selected AGN of Fig.~\ref{fig:downsizing_xray} has so far been
confirmed by almost all recent studies of (2-10 keV) X-ray selected
AGN \citep[see e.g.][]{ueda:03}.

\subsubsection{Optical and Infrared studies of QSOs}

Bright AGN emit a large fraction of their bolometric luminosity in the
optical/UV part of the spectrum (see Chapter ``Active Galactic
Nuclei'' by E. Perlman in this volume).  For Eddington ratios ($\lambda\equiv L_{\rm bol}/L_{\rm
  Edd}$, where $L_{\rm Edd}=4 \pi G M_{\rm BH} m_{\rm p} c /
\sigma_{\rm T} \simeq 1.3 \times 10^{38} (M_{\rm BH}/M_{\odot})$ ergs
s$^{-1}$ is the Eddington luminosity) larger than a few per-cent, the
AGN light out-shines the emission from the host galaxy, resulting in
point-like emission with peculiar blue colors.

Finding efficient ways to select QSO in large optical surveys, trying
to minimize contamination from stars, white dwarfs and brown dwarfs
has been a primary goal of optical astronomers since the realization
that QSO were extragalactic objects often lying at cosmological
distances \citep{schmidt:83,richards:06}.

\begin{figure*}
\centering
\psfig{figure=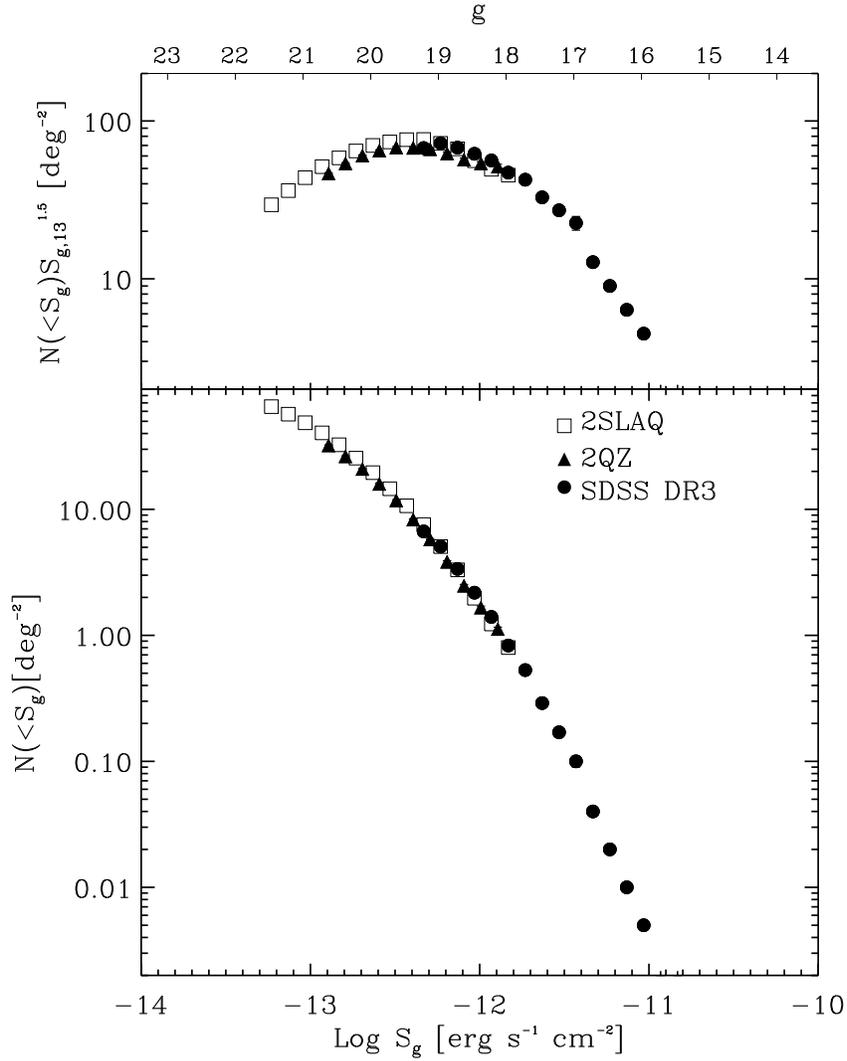,height=15.cm}
\caption{Compilation of (g-band, $\approx 4700 \AA$) quasar number
  counts from the largest recent optical surveys at $0.3<z<2.3$
  (2SLAQ, 2QZ and SDSS, courtesy of G.~Richards and S.~Croom).  g-band
  magnitudes have been converted into CGS flux units ($S_g$) with
  $\log S_g=-0.4(g-20)-12.64$.}
\label{fig:gband_counts}
\end{figure*}

Optical surveys remain an extremely powerful tool to uncover the
evolution of unobscured QSOs up to the highest redshift ($z\sim 6$).
In terms of sheer numbers, the known population of SMBH is dominated
by such optically selected AGN (e.g. more that $10^5$ QSOs have been
identified in the Sloan Digital Sky Survey), essentially due to the
yet unsurpassed capability of ground-based optical telescopes to perform
wide-field, deep surveys of the extra-galactic sky.

Figure~\ref{fig:gband_counts} shows a compilation of (g-band, $\approx
4700 \AA$) quasar number counts from the largest recent optical
surveys (2SLAQ, 2QZ, and SDSS), for objects in the redshift range
$0.3<z<2.2$.  The overall shape is similar to that of the radio AGN
counts, with a steep increase at bright fluxes, followed by a turnover
at around $g \simeq 19$.  A comparison with
Figures~\ref{fig:radio_counts} and \ref{fig:xray_counts} reveals,
however, that such large area QSO surveys reach depths corresponding
to a number density of sources in the sky more than one order of
magnitudes smaller that those probed by the deepest X-ray and radio
surveys.

Thus, the dominant AGN population eludes systematic detection in
optical surveys.  In general terms, the difficulty of optical QSO
surveys to probe deep into the AGN population is due to two major
effects: the first one is the already mentioned issue of 
{\it nuclear obscuration}, dramatically
affecting the UV/optical appearance of AGN; the second is {\it galaxy
  dilution} of the AGN light (and of the broad emission line signature
often used to select quasars). More specifically, let us consider an
AGN with B-band luminosity given by $L_{\rm AGN,B}=\lambda L_{\rm Edd}
f_{\rm B}$, with bolometric correction $f_{\rm B}\approx 0.1$
\citep{richards:06}.  Assuming a bulge-to-black hole mass ratio of
0.001 and a bulge-to-total galactic stellar mass ratio of $(B/T)$, the
contrast between nuclear AGN continuum and host galaxy blue light is
given by:

\begin{equation}
\label{eq:agn_contrast}
\frac{L_{\rm AGN,B}}{L_{\rm
    host,B}}=\frac{\lambda}{0.1}\frac{(M_*/L_{\rm B})_{\rm
    host}}{3(M_{\odot}/L_{\odot})}(B/T)
\end{equation} 

Thus, for typical mass-to-light ratios, the AGN will become
increasingly diluted by the host stellar light at Eddington ratios
$\lambda$ smaller than a few per cent.

High-spatial resolution observations of the numerically dominant
population of Low-Luminosity AGN (LLAGN; see the comprehensive review
of \citealt{ho:08}, and references therein) have so far only been
possible in the very local universe.  At higher redshift, the deepest
multi-wavelength AGN/galaxy surveys to date are starting to probe AGN
luminosities such that the contribution of the host galaxy to the
overall SED cannot be neglected. This compromises the efficiency and
``cleanness'' of AGN selection at optical/IR wavelength, but opens up
the possibility  of studying the connection between nuclear black hole
activity and host galaxy properties. We will come back more extensively
in section~\ref{sec:agn_gal} to the issue of the overall decomposition
of the AGN-galaxy spectral energy distribution in large
multiwavelength surveys.

\begin{figure*}
\centering
\begin{tabular}{cc}
\psfig{figure=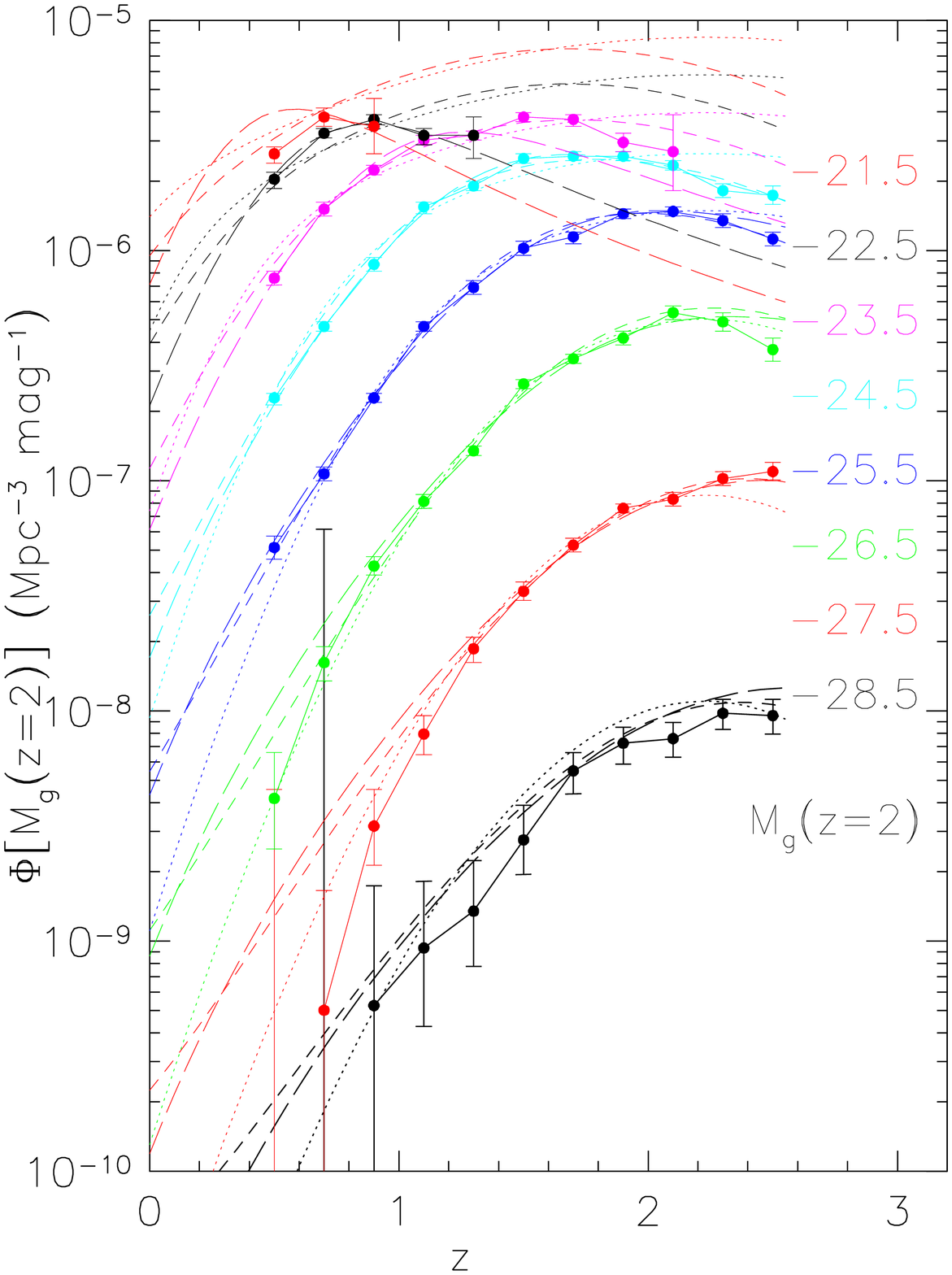,height=9.cm}&
\psfig{figure=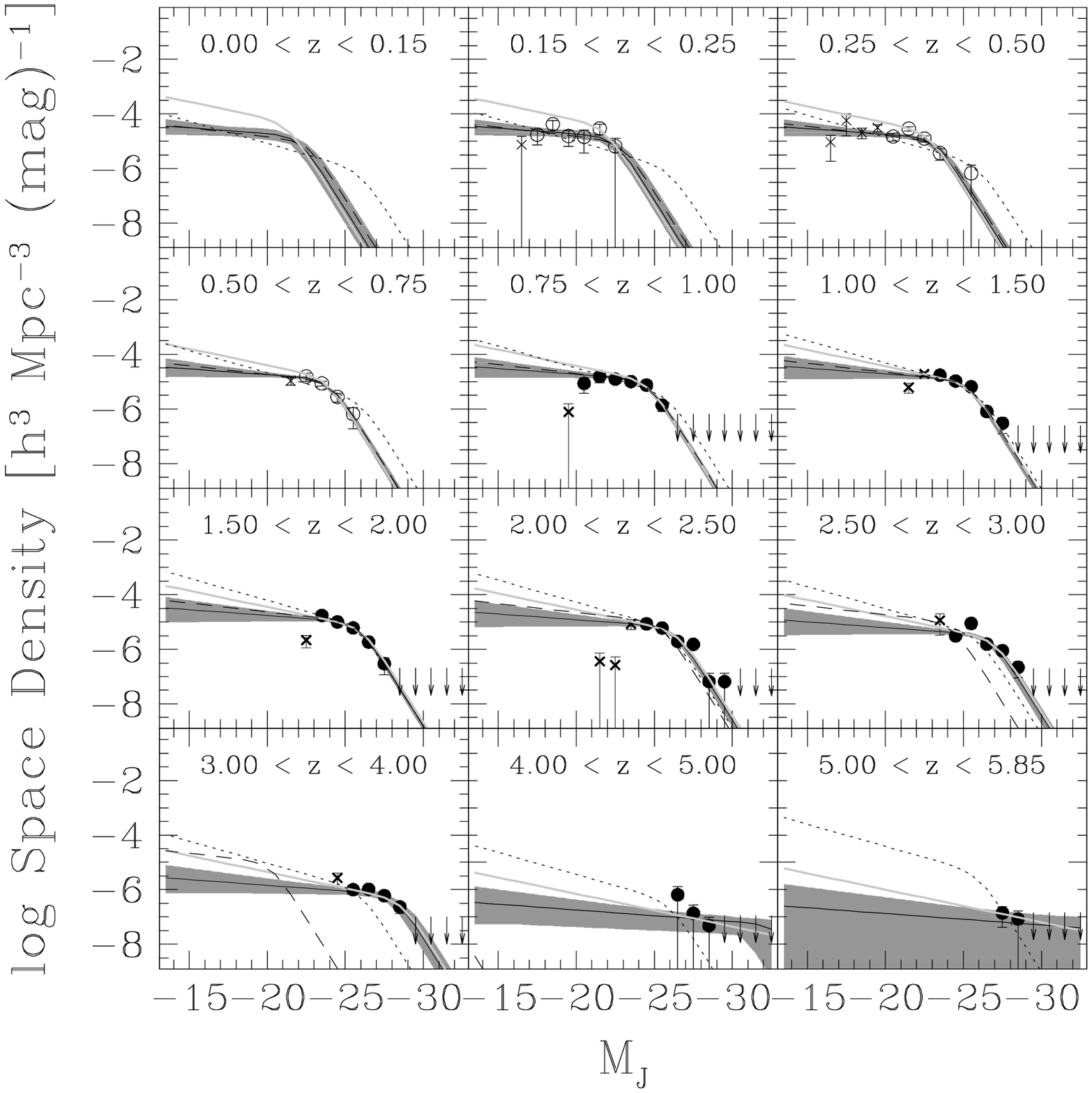,height=9cm}\\
\end{tabular}
\caption{{\it Left}: The combined 2SLAQ and SDSS optical QSO
  luminosity function plotted as a function of redshift for different
  absolute $g$ band magnitude intervals (the brightest at the bottom
  of the plot and the faintest at the top). The measured LF is
  compared to the best fit PLE model (dotted lines), smooth LDDE model
  (long dashed lines) and LADE (Luminosity And Density Evolution) 
  model (short dashed lines). Adopted
  from \citep{croom:09}; {\it Right:} J-band luminosity function of
  mid-IR-selected AGN for several redshift bins.  The crosses show
  points that were not used in the fits.  The best-fit LADE, PLE, and
  pure PDE models are shown by the solid, dashed, and
  dotted line, respectively, although only the LADE model is an
  acceptable fit to the data.  The shaded area shows the 2$\sigma$
  confidence region for the LADE fit.  For reference, the solid light
  gray line shows the best-fit LADE model to a sample from a combined
  IR/X-ray selection.  From
  \citet{assef:11}. \label{fig:downsizing_opt_ir} }
\end{figure*}

As for the general evolution of the optically selected QSO luminosity
function, it has been known for a long time that luminous QSOs were
much more common at high redshift ($z \sim 2$). Nevertheless, it is
only with the aid of the aforementioned large and deep surveys
covering a wide enough area of the distance-luminosity plane that it
was possible to put sensible constraints on the character of the
observed evolution.  The most recent attempts \citep{croom:09} have
shown unambiguously that optically selected AGN do not evolve
according to a simple PLE, but instead more luminous objects peaked in
their number densities at redshifts higher than lower luminosity
objects, as shown in the left panel of
Figure~\ref{fig:downsizing_opt_ir}.

We close this section with a brief discussion of the current status of
IR AGN LF studies.

According to the AGN unification paradigm, obscuration comes from
optically thick dust blocking the central engine along some lines of
sight.  The temperature in this structure, which can range up to 1000K
(the typical dust sublimation temperature), and the roughly isotropic
emission toward longer wavelengths should make both obscured and
unobscured AGNs very bright in the mid- to far-infrared bands.  This
spectral shift of absorbed light to the IR has allowed sensitive
mid-infrared observatories ({\it IRAS, ISO, Spitzer}) to deliver large
numbers of AGN \citep[see, e.g.][]{treister:06}.

Traditionally, the problem with IR studies of AGN evolution, however, 
lies neither in the {\it efficiency} with which growing supermassive black holes can
be found, nor with the {\it completeness} of the AGN selection, which
is clearly high and (almost) independent of nuclear obscuration, but
rather in the level of {\it contamination}.  IR counts are,
in fact, dominated by star forming galaxies at all fluxes.  This, and
the lack of clear spectral signatures in the nuclear, AGN-powered
emission in this band, implies that secure identification of AGN in
any IR-selected catalog often necessitates additional information from
other wavelengths, usually radio, X-rays, or optical spectroscopy.

Indeed, unlike the case of the CXRB, AGN contribute only a small
fraction (up to 2-10\%) of the cosmic IR background radiation
\citep{treister:06}, and similar fractions are estimated for the
contribution of AGN at the ``knee'' of the total IR luminosity
function at all redshifts.

Nonetheless, tremendous progress has been achieved in recent years,
thanks to more refined mid-IR color-color selection criteria
\citep{stern:05}
 which are little affected by contamination, and provide
reliable AGN samples, albeit with some well understood completeness
biases against AGNs that are faint with respect to their hosts, and
z~4.5 Type 1 AGNs.

Thus, deep surveys with extensive multi-wavelength coverage can also be
used to track the evolution of active galaxies in the mid-infrared
\citep[see, e.g.][]{assef:11}.  Strengthening similar conclusions
discussed above from other wavelengths, IR-selected AGN do not appear
to evolve following either the PLE or PDE parametrizations, but
require significant differences in the evolution of bright and faint
sources, with the number density of the former declining more steeply
with decreasing redshift than that of the latter (see the right panel
in Figure~\ref{fig:downsizing_opt_ir}).

%\subsubsection{Gamma-ray AGN and relativistic jets}

\begin{figure*}
\centering
\psfig{figure=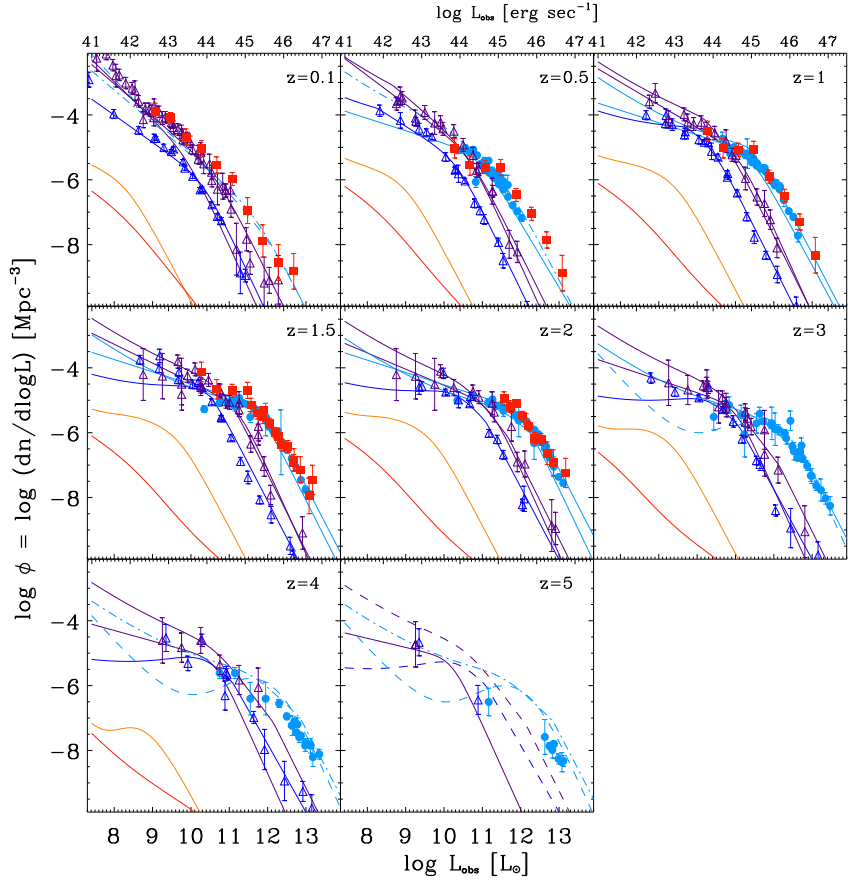,height=12.cm}
\caption{A compilation of luminosity functions observed in various
  energy bands.  The logarithm of the number of AGN per unit comoving
  volume and unit logarithm of luminosity is plotted as a function of
  the observed luminosity (in solar units). Observational points for
  IR (15$\mu$m; filled red 
  squares), B-band (filled blue circles), soft- (0.5-2 keV; empty blue
  triangles) and hard-X-rays (2-10 keV; empty purple triangles) are
  shown alongside published analytic fits for each band (solid lines
  in corresponding colors).
  The best fit radio luminosity functions of steep and flat spectrum
  sources from \citet{massardi:10} are also shown for comparison with
  orange and red thick lines, respectively.  The observed mismatch
  among the various luminosity functions in Fig.~\ref{fig:lf_obs_all}
  is due to a combination of different bolometric corrections and
  incompleteness due to obscuration.  Courtesy of P.~Hopkins}
\label{fig:lf_obs_all}
\end{figure*}

\subsubsection{Bolometric Luminosity functions}
\label{sec:bol_lf}

We have seen in the previous sections how a qualitatively consistent
picture of the main features of AGN evolution is emerging from the
largest surveys of the sky in various energy bands.  Strong (positive)
redshift evolution of the overall number density, as well as marked
differential evolution (with more luminous sources being more dominant
at higher redshift) characterize the evolution of AGN.

Fundamental constraints on the physical evolution of the
accretion-powered emission over cosmological times, like the ones we
will discuss later in \S\ref{sec:cosmology}, require, ideally, a good
knowledge of the {\it bolometric} luminosity function of AGN.  This,
in turn, demands a detailed assessment of selection biases and a
robust estimation of the AGN Spectral Energy Distribution (SED).

A thorough and detailed understanding of the AGN SED as a function of
luminosity (and, possibly, of redshift, but see
section~\ref{sec:sed_vs_z} above) could in principle allow us to
compare and cross-correlate the information on the AGN evolution
gathered in different bands.  As for the accuracy of our knowledge of
the bolometric correction, we refer the reader to the studies of
\citet{marconi:04,richards:06,hopkins:07}.  All of them consistently
demonstrate that a luminosity dependent bolometric correction is
required in order to match type I (unabsorbed) AGN luminosity
functions obtained by selecting objects in different bands.

\begin{figure*}
\centering
\begin{tabular}{cc}
  \psfig{figure=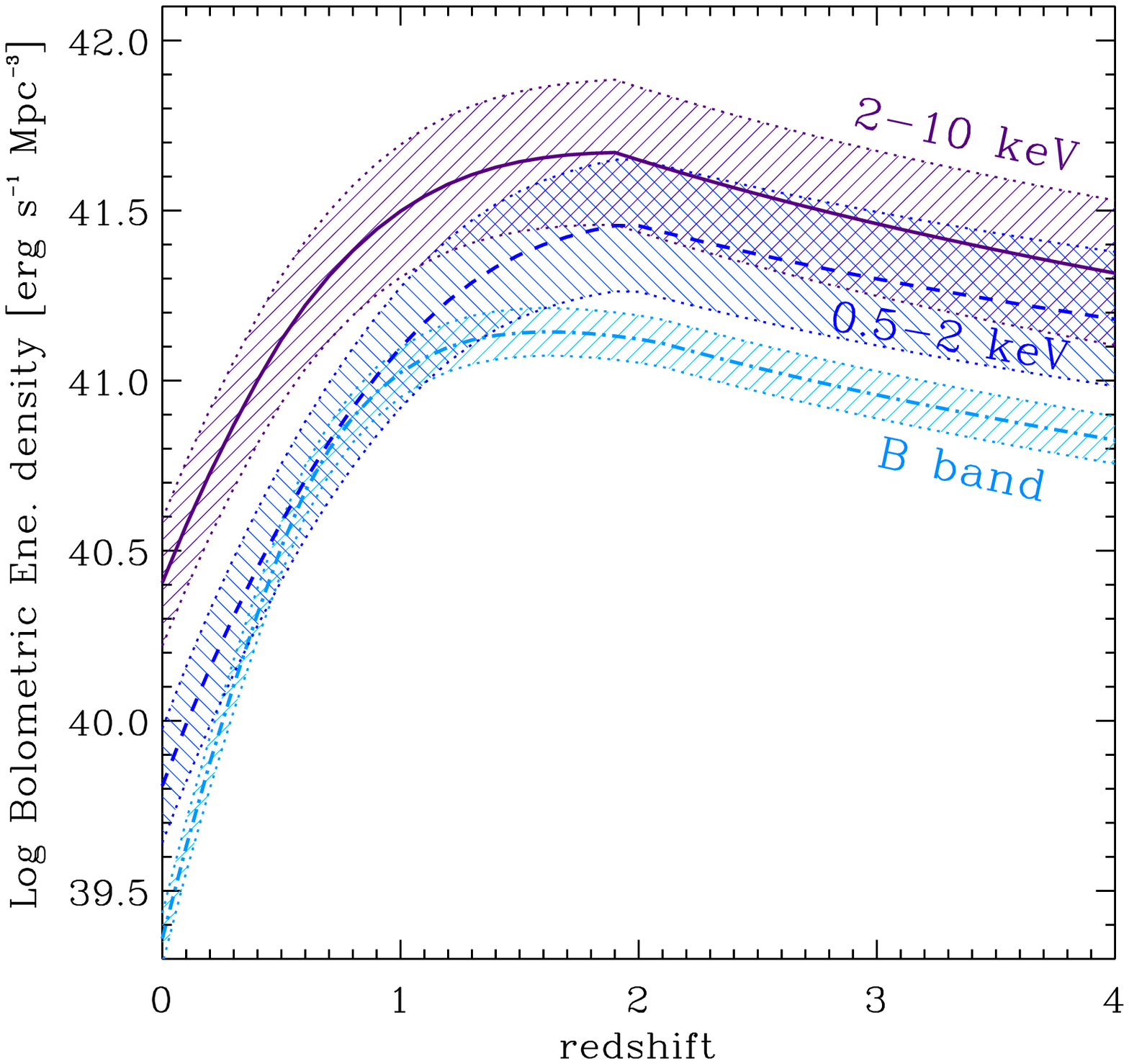,height=7.5cm} &
  \psfig{figure=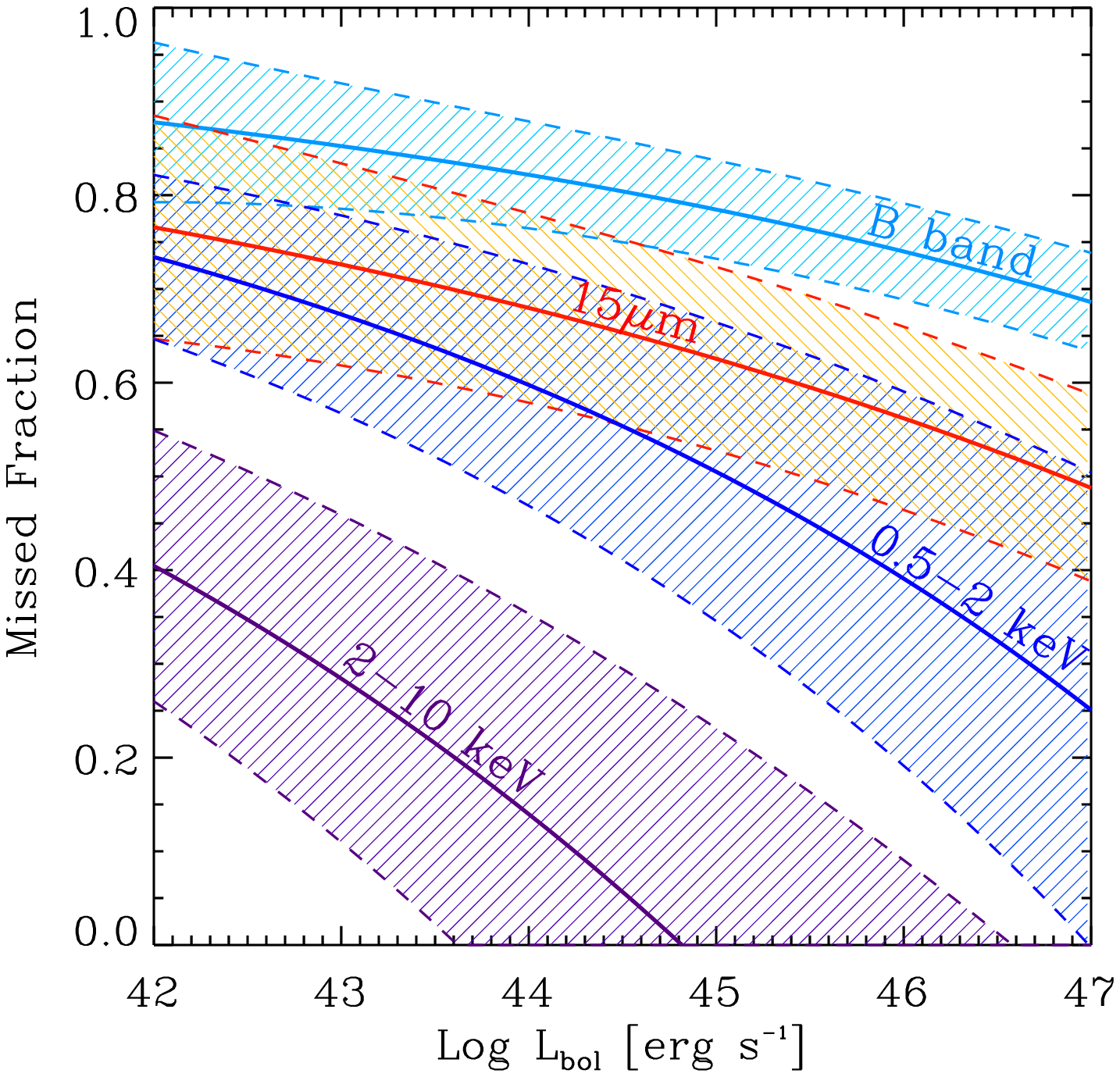,height=7.5cm} \\
\end{tabular}
\caption{{\it Left:} The redshift evolution of the bolometric energy
  density for AGN selected in different bands.  Bolometric corrections
  from \citet{hopkins:07} have been used, and the shaded areas
  represent the uncertainty coming from the bolometric corrections
  only.  {\it Right:} The fraction of AGN missed by observations in
  any specific band as a function of the intrinsic bolometric
  luminosity of the AGN.  Red, light blue, dark blue and purple shaded
  areas correspond to rest-frame mid-IR (15$\mu$m), UV (B-band), soft
  X-rays (0.5-2 keV) and hard X-rays (2-10 keV), respectively.  The
  uncertainty on the missed fractions depend on the uncertainties of
  the bolometric corrections and on the shape of the observed
  luminosity functions only.}
\label{fig:dens_bol}
\end{figure*}

\begin{figure*}
\centering
\psfig{figure=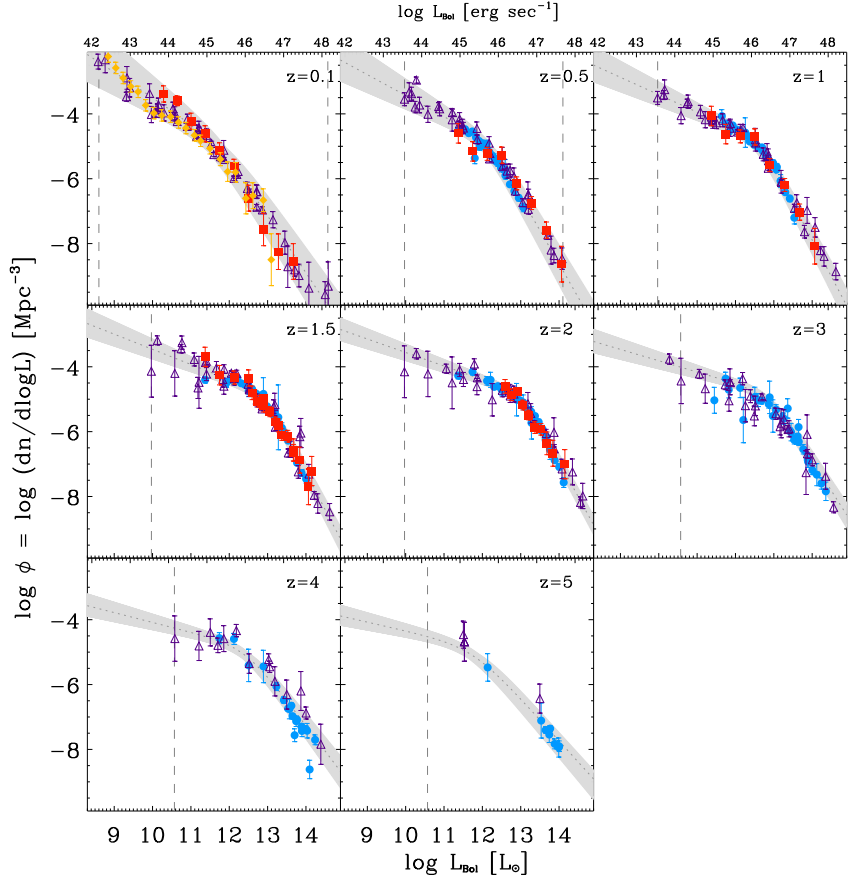,height=12.cm}
\caption{Bolometric AGN luminosity function (gray band) as a function
  of redshift, as calculated by \cite{hopkins:07}. The different
  symbols and colors refer to different bands from which data have
  been extracted: solid blue circles are optical data, filled red
  squares IR (at 15 $\mu$m), blue and purple triangle are soft and
  hard X-ray data, respectively, and the filled orange diamonds are
  luminosities from emission lines. The vertical dashed lines bracket
  the observational limits. We refer the reader to \cite{hopkins:07}
  for a more detailed description of the data and methodology used to
  extract the bolometric luminosity function. Data points courtesy of
  P. Hopkins}
\label{fig:lf_bol_all}
\end{figure*}

Summarizing the discussion of the previous sections,
Figure~\ref{fig:lf_obs_all} shows a compilation of luminosity
functions observed at various wavelengths.  The observed mismatch
among the various LF observed at all redshift 
is due to a combination of different
bolometric corrections and incompleteness due to obscuration.  In
fact, adopting a general form of luminosity-dependent bolometric
correction, and with a relatively simple parametrization of the effect
of the obscuration bias on the observed LF, \citet{hopkins:07} were
able to project the different observed luminosity functions in various
bands into a single bolometric one, $\phi(L_{\rm bol})$
(Figure~\ref{fig:lf_bol_all}).  As a corollary from such an exercise, 
we can then provide
a simple figure of merit for AGN selection in various bands by
measuring the bolometric energy density associated with AGN selected
in that particular band as a function of redshift. We show this in the
left panel of Figure~\ref{fig:dens_bol} for four specific bands (hard
X-rays, soft X-rays, UV, and mid-IR). From this, it is obvious that
the reduced incidence of absorption in the 2-10 keV band makes the 
hard X-ray surveys recover a higher fraction of the accretion power
generated in the universe than any other method.  

While optical QSO surveys miss more than three quarters of all AGN of
any given $L_{\rm bol}$, hard X-ray selection only fails to account
for about one third (up to 50\%) of all AGN, the most heavily obscured
(Compton Thick) ones, as shown in the right panel of
Figure~\ref{fig:dens_bol}.  A common feature apparent from such a figure is
that the effects of obscuration appear to be more severe at lower
intrinsic luminosities, an observational fact that has been discussed
previously in the context of X-ray surveys of AGN (see
section~\ref{sec:sed_vs_z} above).  It is important to note that the
high missed fraction for mid-IR selected AGN is a direct consequence
of the need for (usually optical) AGN identification of the IR
sources, so that optically obscured active nuclei are by and large
missing in the IR AGN luminosity functions considered here.

\begin{figure*}
\centering
\psfig{figure=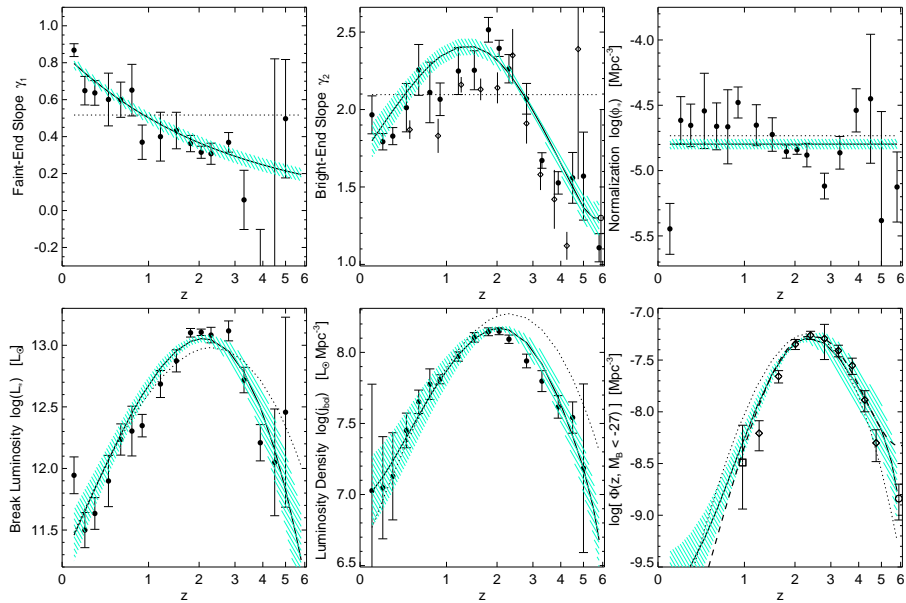,height=8.cm}
\caption{Best-fit AGN bolometric LF double power-law parameters as a
  function of redshift. Symbols show the best-fit values to data at
  each redshift, dotted lines the best-fit PLE model, and solid lines
  the best-fit full model (a luminosity and density evolution one).
  Although PLE is appropriate for a lowest order fit, both the bright-
  and faint-end slopes evolve with redshift to high significance. The
  bottom right panel shows the predicted number density of bright
  optical quasars from the full fit (solid line), compared to that
  observed ones.  From \citet{hopkins:07}.}
\label{fig:lf_bol_param}
\end{figure*}

Figure~\ref{fig:lf_bol_param} shows the evolution of the parameters of
the analytic fit to the bolometric LF data.  They encompass our global
knowledge of the evolution of accretion power onto nuclear black holes
throughout the history of the universe. The three bottom panels reveal
the overall increase in AGN activity with redshift, up to $z\approx
2$, and the mirroring high-redshift decline. At the center, the total
integrated luminosity density evolution mark the epochs of rapid
build-up of the SMBH mass density. On the lower left, the evolution in
the break luminosity $L_{\rm bol,*}$ indicates that the ``typical''
accreting black holes was significantly more luminous at $z\approx 2$
than now, a different way of looking at AGN ``downsizing''. This is
accompanied by a progressive steepening of the faint end slope of the
LF (upper left panel): low-luminosity AGN become more and more
dominant in the overall number density of AGN as time progresses.\

Such a detailed view of the evolution of active galactic nuclei, with
its distinctive signatures of ``downsizing'', has lent additional
support to the notion that the lives of growing black holes must be
intimately linked to those of their host galaxies.  Indeed, both
galaxies and black holes show signs of a similar differential
evolution.  The very term ``downsizing'' was first used by
\citet{cowie:96} to describe the finding that actively star-forming
galaxies at low redshift have smaller masses than actively
star-forming galaxies at $z \sim 1$.  It has come to identify, in the
current cosmology jargon, a variety of possibly distinct phenomena,
not just related to the epoch of star formation, but also to that of
star formation quenching, or galaxy assembly (see Chapter
``Galaxies in the Cosmological Context'' by G. De Lucia in this same volume).  Our current understanding
of AGN evolution, encapsulated in the observable evolution of their
bolometric luminosity function, emphatically suggests that growing nuclear black
holes take part in this global process of structure formation.

\subsection{AGN Clustering and the Large scale structure of the
  universe}
\label{sec:clustering}

The bolometric luminosity function of AGN provides the basic tool to
describe the differential evolution of growing black holes.  Some key
properties of AGN, however, remain impossible to determine on the sole
basis of the observed LF.  As it is increasingly difficult to measure
black hole masses at high redshift, and currently only possible for
bright, un-obscured broad line QSOs, we do not
have robust, direct, observational constraints on the distribution of
AGN Eddington ratios beyond the local universe.  The Eddington ratio
distribution, in turn, depends on the details of the average AGN
lightcurves which could reveal important details of the physical
processes driving the accreting gas towards the black hole.

The spatial distribution of QSOs (clustering) in the sky could provide
such an alternative method to estimate lifetimes
(\citealt{martini:01}; see also Chapter ``Large scale structure of
the universe'' by A. Coil in this volume).  In the current $\Lambda$CDM paradigm for structure
formation, more clustered objects are rarer and live in more massive
dark matter structures (or ``halos'').  Thus, if AGN are strongly
clustered, their hosts must be rare objects, too, and the effective
AGN lifetime must be long, in order for such a rare ``parent''
population to account for the total AGN luminosity density observed.
If, on the other hand, their clustering is comparable to the
clustering of smaller, less massive, dark matter halos, their host are
more common and their luminous phases must therefore have short
duration.
 
A commonly used technique for measuring the spatial clustering of a
class of objects is the two-point correlation function $\xi(r)$, which
measures the excess probability $dP$ above a random distribution of
finding an object in a volume element $dV$ at a distance $r$ from
another randomly chosen object:
\begin{equation}
  \label{eq:corr_func}
  dP = n[1 + \xi(r)]dV \, 
\end{equation}
where $n$ is the mean number density of objects.  In the scale range
between a few tens of kpc and a few tens of Mpc, for most classes of
astronomical objects $\xi(r)$ can be described by a single power-law:
\begin{equation}
  \label{eq:clustering}
  \xi(r)=\left(\frac{r}{r_0}\right)^{-\gamma} \, 
\end{equation}
where $r_0$ is the {\it correlation, or clustering, length}, defined as the scale at
which the two-point correlation function is equal to unity.  

Unfortunately, a direct comparison of the measured clustering length
of AGN with that expected for dark matter halos of different masses is
hampered by the fact that, according to current theories of structure
formation, galaxies (and their nuclear black holes) do not follow the
distribution of the underlying matter, but form in the high-density
peaks of the dark matter field.  The {\it bias} of any astrophysical
population $X$ is defined as the (square root of the) ratio between
the two-point autocorrelation functions of population $X$ and of the
dark matter (DM) halos: $b_{\rm X, DM}(r)\equiv \sqrt{\xi_{\rm
    X}(r)/\xi_{\rm DM}(r)}$.

Many groups have now been able to measure the clustering of AGN at
different luminosities, bands, scales and redshifts (see, e.g.,~the
recent review of \citealt{shankar:09} for a complete list of
references).  Overall, the clustering length of quasars appears to be
an increasing function of redshift, but does not depend strongly on
luminosity.

\begin{figure*}
\centering
\psfig{figure=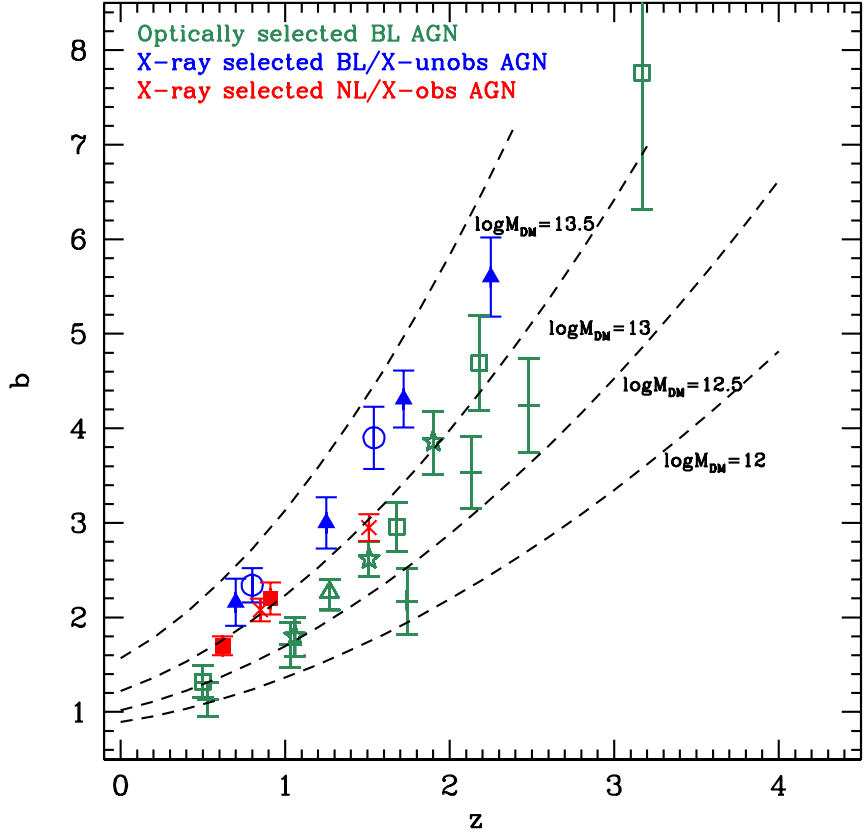,height=12.cm}
\caption{Bias parameter as a function of redshift for various AGN
  surveys. The range of the bolometric luminosity probed is given in
  parenthesis: (i) optically selected BL AGN: green-crosses (2dF;
  $45.3<\log L_{\rm bol}<46.7$), green-stars (2dF; $45.5<\log L_{\rm
    bol}<47.4$), green-open squares (SDSS; $45.6<\log L_{\rm
    bol}<46.9$) and green-open triangles (SDSS; $\log L_{\rm bol}\sim
  46.5$); (ii) X-ray selected unobscured AGN: blue triangles and blue
  open-circles (XMM-COSMOS; $45<\log L_{\rm bol}<45.7$) and (iii)
  X-ray selected obscured AGN: red squares and red crosses
  (XMM-COSMOS; $44.1<\log L_{\rm bol}<44.6$).  See \citet{allevato:11}
  for a similar figure and the full list of references for the
  observed data points.  The dashed lines show the expected bias
  evolution of typical DM halo masses.}
\label{fig:agn_bias}
\end{figure*}

As shown in Fig.~\ref{fig:agn_bias}, the bias of optically selected
(broad line) AGN increases with redshift following an evolution at
approximately constant dark matter halo mass (since halos of a fixed
mass are progressively more clustered towards higher redshift), in the
range $\log M_{\rm DM} \simeq 12.5 \div 13$h$^{-1}M_{\odot}$ at
redshifts $z < 3.5$.

Instead, X-ray selected objects \citep[]{allevato:11}, both obscured
and unobscured, reside in 
more massive DM structures at all redshifts $z < 2.25$, with a typical
mass of the hosting halos constant over time in the range $log M_{\rm
  DM} \simeq 13 \div 13.5$h$^{-1}M_{\odot}$.

By combining the number density of AGN with that of the hosting dark
matter halos, one can estimate an AGN duty cycle, and a corresponding
average lifetime.  The observed biases of the rare, luminous broad
lined quasars imply timescales of the order of $10^7 \div 10^8$ years,
increasing with redshift, as the massive halos typically hosting AGN
become increasingly rare.

For X-ray selected AGN, a larger duty cycle is inferred, which
translates into an AGN lifetime of $\sim 0.1 \div 1$ Gyr, about
 one order of magnitude longer than that estimated for
optically bright QSOs at the same redshift. This is mainly due to the
higher number density and higher bias of AGN found in X-ray selected
samples.

Numerical simulations of merger-induced AGN activity in a cosmological
context have shown \citep{bonoli:09} that the clustering of optically
selected quasars is well explained by a model in which these objects
are triggered by major merger events.  The difference between
optically selected and the (lower-luminosity) X-ray selected AGN of
Figure~\ref{fig:agn_bias} might suggest that X-ray selected AGN are
triggered by different (secular) processes which may be capable of
fueling luminous AGN in the gas-rich environment of star-forming
galaxies at high redshift.  The same models also predict an increase
in AGN duty cycle for the brightest quasars at halo masses larger than
$10^{12} M_{\odot}$, but fail to reproduce the large biases for less
luminous X-ray selected AGN at $z\sim 1$, possibly pointing (again)
towards the need for a larger variety of AGN triggering mechanisms for
this class of objects.

\section{Cosmology I: The growth of supermassive black holes in
  galaxies}
\label{sec:cosmology}

As we have discussed in the previous section, the strong cosmological
evolution of the quasar population was recognized early on by
observers in essentially all bands of the electromagnetic spectrum.
In the early 1990s, deep optical surveys of star-forming galaxies
began to probe the cosmological evolution of the rate at which stars
are formed within galaxies, thus providing robust constraints for
models of galaxy formation and evolution (the so-called Lilly-Madau
plot; \citealt{madau:96}).  It was soon clear that QSOs luminosity
density and Star Formation Rate (SFR) density evolved in similar
fashion, being much higher in the past, with a possible (very broad)
peak at $z\approx 2$ \citep{boyle:98}.

In the previous section we have traced the history of the study of AGN
luminosity functions in various spectral bands, closing with an
assessment of our current understanding of the bolometric luminosity
function evolution.  A reliable census of the bolometric energy output
of growing supermassive black holes (see, e.g., the central bottom
panel of Figure~\ref{fig:lf_bol_param}) allows a more direct estimate
of the global rate of mass assembly in AGN, and an interesting
comparison with that of stars in galaxies.  Together with the tighter
constraints on the ``relic'' SMBH mass density in the local universe,
$\rho_{\rm BH,0}$, provided by careful application of the scaling
relations between black hole masses and host spheroids, this enables
meaningful tests of the classical 'Soltan argument' \citep{soltan:82},
according to which the local mass budget of black holes in galactic
nuclei should be accounted for by integrating the overall energy
density released by AGN, with an appropriate mass-to-energy conversion
efficiency.

Many authors have carried out such a calculation, either using the
CXRB as a ``bolometer'' to derive the total energy density released by
the accretion process \citep{fabian:99}, or by considering evolving
AGN luminosity functions \citep{yu:02,marconi:04,merloni:08}.  Despite
some tension among the published results that can be traced back to
the particular choice of AGN LF and/or scaling relation assumed to
derive the local mass density, it is fair to say that this approach
represents a major success of the standard paradigm of accreting black
holes as AGN power-sources, as the radiative efficiencies needed to
explain the relic population are within the range $\approx 0.06 \div
0.20$, predicted by standard relativistic accretion disc theory
\citep{novikov:73}.

In this section, we begin with a schematic account of the current
constraints on the black hole mass density growth, and discuss some
recent attempts to compare it on a quantitative level with the
observed growth of the galaxy population.  This will be followed by
the (related) discussion of the possible evolution of the scaling
relations.

\subsection{A global view of the accretion history of the universe}
\label{sec:global_growth}

Under the standard assumption that black holes grow mainly by
accretion, their cosmic evolution can be calculated from the
bolometric luminosity function of AGN $\phi(L_{\rm bol},z)$, where
$L_{\rm bol}=\epsilon_{\rm rad} \dot M c^2$ is the bolometric
luminosity produced by a SMBH accreting at a rate of $\dot M$ with a
{\it radiative} efficiency $\epsilon_{\rm rad}$.  The non-negligible
fraction of the AGN population which is unaccounted for in current
surveys, the so-called Compton thick AGN (see \S\ref{sec:xray_rveys}
above), is usually included in the bolometric luminosity function 
by assuming a redshift-invariant column density
distribution as measured in the very local universe and an overall
number density of heavily obscured AGN that fits the CXRB.

The total, integrated mass density in supermassive black holes can
then be computed as a function of redshift:
\begin{equation}
  \label{mer_eq:rhobh_z}
  \frac{\rho_{\rm BH}(z)}{\rho_{\rm BH,0}} =
  1-\int_0^{z}\frac{\Psi_{\rm BH}(z')}{\rho_{\rm
      BH,0}}\frac{dt}{dz'}dz',
\end{equation}
where the black hole accretion rate (BHAR) density is given by:
\begin{equation}
\label{mer_eq:bhar}
\Psi_{\rm BH}(z)=\int_0^{\infty} \frac{(1-\epsilon_{\rm rad})L_{\rm
    bol}}{\epsilon_{\rm rad} c^2}\phi(L_{\rm bol},z)dL_{\rm bol}
\end{equation}
and
\begin{equation}
  \frac{dt}{dz}=-\left[(1+z)H_0\sqrt{(1+z)^3 \Omega_m +
      \Omega_{\Lambda}}\right]^{-1} 
\end{equation}
The exact shape of $\rho_{\rm BH}(z)$ and $\Psi_{\rm BH}(z)$ then
depends only on the local black hole mass density $\rho_{\rm BH,0}$
and the (average) radiative efficiency $\epsilon_{\rm rad}$.

\begin{figure*}
\centering
\begin{tabular}{cc}
\psfig{figure=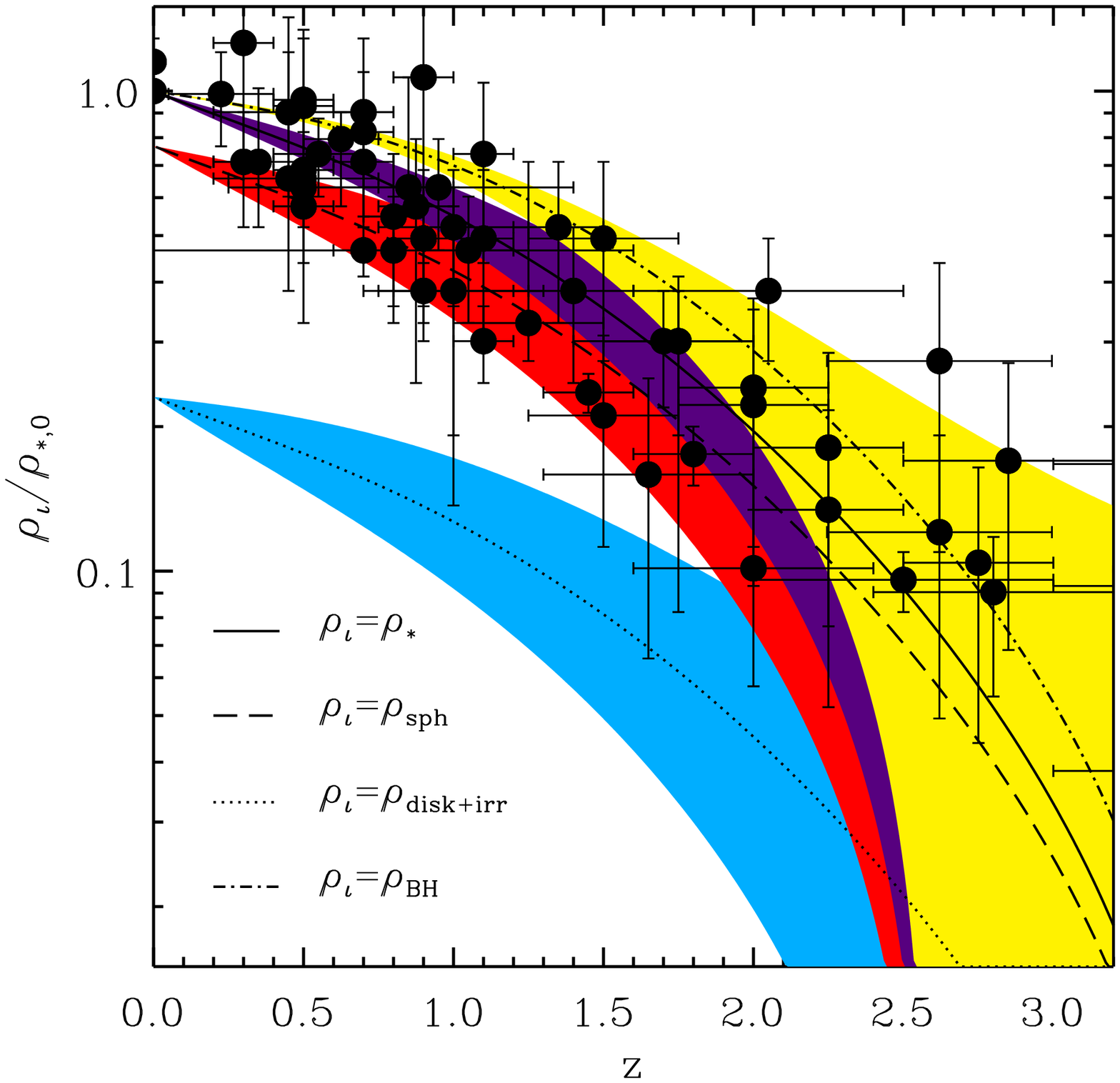,height=7.8cm}&
\psfig{figure=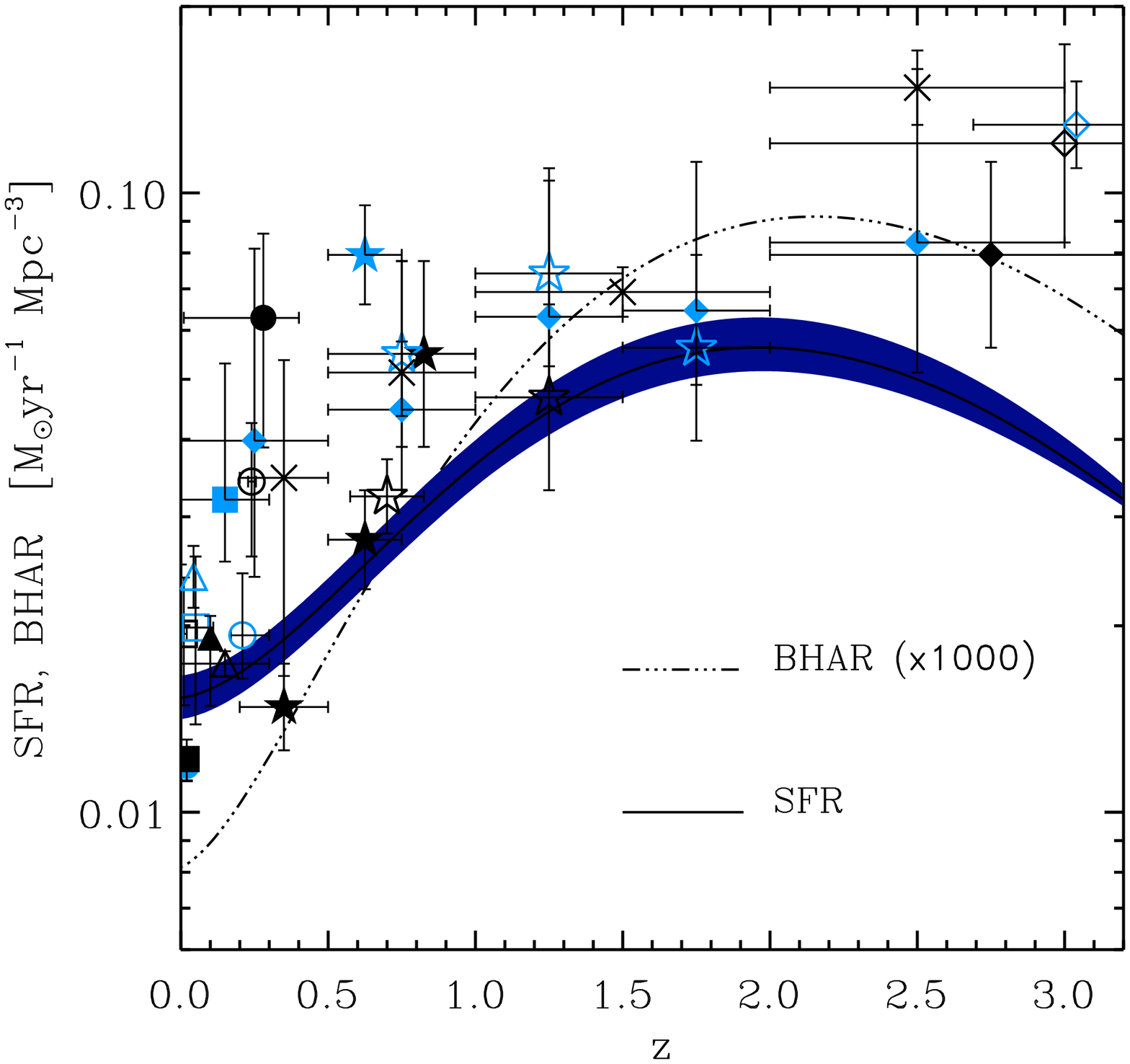,height=7.4cm}\\
\end{tabular}
\caption{{\it Left}: Evolution of the stellar mass density as a
  function of redshift (black points, observations), where the density
  is given as a ratio to the local value, $\rho_{*,0} = 5.6 \times
  10^8 M_{\odot}$ Mpc$^{-3}$ \citep{cole:01}.  Shaded areas represent
  1-sigma confidence intervals of the model fits.  Solid black line
  with purple shaded area shows the best joint fit from
  eq.~(\ref{mer_eq:rhostar}) and eq.~(\ref{eq:sfr}) to both stellar
  mass and SFR density points (left and right panels), while the
  dot-dashed line with yellow shaded area marks the normalized
  evolution of the SMBH mass density only.  The slight offset between
  the two is compensated by a change in the normalization of the
  average black hole to spheroid mass ratio with redshift (see text
  for details).  The dashed (dotted) line with red (blue) shaded area
  shows the relative growth of the mass density in spheroids (discs).
  Values of $\lambda_0=0.3$ \citep{fukugita:04} and $\rho_{\rm
    BH,0}=4.2 \times 10^5 M_{\odot}$ Mpc$^{-3}$ \citep{shankar:09} are
  adopted here.  {\it Right}: The corresponding best-fit relation for
  the SFR density evolution, from eq.~(\ref{eq:sfr}) is shown with a
  solid line and dark blue shaded area.  The dash-triple-dotted line
  is (1000 times) the black hole accretion rate density $\Psi_{\rm
    BH}(z)$ (BHAR). It appears that the BHAR declines slightly faster
  than the SFR, another way to emphasize the need of an evolution in
  the average $M_{\rm BH}/M_{\rm sph}$ ratio.}
\label{fig:rhobh}
\end{figure*}

We can then link the growth of SMBH from eq.~(\ref{mer_eq:rhobh_z}) to
the growth of stellar mass in galaxies.  To do so, we will use the
\citet{hopkins:07} bolometric LF of AGN (see section~\ref{sec:bol_lf}
above)\footnote{As discussed in \citet{marconi:04}, in order to
  correctly estimate the total bolometric output of an AGN, care
  should be taken in avoiding double counting of the IR reprocessed
  emission.  This appears not have been done in \citet{hopkins:07}, so
  we correct the bolometric luminosities by 30\% to account for this.}.
Because local SMBH are observed to correlate with spheroids only, we
introduce the parameter $\lambda(z)$, the ratio of the mass in disks
and irregulars to that in spheroids at any redshift, so that the total
stellar mass density can be expressed as: $\rho_{*}(z)=\rho_{\rm
  sph}(z)+\rho_{\rm disk+irr}(z)=\rho_{\rm sph}(z)[1+\lambda(z)]$.

We can now assume that $\lambda(z)$ evolves according to $\lambda(z) =
\lambda_0 (1+z)^{-\beta}$, where $\lambda_0$ is the value of the disk
to spheroid mass density ratio in the local universe.  Also, we assume
that the mass density of spheroids and supermassive black holes evolve
in parallel, modulo a factor $(1+z)^{-\alpha}$, obtaining a prediction
for the observable stellar mass density evolution as traced by SMBH
growth:
\begin{equation}
\label{mer_eq:rhostar}
\rho_{*}(z)={\cal A}_{0}\rho_{\rm
  BH}(\epsilon_{\rm rad},z)(1+z)^{-\alpha}[1+\lambda_0(1+z)^{-\beta}]
\end{equation}
where ${\cal A}_{0}$ is the constant of proportionality in the SMBH
mass--spheroid mass relation.  By taking the derivative of
eq.~(\ref{mer_eq:rhostar}), accounting for stellar mass loss, an
expression is also found for the corresponding star formation rate
(SFR) density evolution:
\begin{equation}
 d\rho_{*}(z)/dt = \Psi_{*}(z) - \int_{z_i}^{z}
 \Psi_*(z')\frac{d\chi[\Delta t(z'-z)]}{dt}\frac{dt}{dz'}dz',
\label{eq:sfr}
\end{equation} 
where $\chi[\Delta t(z'-z)]$ is the fractional mass loss that a simple
stellar population experiences after a time $\Delta t$ (corresponding
to the redshift interval $(z'-z)$)\footnote{An analogous term for
  $\rho_{\rm BH}$, due to the ejection of SMBHs from galaxy halos
  after a merger event, is much more difficult to estimate and is
  neglected here.} and $z_i$ is the redshift of (instantaneous)
formation of the first stellar populations; in practical terms, fixing
any $z_i>4$ will not substantially alter the results of such kind of computations.

With these expressions we fit observational data points of both
$\rho_*(z)$ and SFR$(z)$. For each choice of $\rho_{\rm BH,0}$,
$\lambda_0$, and of the critical accretion rate $\dot m_{\rm cr}$, the
fitting functions depend only on three parameters: $\alpha$, $\beta$
and the radiative efficiency $\epsilon_{\rm rad}$.  One example of
such fits is shown in Fig.~\ref{fig:rhobh} for the specific case
$\rho_{\rm BH,0}=4.2\times 10^5 M_{\odot}$ Mpc$^{-3}$
\citep{shankar:09}, and $\lambda_0=0.3$ \citep{fukugita:04}.

Because the drop in the AGN integrated luminosity density at low $z$
is apparently faster than that in the SFR density, the average black
hole to spheroid mass ratio must evolve (slightly) with lookback time
($\alpha>0$).  This result is independent of the local black hole mass
density, and independent of $\lambda_0$.  For the particular example
shown here, the average radiative efficiency turns out to be
$\epsilon_{\rm rad}=0.08^{+0.01}_{-0.02}$, while we obtain
$\alpha=0.35^{+0.22}_{-0.3}$ (both shown with 3-$\sigma$ confidence
bounds).  At face value, this would imply a very mild evolution of the
average $M_{\rm BH}/M_{\rm sph}$ mass ratio.  We will discuss in
section~\ref{sec:scaling_relation_evol} how these constraints compare
with recent efforts to directly measure the ratio of black hole to
host galaxy mass at high redshift.

This simple exercise should make clear that the available constraints
on SMBH growth from the observed bolometric LF are robust enough to
provide interesting non trivial insight into the cosmological
co-evolution of AGN and galaxies.

\subsubsection{The Evolution of the SMBH mass function}
\label{sec:mf_evol}

Despite the relative successes of 'Soltan argument'--like calculations
of the integral evolution of the SMBH mass density, it is obvious that
a much greater amount of information is contained in the {\em
  differential} distributions (mass and luminosity functions).  We
will now discuss attempts to use this information to constrain the
evolution of the mass function of SMBH.

As opposed to the case of galaxies, where the direct relationship
between the evolving mass functions of the various morphological types
and the distribution of star forming galaxies is not straightforward
due to the never-ending morphological and photometric transformation
of the different populations, the situation in the case of SMBH is
much simpler.  For the latter case, we can assume their evolution is
governed by a continuity equation \citep[][and references
therein]{merloni:08}, where the mass function of SMBH at any given
time can be used to predict the mass function at any other time,
provided the distribution of accretion rates as a function of black
hole mass is known.  Such an equation can be written as:
\begin{equation}
\label{eq:continuity}
\frac{\partial \psi(\mu,t)}{\partial t} +
\frac{\partial}{\partial \mu}\left( \psi(\mu,t) \langle \dot M
  (\mu,t)\rangle \right)=0
\end{equation}
where $\mu=\log\, M$ is the black hole mass in solar units, and
$\psi(\mu,t)$ is the SMBH mass function at time $t$.  $\langle \dot M
(\mu,t) \rangle$ is the average accretion rate of SMBH of mass $M$ at
time $t$ and can be defined through a ``fueling'' function,
$F(\dot\mu,\mu,t)$, which describes the distribution of accretion
rates for objects of mass $M$ at time $t$
\begin{equation}
  \langle \dot M(\mu,t)\rangle
  = \int \dot M F(\dot\mu,\mu,t)\, \mathrm{d}\dot\mu
\end{equation}

Such a fueling function is not a priori known, and observational
determinations thereof have been possible in any robust sense only for
the extremes of the overall population.  However, the AGN fueling
function can be derived by inverting the integral equation that
relates the luminosity function of the population in question with its
mass function.  And so, we can write:
\begin{equation}
  \label{eq:filter}
  \phi(\ell,t)=\int F(\ell-\zeta,\mu,t) \psi(\mu,t)\; \mathrm{d}\mu
\end{equation}
with the definitions $\ell \equiv \log\, L_{\rm bol}$ and $\zeta
\equiv \log\, (\epsilon_{\rm rad}c^2)$, with $\epsilon_{\rm rad}$ the
radiative efficiency, here assumed to be constant.

\begin{figure*}
\centering
\psfig{figure=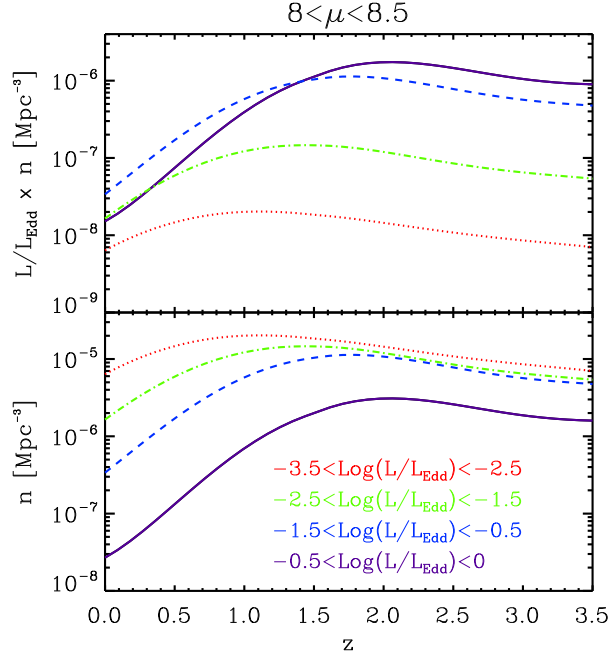,height=9.5cm}
\caption{{\it Bottom}: Evolution of the number density (objects per
  comoving Mpc) as a function of redshift for black holes of constant
  mass (in the range $8<\log (M_{\rm BH}/M_{\odot})<8.5$) at different
  Eddington ratios.  Solid (purple), dashed (blue), dot-dashed (green)
  and dotted (red) lines correspond to intervals of Eddington ratio
  ranging from 1 to $10^{-3.5}$.  {\it Top:} The product of average
  Eddington ratio times number density vs.~redshift.  Despite being
  numerically sub-dominant, rapidly accreting black holes (i.e., those
  with $L/L_{\rm Edd}>3$\%) clearly dominate the mass assembly of SMBH
  in this range of masses.}
   \label{fig:mdot_dist_8}
\end{figure*}

Using this approach, eq.~(\ref{eq:continuity}) can be integrated
backwards from $z=0$, where we have simultaneous knowledge of both the
mass function, $\psi(\mu)$, and the luminosity function, $\phi(\ell)$,
thus evolving the SMBH mass function {\em backwards} in time, up to
where (i) reliable estimates of the AGN luminosity functions are
available, and (ii) the accumulated error in the mass function becomes
of the order of the mass function itself.

The first thing to notice from such an approach is that the different
shapes of the observed SMBH mass function $\psi(\mu)$ (that decays
exponentially at high masses) and AGN LF $\phi(\ell)$ (well described
by a double power-law) necessitate a broad distribution of accretion
rate \citep{merloni:08}: AGN, as a population, cannot be simply
characterized by an on-off switch at fixed Eddington ratio. Instead,
integration of eq.~(\ref{eq:continuity}) gives insight on the relative
importance of massive black hole growth at different accretion rates.

Figure~\ref{fig:mdot_dist_8} shows the number density evolution as a
function of redshift for black holes of constant mass (in the range
$8<\log (M_{\rm BH}/M_{\odot})<8.5$) at different Eddington ratios.
Numerically, the AGN population is always dominated by slowly
accreting objects, but the observed flattening of the bolometric LF
shape (see Fig.~\ref{fig:lf_bol_param}) implies that the relative
number of rapidly accreting black holes increases significantly with
redshift.  In terms of grown mass, however, high-Eddington-ratio AGN
strongly dominate the budget, as shown in the top panel of
Fig.~\ref{fig:mdot_dist_8}: most of the mass of a typical $\approx
10^8 M_{\odot}$ black hole has been accumulated in (short-lived)
episodes of rapid accretion, between a few and a few tens of per cent
of the Eddington luminosity.

\begin{figure}
\centering
\begin{tabular}{cc}
\resizebox{!}{0.44\textwidth}{\includegraphics{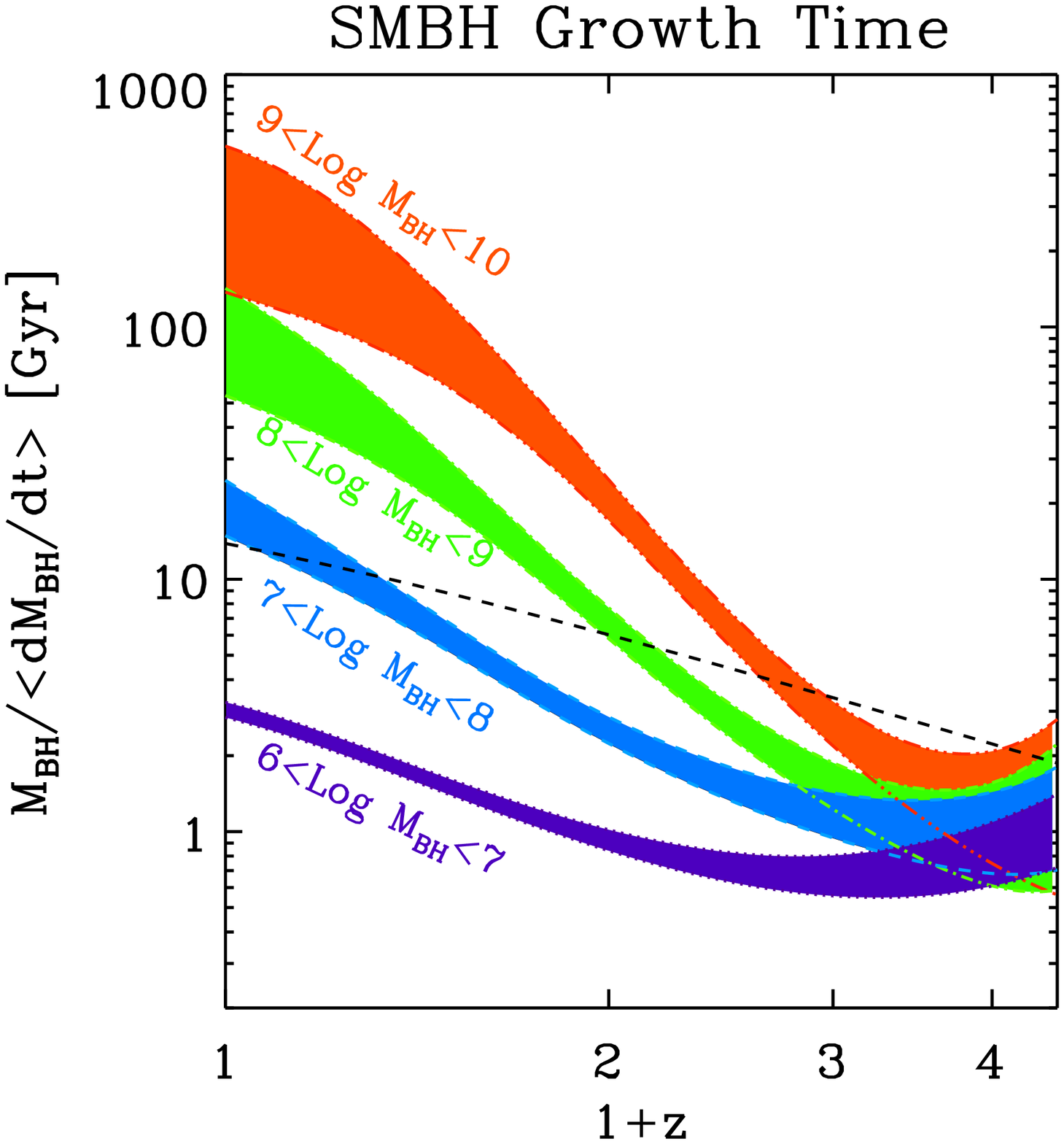}}
\resizebox{!}{0.46\textwidth}{\includegraphics{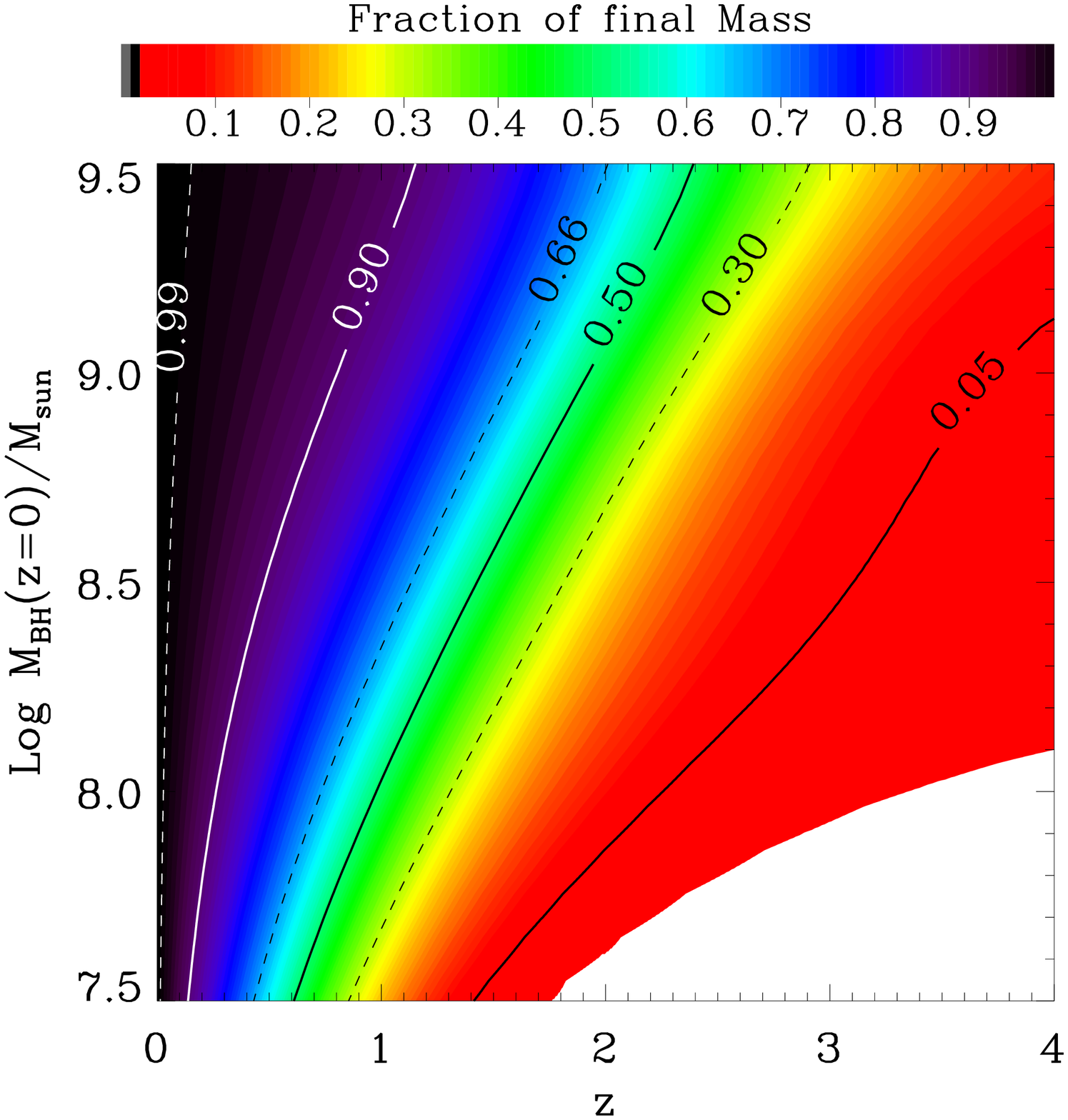}}
\end{tabular}
\caption{{\it Left}: Average Growth time of Supermassive Black Holes
  (in years) as a function of redshift for different black hole mass
  ranges.  The dashed line marks the age of the universe; only black
  holes with instantaneous growth time smaller than the age of the
  universe at any particular redshift can be said to be effectively
  growing.  {\it Right}: the fraction of the final black hole mass
  accumulated as a function of redshift and final (i.e. at $z=0$) mass
  is plotted as contours.}
   \label{fig:xmm}
\end{figure}

The specific instantaneous ratio of black hole mass to accretion rate
as a function of SMBH mass defines a timescale, the so-called {\it
  growth time}, or mass doubling time.  The redshift evolution of the
growth time distribution can be used to identify the epochs when black
holes of different sizes grew the largest fraction of their mass:
black holes with growth times longer than the age of the universe are
not experiencing a major growth phase, which must have necessarily
happened at earlier times.

Figure~\ref{fig:xmm} (left) shows that, according to this simple
estimate, while at $z < 1$ only black holes with masses smaller than
$10^7 M_{\odot}$ are experiencing significant growth, as we approach
the peak of the black hole accretion rate density ($z \sim 1.5-2$), we
witness the rapid growth of the {\em entire} SMBH population.

Solutions of the continuity equation also allow one to trace the
growth of black holes of a given final (i.e., at $z=0$) mass.  The
right hand side panel of Fig.~\ref{fig:xmm} shows that, for the most
massive black holes ($>10^9 M_{\odot}$), half of the mass was already
in place at $z\sim 2$, while those with $M(z=0)<10^8 M_{\odot}$ had to
wait until $z\sim 1$ to accumulate the same fraction of their final
mass.

\subsection{The AGN-galaxy connection}
\label{sec:agn_gal}

The very existence of scaling relations between black holes and their
host galaxies, and the broad accretion rate distributions of AGN
derived from the continuity equation approach imply that, as observed
throughout the electromagnetic spectrum, growing black holes will
display a large range of ``contrast'' with the host galaxy light.

The most luminous QSO, accreting at the highest Eddington ratios, will
be able to outshine the stellar light from the galaxy, while less
luminous, Seyfert-like AGN will have a global SED with a
non-negligible contribution from the host (see also eq.~\ref{eq:agn_contrast}
above). At high redshift, when it
becomes increasingly difficult to spatially separate the nuclear
emission, unbiased AGN samples will have optical-NIR colors spanning a
large range of intermediate possibilities.

\begin{figure}
  \centering
  \resizebox{!}{0.7\textwidth}{\includegraphics{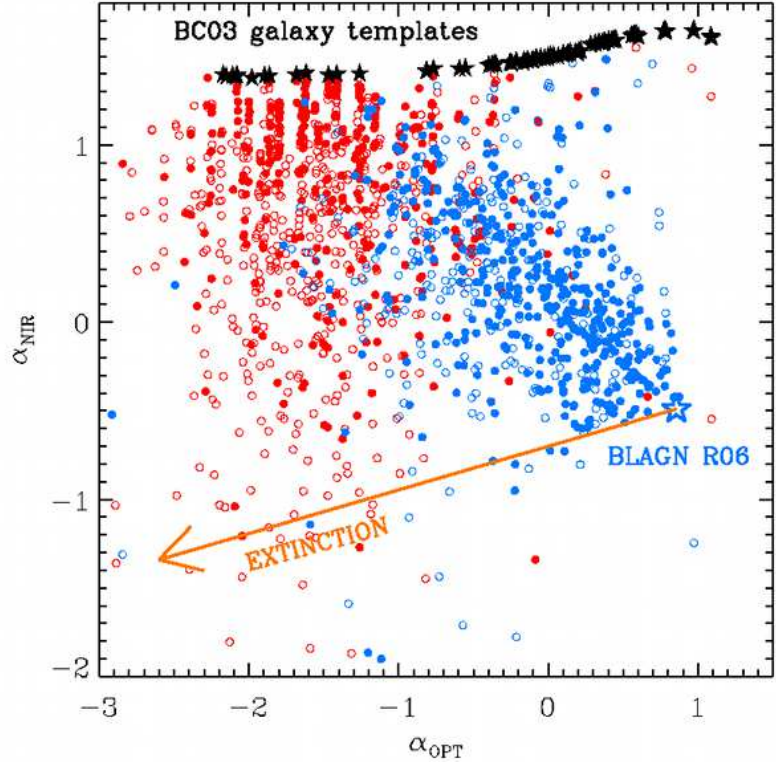}}
  \caption{Observed rest-frame SED slopes in the optical ($\alpha_{\rm
      OPT}$, between 0.3 and 1 $\mu$m) and NIR ($\alpha_{\rm NIR}$
    between 1 and 3 $\mu$m) for all (~1650) X-ray selected AGN in the
    COSMOS survey.  Blue filled circles denote spectroscopically
    confirmed type 1 (Broad Lined) AGN, blue empty circles denote
    candidate type 1 AGN from the photo-z sample.  Red filled circles
    are spectroscopically confirmed type 2 (Narrow lined) AGN, empty
    red circles are candidate type 2 AGN from the photo-z sample.  The
    empty blue star marks the colors of a pure intrinsic type 1 quasar
    SED (from \citealt{richards:06}), while black stars are the loci
    of synthetic spectral templates of galaxies, with increasing
    levels of star formation form the left to the right.  Nuclear
    obscuration, parametrized with a Calzetti extinction law, moves
    every pure type 1 AGN along the direction of the orange arrow.}
   \label{fig:sed_cosmos}
\end{figure}

Figure~\ref{fig:sed_cosmos} nicely illustrates this point.  It is
taken from the analysis of an X-ray selected sample of AGN in the
COSMOS field \citep{brusa:10}, the largest fully identified and
redshift complete AGN sample to date.  It displays the slope of the
rest-frame SED in the optical ($\alpha_{\rm OPT}$, between 0.3 and 1
$\mu$m) and NIR ($\alpha_{\rm NIR}$ between 1 and 3 $\mu$m).  Pure
QSOs, i.e., objects in which the overall SED is dominated by the
nuclear (AGN) emission have a typical dip in the NIR region, and would
lie close to the empty blue star in the lower right corner (positive
optical slope and negative NIR slope).  The location of the X-ray
selected AGN in Figure~\ref{fig:sed_cosmos} shows instead that, in
order to describe the bulk of the population, one needs to consider
both the effects of obscuration (moving each pure QSO in the direction
of the orange arrow) and an increasing contribution from galactic stellar
light (moving the objects towards the black stars in the upper
part of the diagram).

This demonstrates that current multi-wavelength extragalactic surveys
are sensitive enough to disentangle the complex interplay between
nuclear and galaxy light in the SED of more typical AGN.  It is no
coincidence that such surveys are beginning to probe the details of
the co-evolution of black holes and host galaxies on an
object-by-object basis.  In the following section we will briefly
discuss how one can use such information to observationally trace the
evolution of the scaling relations between nuclear SMBH and their host
galaxies.

\subsubsection{Redshift evolution of the scaling relations}
\label{sec:scaling_relation_evol}

{\em Local} scaling relations between black hole mass and structural
properties of their (spheroidal) hosts have been unable to
unambiguously determine the physical nature of the SMBH-galaxy
coupling.  A large number of theoretical models for the AGN-galaxy
interaction responsible for establishing, for example, the
$M-\sigma_{*}$ relation, have been proposed, all tuned to reproduce
the $z=0$ observations.  One obvious way out of this impasse is the
study of their {\em evolution}.

In recent years, a number of groups have employed different techniques
to detect signs of evolution in any of the locally observed scaling
relations.  Only type 1 AGN, with un-obscured broad line region allow
a simple direct estimate of BH masses, via the so-called ``virial'' or
empirically calibrated ``photo-ionization'' method
\citep{peterson:04}.  Based on existing samples of broad line QSOs,
most efforts have been devoted to the study of the $M_{\rm
  BH}-\sigma_{*}$ relation.  For example, \citet{salviander:07} have
used narrow nebular emission lines ([OIII], [OII]) excited by the AGN
emission in the nuclear region of galaxies as proxies for the central
velocity dispersion, and compared these to the black hole mass
estimated from the broad line width of QSOs from $z\sim 0$ to $z\sim
1$. In this case, a large scatter has been found in the relation
between $M_{\rm BH}$ and $\sigma_{*}$.

An alternative path is to study carefully selected samples of
moderately bright AGN in narrow redshift ranges, where the host's
stellar velocity dispersion can be measured directly from the
absorption lines in high signal-to-noise spectra.  These studies also
found evidence of (strong) positive evolution of the $M_{\rm BH}$ to
$\sigma_{*}$ ratio compared to the local value (see
\citealt{bennert:11}, and references therein).  This method, although
promising and reliable, is quite inefficient and telescope-time
consuming: secure detection of spectral absorption features in massive
ellipticals at $1< z < 2$ require hundreds of hours of integration
time on an 8-meter class telescope.

When a good sampling of the AGN SED is instead available, rather than
high-resolution, high signal-to-noise spectra, it is possible to try
to decompose the overall spectral energy distribution into a nuclear
and a galaxy component, and derive in this way the physical properties
of the host galaxies of un-obscured AGN whose SMBH masses can be
estimated from their broad lines \citep{merloni:10}.

Other groups have chosen to try to derive information on the host mass
of broad line AGN using multi-color image decomposition
techniques. Due to the severe surface brightness dimming effects,
employing these techniques for high-redshift QSOs becomes increasingly
challenging, unless gravitationally lensed QSOs are selected.  
In all cases, very deep, high resolution optical
images ({\it HST}) are necessary to reliably disentangle the nuclear
from the host galaxy emission.

\begin{figure*}
\centering
\begin{tabular}{cc}
\resizebox{!}{0.415\textwidth}{\includegraphics{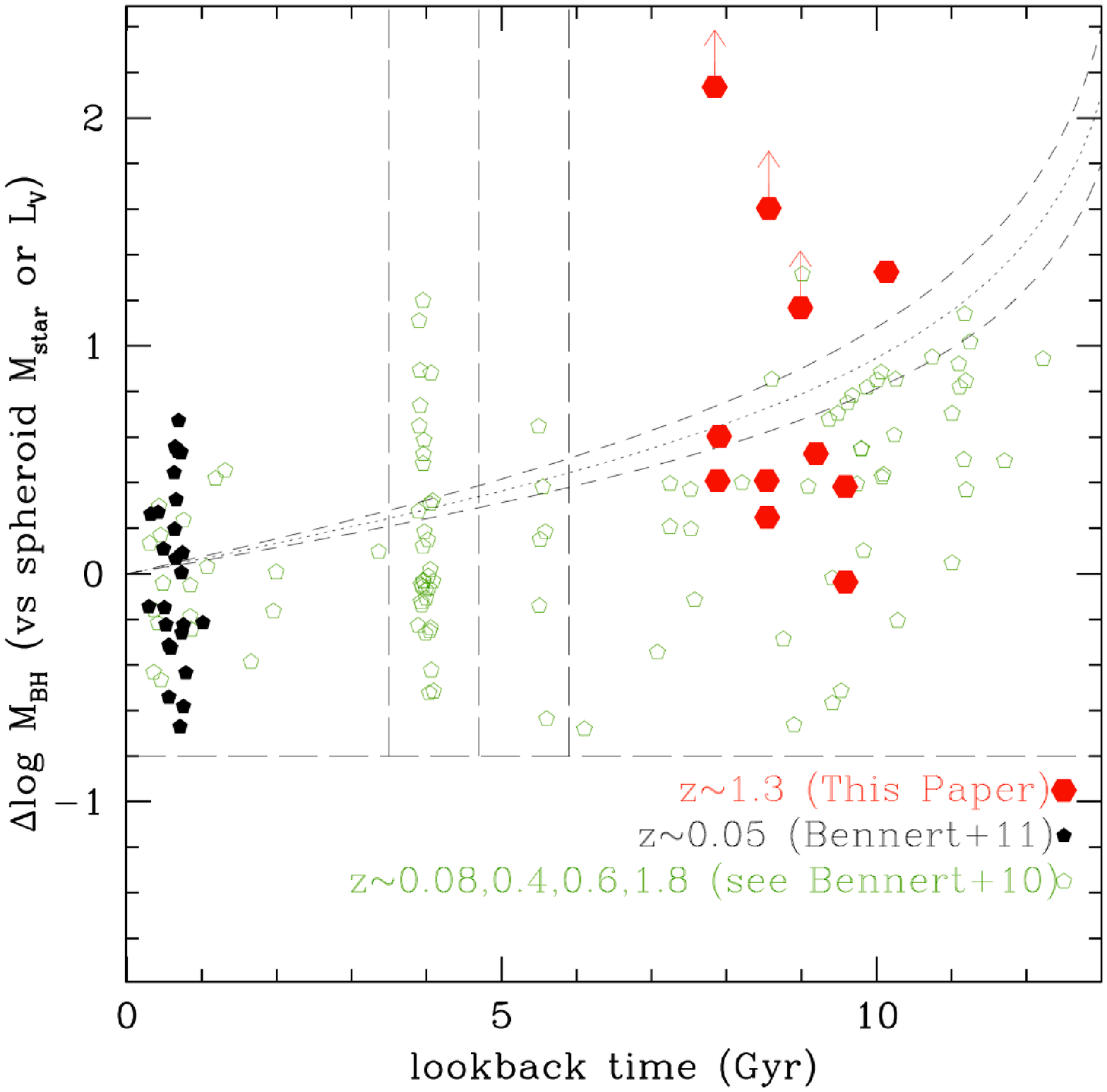}}
\hspace*{16pt}
\resizebox{!}{0.415\textwidth}{\includegraphics{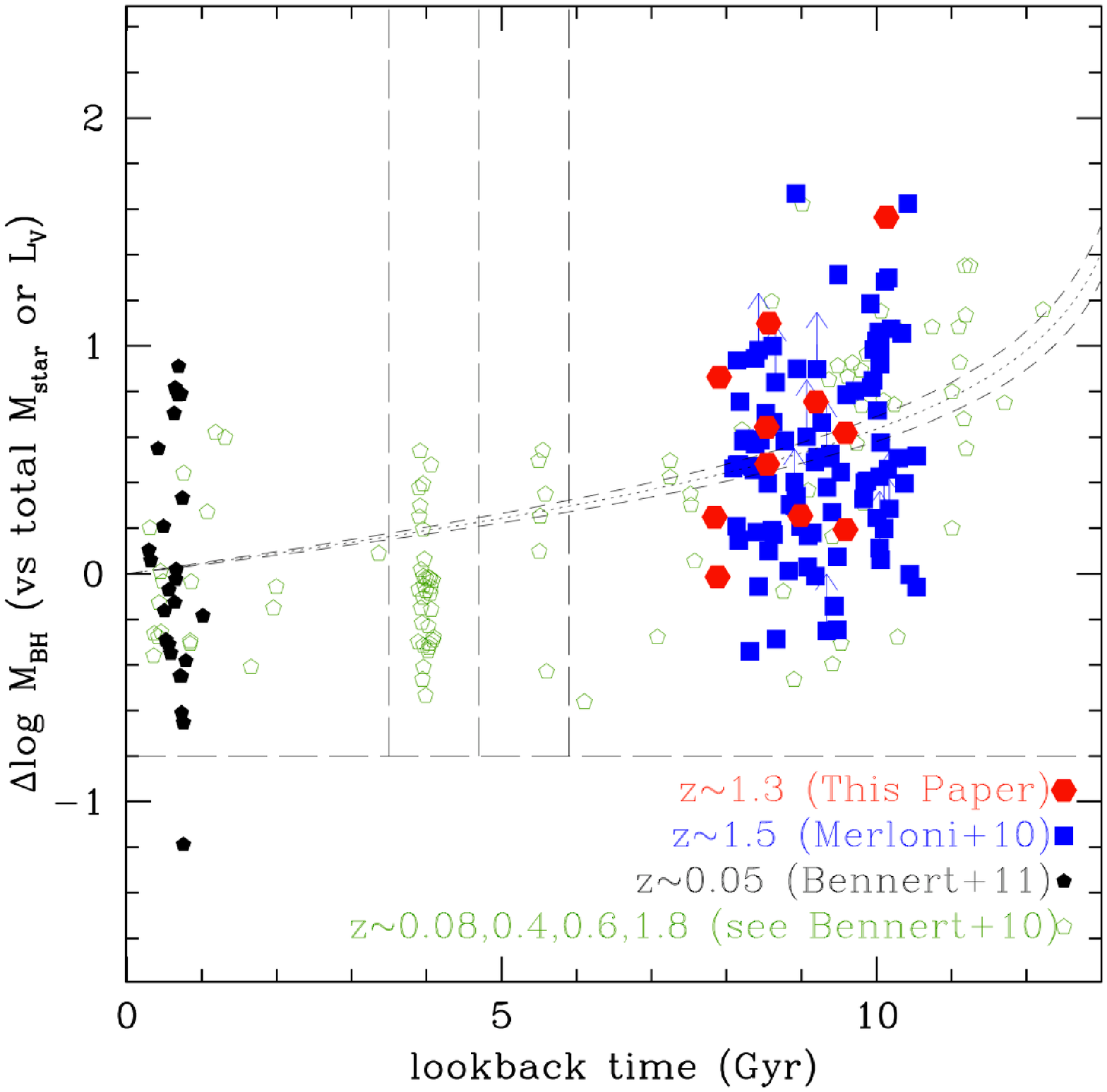}}
\end{tabular}
\caption{{\it Left:} Offset in $\log M_{\rm BH}$ as a function of
  constant spheroid host galaxy mass (red filled pentagons) with
  respect to the fiducial local relation of AGNs (black filled
  circles).  The offset as a function of constant stellar spheroid
  luminosity is over-plotted (green open symbols), corresponding to
  AGNs at different redshifts. The best linear fit derived here is
  over-plotted as dotted line $M_{\rm BH}/M_{\rm sph} \propto
  (1+z)^{2.1\pm0.3}$; dashed lines: 1$\sigma$ range.  {\it Right}: The
  same as in the left panel as an offset in $\log M_{\rm BH}$ as a
  function of constant total host galaxy mass.  The lines correspond
  to $M_{\rm BH}/M_{\rm host} \propto (1+z)^{1.41\pm0.12}$.  From
  \citet{bennert:11}, where a comprehensive list of references for the
  observational data points can also be found.}
\label{fig:dmbh_z}
\end{figure*}

The main result of these various investigations is that our estimates
of the type--1 AGN host physical parameters are {\it inconsistent}
with the hypothesis that they lie on the $z=0$ scaling relation (see
Figure~\ref{fig:dmbh_z}).  At high redshift, bigger black holes are
hosted in galaxies of a given mass as compared to what we observe
locally.  The best linear fit to the ensemble of observations shown in
Figure~\ref{fig:dmbh_z} is $M_{\rm BH}/M_{\rm sph} \propto
(1+z)^{2.1\pm0.3}$ for the black-hole-to-spheroid mass ratio and
$M_{\rm BH}/M_{\rm host} \propto (1+z)^{1.41\pm0.12}$ for the
black-hole-to-total host stellar mass ratio.

However, the objects for which this study can be made are selected
essentially on the basis of the nuclear (AGN) luminosity, and on the
detectability of broad emission lines, clearly leading to a bias
towards more massive black holes, similar to Malmquist bias for
luminosity selected samples of standard candles.  Can such a bias be
responsible for the observed trends?

Let us consider in detail the effects on the observed systems of a
given intrinsic scatter, $\sigma_{\mu}$, in the $M_{\rm BH}$ --
$M_{*}$ scaling relation.  Any non-zero $\sigma_{\mu}$ implies that
there is a range of possible masses $\log M_{*}\pm \sigma_{\mu}$ for
each object of a given black hole mass $M_{\rm BH}$, where we have
assumed, for simplicity, a symmetric scatter in the relation.  If the
number density of galaxies is falling off rapidly in the interval
$\log M_{*}\pm \sigma_{\mu}$, it will then be more likely to find one
of the more numerous small mass galaxies associated with the given
black hole, and therefore a larger ratio $M_{\rm BH}/M_{*}$.  Thus,
given a distribution of galaxy masses, and provided that the scatter
$\sigma_{\mu}$ is not too large, the logarithmic offset of each point
from the correlation, assumed to be held fixed to the local
determination, is given by:
\begin{equation}
  \label{eq:bias_mbh}
  \Delta \log (M_{\rm BH}/M_{*}) = 0.67 \times \Delta \log M_{\rm BH}
  \approx \sigma^2_{\mu} \left(\frac{d  \log \phi}{d \log
      M_{*}}\right)_{{\rm
      \log} M_{*}=(\mu-A)/B},
\end{equation} 
where $\mu \equiv \log M_{\rm BH}$ and $(A,B)=(1.12,-4.12)$ are slope
and intercept of the local scaling relation between BH and host galaxy
masses \citep{haering:04}.

One can estimate observationally the logarithmic derivative of the
galaxy mass function $\frac{d \log \phi}{d \log M_{*}}$.  At $z\approx
2$, the offset expected from such a bias is of the order of 0.25 dex,
if $\sigma_{\mu}=0.5$ and increases to about 0.5 dex for
$\sigma_{\mu}=0.7$.  The average offset shown in Fig.~\ref{fig:dmbh_z}
is clearly in excess of what is expected in the most extreme case of
large intrinsic scatter in the local relation, estimated to be less
than 0.5 dex \citep{gultekin:09}.  The data point towards an evolution
of the scaling relation, either in normalization or in scatter (or a
combination of both).
 
What are the implications of these findings for our understanding of
the cosmological co-evolution of black holes and galaxies?

In the next section, we will discuss in more detail a number of
physical processes by which AGN can regulate the growth of their host
galaxies, thereby affecting any observable evolution of the scaling
relations. We will see how, from the physical point of view, a clear
distinction has to be made between two modes of AGN feedback.

The first one is associated to the numerous, long-lived, LLAGN, with
emitted power dominated by the kinetic energy of their jets and
outflows.  It becomes increasingly important for very massive holes at
low redshift (see section~\ref{sec:mf_evol} above).  Many models of
galaxy formation invoke such a feedback mechanism in order not to
over-produce very massive galaxies in the largest virialized DM halos
at low redshift \citep{croton:06}.

It is not clear, however, how such a feedback mode can effectively
couple the SMBH mass with the structural properties of their galactic
hosts and give rise to the observed scaling relations.  For such a
task, modelers have instead turned to feedback modes associated to
the phases of fast SMBH growth in bright QSO.

In all feedback models in which the black hole energy injection is
very fast (explosive), if strong QSO feedback is responsible for
rapidly terminating star formation throughout the entire bulge
\citep{dimatteo:05}, QSOs and, in general, type--1 AGN are associated
with the final stage of bulge formation.  Then, very little evolution,
as well as very little scatter, is expected for the scaling relations,
and it is very hard to produce any positive offset like the one
observed.

The physics of such (``quasar'') mode of AGN feedback remain elusive,
as it remains the issue of whether the energy release by the
associated process of rapid black hole growth is indeed responsible
for halting the conversion of gas into stars on galactic (kpc) scales,
or whether it is only responsible for a milder form of
``self-regulation'' by cutting off its own gas supply on nuclear (pc)
scales \citep{hopkins:09}.

\section{Cosmology II: AGN feedback}
\label{sec:feedback}

The phenomenological investigation presented in \S~\ref{sec:cosmology}
above
leaves open the fundamental question about the physical origin of such
a clear, parallel differential growth of both the black holes and the
galaxy population.

From the discussion of AGN activity in Chapter ``Active Galactic
Nuclei'' by E. Perlman in this same
  volume, it should be immediately apparent that black hole growth
is often, if not always, accompanied by the release of enormous
amounts of energy, in the form of radiation, outflows, and
gravitational waves.

Black holes accreting at high rates in the so-called radiative (or
quasar) mode will release of order 10\% of the accreted rest mass
energy as radiation.  They can also drive broad (un-collimated)
outflows, again described in more detail in chapter ``Active
Galactic Nuclei'' by E. Perlman in this book, and in about 10\% of bright AGN, radio emitting,
relativistic jets are observed (the quasar 3C273 seems to be such an
object that is accreting efficiently {\it and} making powerful jets at the
same time).

But even black holes in the so-called inefficient accretion regime,
where cooling is dominated by advective processes rather than
radiation, can drive powerful, collimated outflows in the form of
relativistic jets.  Perhaps the best example of such a powerful ``low
efficiency'' black hole is the radio galaxy Virgo A, which is the
product of inefficient accretion onto the supermassive black hole at
the center of M87, which itself is located in the center of the Virgo
cluster.  See \S ``Active Galactic Nuclei'' by E. Perlman in this book
 for a detailed discussion
of the properties of the M87 jet.

This energy will be released directly into the environment from which
the black hole grows: the cooling, possibly star-forming gas in the
central galaxy.  Any transfer of energy to the gas should thus reduce
the rate at which gas cools and forms stars.  While the direct link
between star formation through cooling in the centers of galaxies and
black hole growth through accretion is not fully established, it is
easy to imagine how such an energy deposition can reduce the rate of
accretion onto the central black hole as well.

This process of cooling-induced black hole activity can therefore be
considered as a negative feedback loop, in that {\em increased}
accretion activity acts to {\em decrease} the large scale gas supply
to the black hole.  The impact on star formation might be coincidental
(if black hole growth is unrelated to the actual star formation rate)
or fundamental (if black hole growth is mediated or directly fueled by
star formation, for example, through direct accretion of stars).

Furthermore, while the direct link between black hole growth and star
formation suggested by the $M$--$\sigma_{*}$ relation is hidden, and
evidence for the suggested underlying feedback process {\em on stars}
is largely circumstantial (as will be discussed below), some important
clues can be derived by tracing the evolution of the feedback energy
released by growing black holes as a function of black hole mass and
redshift.

In this section, we describe in some detail how such an inventory can
be made and how feedback itself operates.  We will focus primarily on
the information about the properties of black holes that can be
extracted from observations of feedback, with other chapters
discussing the role of black hole growth on the formation of structure
in more detail (see \S ``Clusters of Galaxies'' by R. Bower in this book).

\subsection{Evidence and arguments for feedback}

The process of accretion is a multi-scale phenomenon: The range from
the place of capture, where the gas first enters the sphere of
influence of the black hole, to the event horizon of the black hole
spans roughly seven orders of magnitude in scale --- too much to
simulate in one big simulation for even just one dynamical time on the
outer scale.

Yet, as extreme as this range in scales may be, the process of
feedback can cover another 5 orders of magnitude more in scale: From
the scales of the horizon (about AU-size for a typical central black
hole in a typical, $L_*$, galaxy) to scales of entire galaxy
clusters (several hundred kiloparsec): a dynamic range of twelve
orders of magnitude.

Given that our understanding of accretion is still developing, and
that our understanding of jet formation is, at best, elementary, it
should not come as much of a surprise that our understanding of AGN
feedback is mostly limited to fairly crude statements about energy
input and global heating efficiencies from a theoretical perspective.

The best {\em observational} evidence for feedback is not on galaxy
scales at all, but on the largest spatial scales on which we can
expect black holes to have any meaningful influence: In the centers of
galaxy clusters.  The reason for this is twofold:

\begin{itemize}
\item{First, the angular scales on which feedback in galaxy clusters
    unfolds are readily resolvable by telescopes in all bands of the
    electromagnetic spectrum.}

\item{Second, the signatures of AGN feedback in galaxy clusters are
    easy to identify from X-ray and radio imaging, as we will discuss
    momentarily.}
\end{itemize}

Consequently, we have developed a fairly mature picture of how AGN
feedback works on the very largest scales and have even successfully
simulated the feedback processes in computers.

On smaller scales, the evidence for feedback becomes increasingly
circumstantial.  Thus, while the link between black hole and galaxy
properties may be the most fundamental expression of direct coupling
of their growth processes, it is also the most elusive in terms of
direct evidence for this coupling.

A number of reasons conspire to limit our observational insight into
galactic scale feedback:

\begin{itemize}
\item{The angular scales of this process are inherently small, given
    that the feedback must be happening in the centers of galaxies.}

\item{While cluster evolution is happening in the current epoch, at
    low redshift, galaxy growth happens at higher redshift - during
    the star formation epoch.  This is especially true for the
    galaxies that seem to require AGN feedback the most.}

\item{Both star formation and rapid black hole growth tend to cloak
    themselves in dust extinction and photo-electric absorption.  It
    may be that the smoking guns of feedback are mostly hidden behind
    Compton-thick X-ray absorbers and many magnitudes of dust
    extinction.}
\end{itemize}

Finally, the tight connection between black hole growth and star
formation suggested by the $M - \sigma_{*}$ relation and by the
similarity in the redshift evolution of both populations seems to
imply that stars and black holes grow roughly {\em simultaneously}.
For feedback to have a strong impact on star formation and at the same
time couple the mass of growing black holes to the mass of stars in a
galaxy, one would expect rapid black hole growth to be concurrent with
episodes of star formation.

This would imply that feedback on star formation would have occurred
during the quasar phases, and since most quasars are radio quiet, this
suggests that at least part of the feedback on star formation operates
through a different channel than the readily observable ``radio
galaxy'' feedback on cluster scales at low redshift.  Given that this
feedback must have occurred during the quasar epoch, it is commonly
referred to as ``quasar mode'' feedback.

In fact, the most convincing ``evidence'' for such a mode comes not
from actual observations of black holes but from semi-analytic models
of galaxy formation: In order to explain the galaxy luminosity
function and galaxy color distribution, modelers have to assume {\em
  two} types of feedback: One that disperses and heats the
star-forming gas at the end of a star-formation cycle (generally
triggered by mergers), effectively halting star formation --- this is
the ``quasar mode'' --- and one that maintains the gas in typical
elliptical galaxies in its tenuous, hot state --- this is the
``radio'' or ``maintenance mode'' \citep{springel:05,croton:06}.

However, these models say nothing about the actual physical mechanism
of feedback: they assume quasi-spherical heating of the gas in both
``quasar'' and ``radio'' mode, and the only thing that distinguishes
them is the prescription of how the black hole accretes (whether from
cold or hot gas).  The more appropriate naming convention is thus
``cold'' and ``hot'' mode accretion.  

Thus, the circumstantial evidence for ``cold'' mode feedback does not
answer the question of whether the heating/dispersal occurs as a
result of winds, jets, or radiation released by the accreting black
hole.  Given that slowly growing black holes are radiatively
inefficient and universally seem to be radio-active \citep{ho:08}, it
has generally been assumed that any feedback from black holes in the
``hot'' mode must be in the form of jets.

\subsection{Feedback in galaxy clusters}
\label{sec:clusterfeedback}
It is instructive to begin by discussing the obvious examples of AGN
feedback.  This will inform our discussion of the possible influence
of AGN on the process of star formation on galactic scales.  In
particular, from a discussion of radio galaxy feedback on cluster
scales it is possible to draw quantitative conclusions about
``radio-mode'' or ``hot-mode'' feedback by jets from slowly growing
black holes.  For a more detailed discussion of feedback in galaxy
clusters, see \citet{mcnamara:07}.

\begin{figure}[t]
\begin{center}
  \resizebox{!}{0.315\textwidth}{\includegraphics{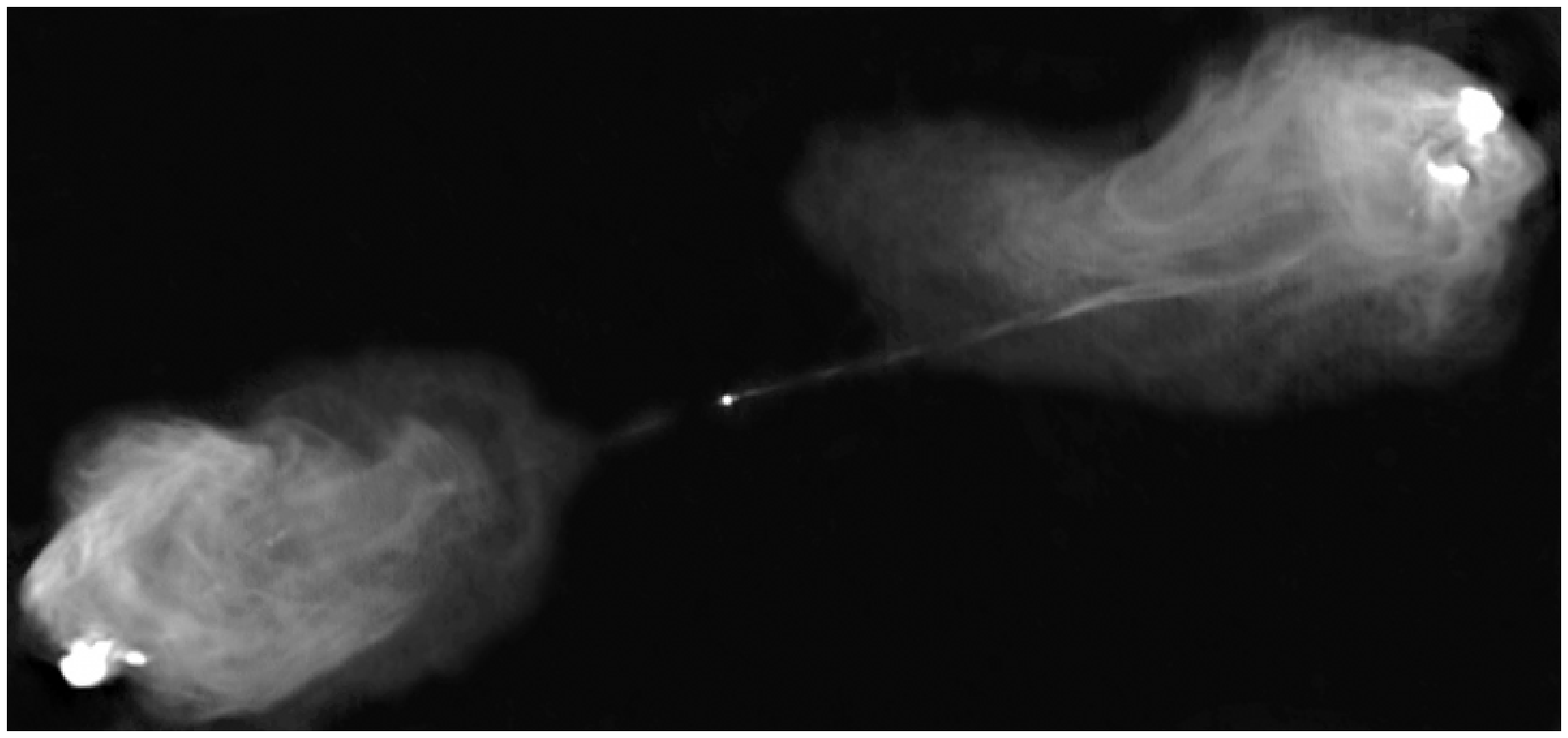}}
  \resizebox{!}{0.315\textwidth}{\includegraphics{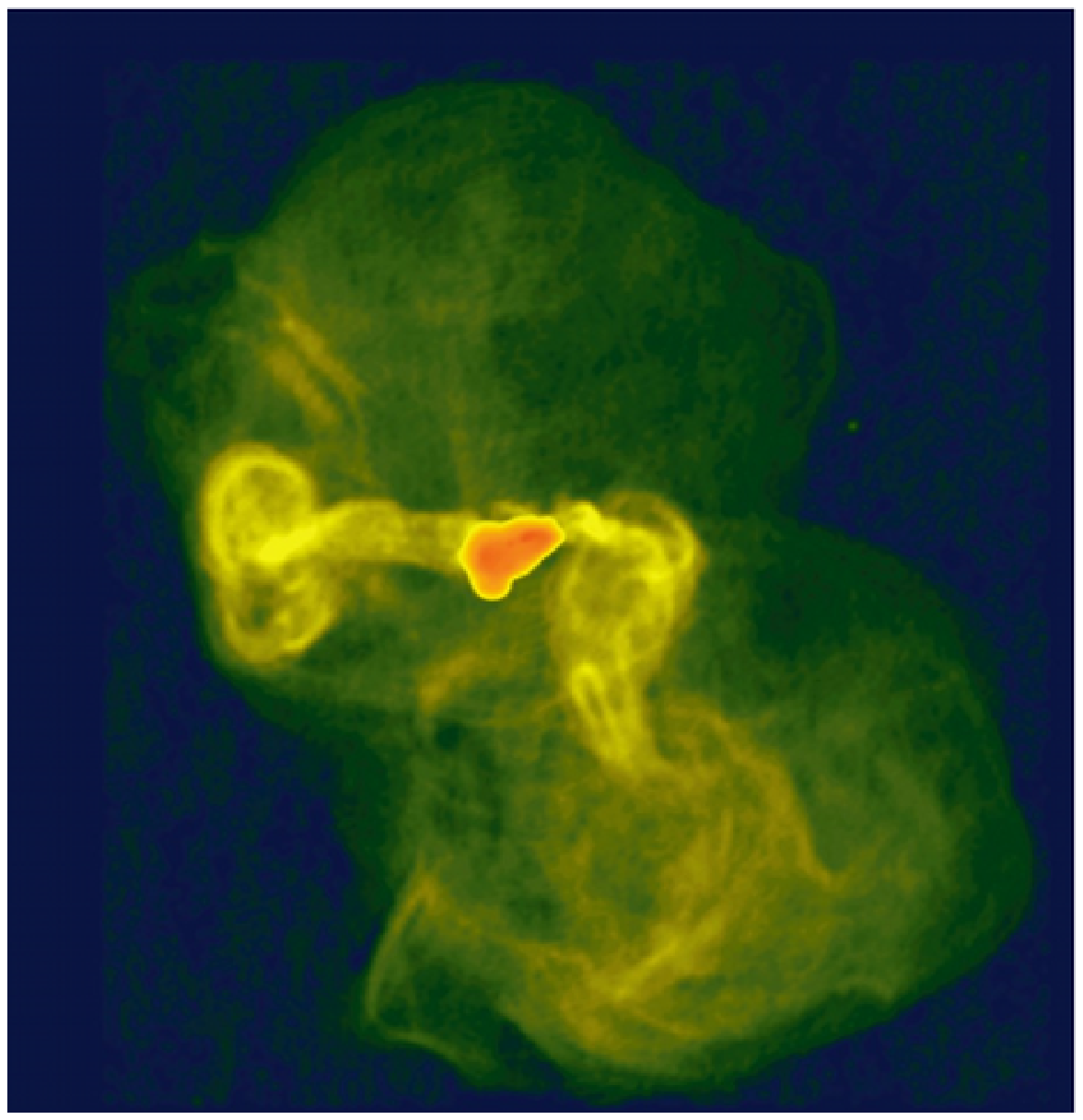}}
\end{center}
\caption{{\it Left}: The FRII radio galaxy Cygnus A, observed by the
  VLA at 6cm; image scale: $150\,{\rm kpc}\times 85\,{\rm kpc}$; {\em
    right:} The large scale structure of the FR I radio galaxy Virgo
  A, observed by the VLA at 90cm.  The relativistic inner jet of M87
  is contained in the over-exposed central radio lobes; image scale:
  $80\,{\rm kpc}\times 80\,{\rm kpc}$.}
\label{fig:radio}
\end{figure}

On a basic level, the importance of feedback was already apparent with
the discovery of powerful radio galaxies in the 1960s and onward
(though the relevance of the black hole in this context took longer to
establish): Radio galaxies, like the example of Cygnus A shown in
Fig.~\ref{fig:radio}, exhibit diffuse ``lobes'' of synchrotron
emission (see chapter ``Active Galactic Nuclei'' by E. Perlman in this
same volume), on
scales of tens and even hundreds of Kiloparsec.  In other words: The
action of a jet from the central black hole deposits magnetized
relativistic plasma into the surrounding medium.

Simply summing up the entire synchrotron radiation and making
reasonable assumptions about the shape and volume filling fraction of
the emitting regions, it is straight forward to derive lower limits on
the total energy needed to explain the radio emission.  In some cases,
the minimum energy derived could be enormous: \citet{perley:84} found
that Cygnus A required at least $E_{\rm min} \gtrsim 10^{60}\,{\rm
  ergs}$ pumped into the lobes by the central black hole.  For
perspective: This is of the same order as the gravitational binding
energy of the Milky way.

If jets could release this much energy in relativistic gas into the
environments of black holes, it would be hard to imagine how the
environment could {\em not} be strongly affected.

\begin{figure}[t]
\resizebox{!}{0.3\textwidth}{\includegraphics{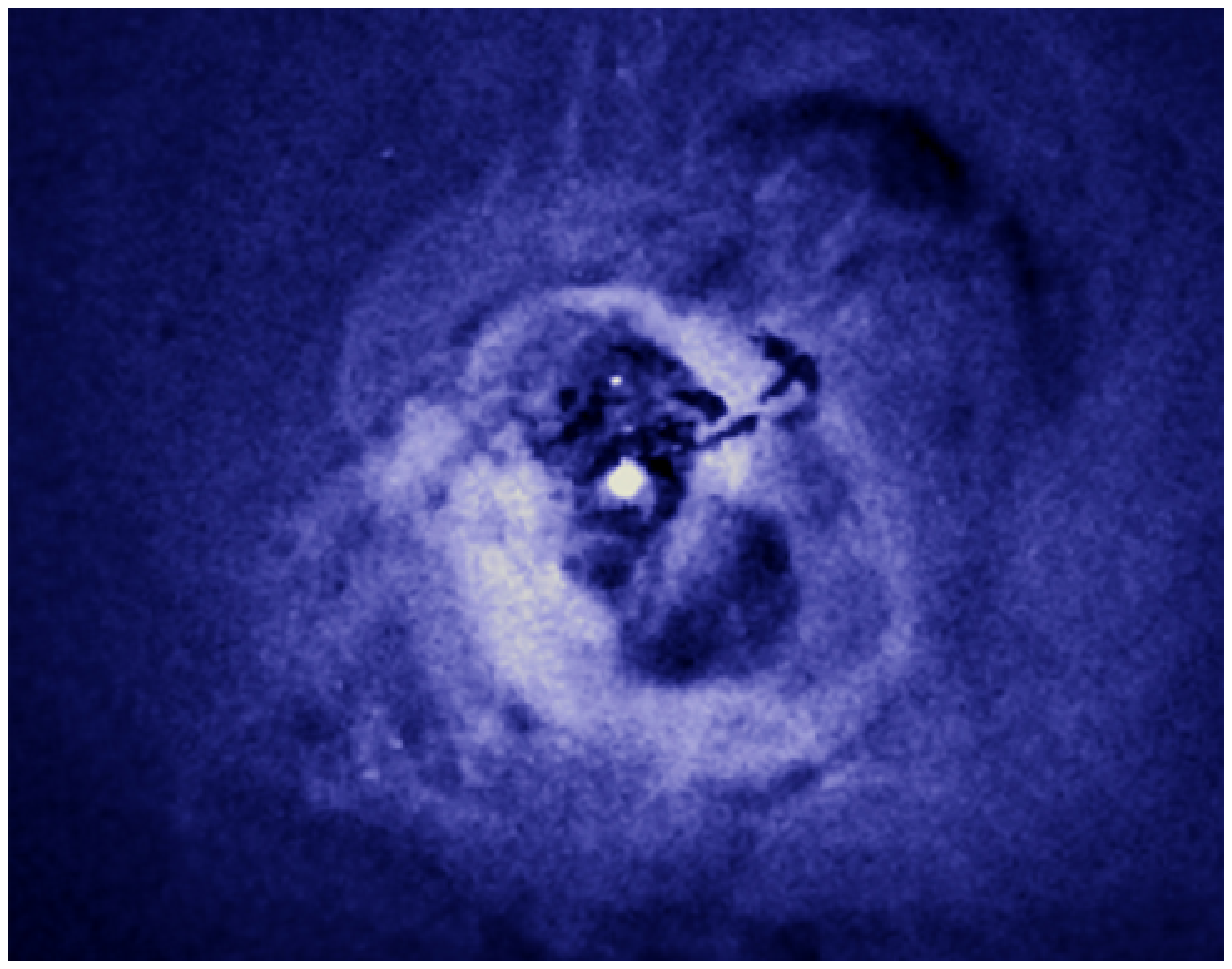}}
\resizebox{!}{0.3\textwidth}{\includegraphics{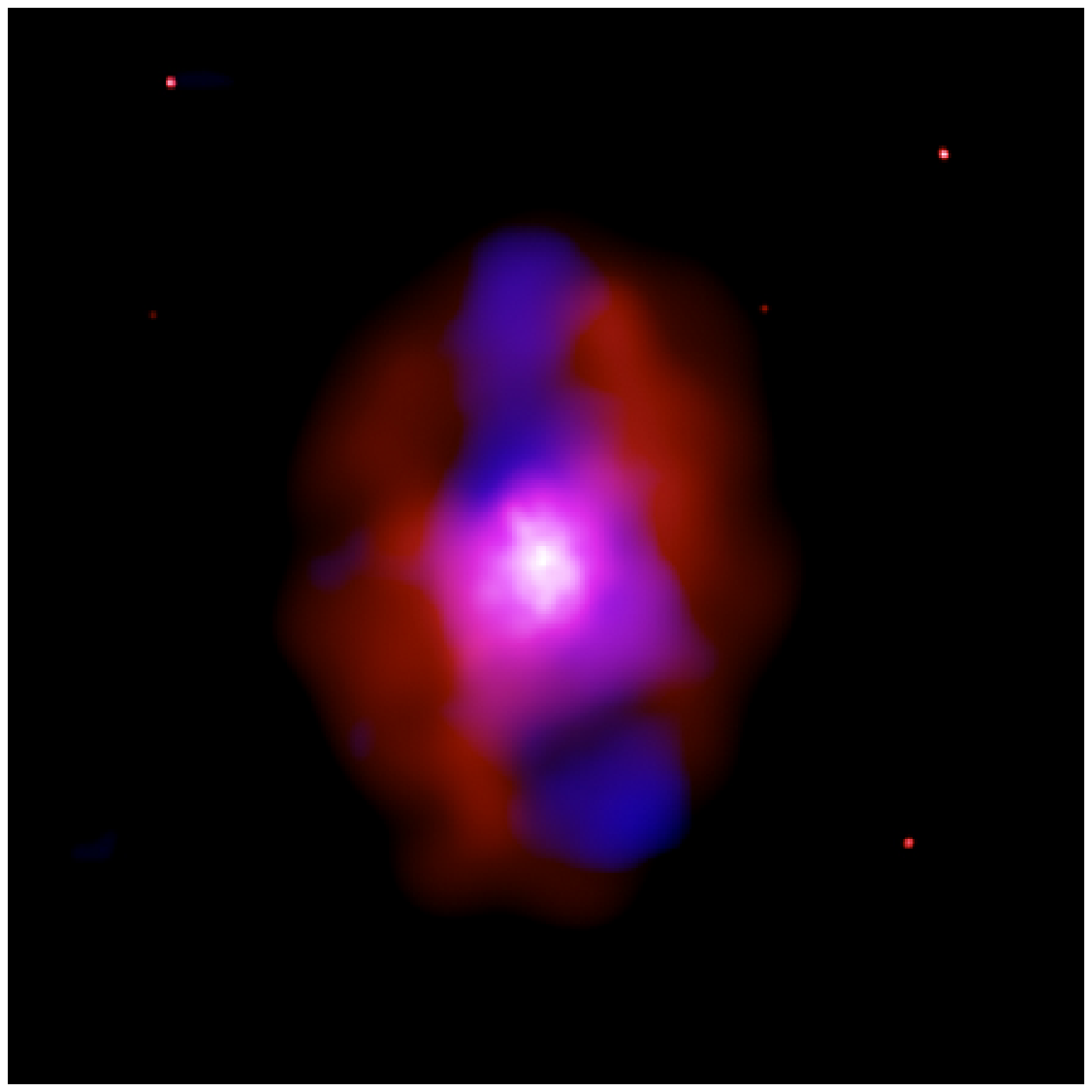}}
\resizebox{!}{0.3\textwidth}{\includegraphics{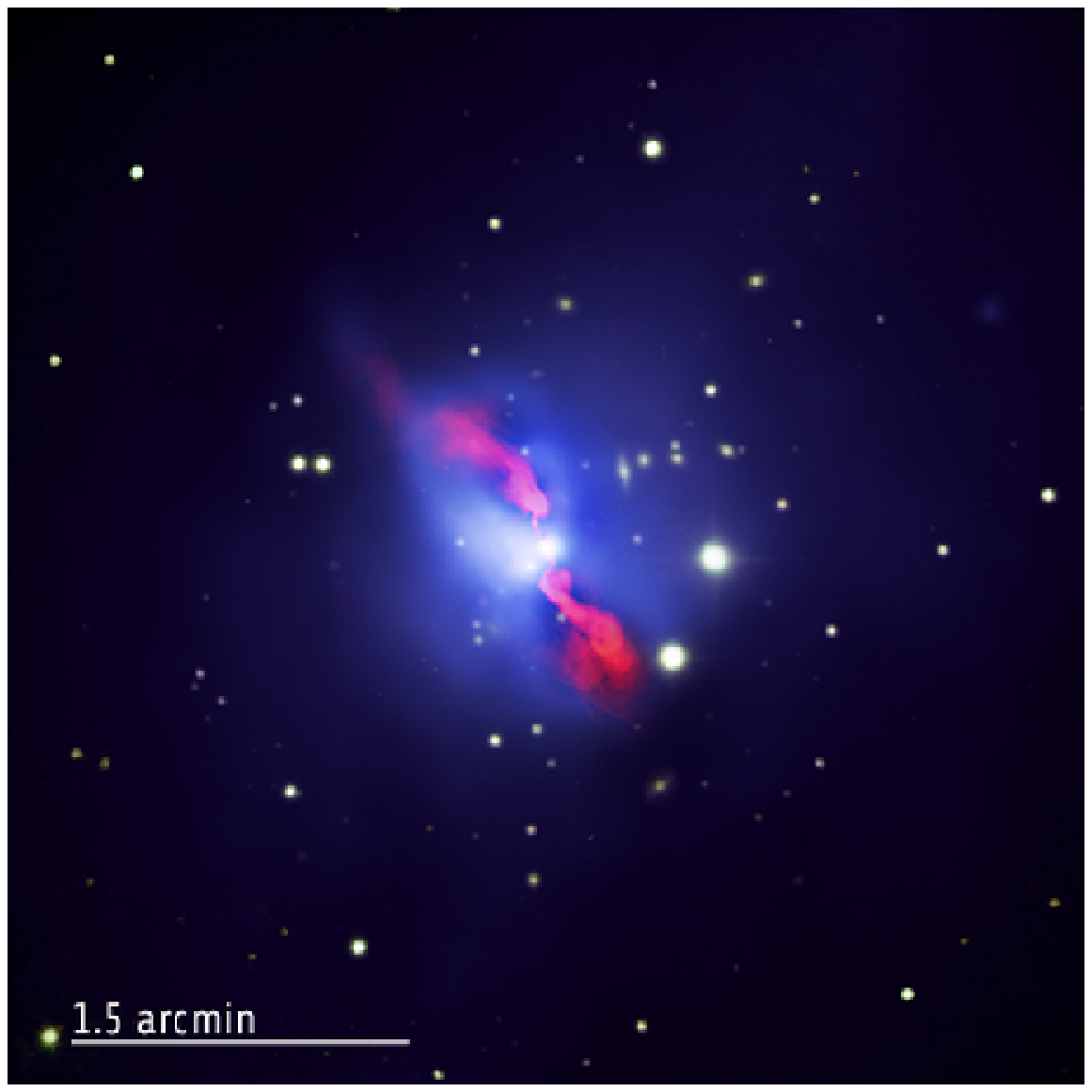}}
\caption{{\it Left}: Deep {\em Chandra} observation of the Perseus
  cluster \citep{fabian:06}; {\em middle:} {\em Chandra} image of
  MS0735 (red) and VLA (blue), adopted from \citet{mcnamara:05}; {\em
    right:} {\em Chandra} image of Hydra A (blue) and 6cm VLA radio
  image (red), adopted from NASA \citep{kirkpatrick:09}; image scale:
  $80\,{\rm kpc}\times 80\,{\rm kpc}$.}
\label{fig:cavities}
\end{figure}

The first {\em direct} evidence for feedback on the gas surrounding
the black hole came with the arrival of high-resolution X-ray imaging:
using {\it ROSAT} data, 
\citet{boehringer:93} discovered that the radio galaxy Perseus A
(powered by the supermassive black hole in the central cluster galaxy
NGC 1275) excavates large cavities in the hot, X-ray emitting thermal
gas that fills the Perseus cluster.  The gas is pushed aside into
dense shells and the excavated X-ray cavities are filled with radio
emission by the lobes of the radio galaxy (see
Fig.~\ref{fig:cavities})

Similarly, the privileged view we enjoy of the nearby radio galaxy
Virgo A (also known as M87) allowed a uniquely detailed study of its
multi-scale emission well before the idea of feedback had taken hold.
The inner (roughly kpc) jet of M87 is discussed in some detail in
chapter ``Active Galactic Nuclei'' by E. Perlman in the same volume. 
However, lower frequency
observations revealed a much richer picture on scales just outside of
the visible galaxy, still in the very center of the Virgo cluster
\citep{owen:00}: Curling and twisting strands of radio emission,
connecting the nucleus to a set of radio lobes about 20 kpc in radius,
and misaligned with respect to the central jets by about 90 degrees on
the sky\footnote{Part of this misalignment could be due to projection,
  of course, given that at least the inner jet is directed fairly
  close to the line-of-sight.}.

Both Perseus and Virgo are cool core clusters, and in particular
Perseus had long been considered a prototypical example of a cooling
flow.  That is, the radiative cooling time in the center of the
cluster is shorter than the age of the cluster.  In a
quasi-hydrostatic model of a cluster (inward gravity, mostly provided
by the dark matter contribution, balanced by an outward thermal
pressure gradient), this would imply that the cluster must be
contracting on a Kelvin-Helmholtz timescale, with gas at the center
cooling rapidly to star-forming temperatures.  Cooling gas from
further out in the cluster would replace the gas in a slow, sub-sonic
inflow \citep[for a review, see][and references therein]{fabian:94}.

Even before the era of {\em Chandra} and {\em XMM} it was already
apparent that this simple picture of ongoing inflow of cooling gas did
not accurately describe cool core clusters: The implied rates for star
formation were an order of magnitude higher than the observed rates.

Radio surveys of cluster centers revealed that essentially all
traditional ``cooling flow'' clusters had active radio galaxies in
their centers \citep{burns:90}.  Generally, these radio sources are
Fanaroff-Riley type I sources (henceforth abbreviated as FR I; see
chapter ``Active galactic nuclei'' by E. Perlman in this book for a discussion of radio
source morphology), though it is not entirely clear whether this is
due to lower average source power compared to field FR II galaxies or
due to the increased gas density in clusters (frustrating source
evolution and possibly leading to increased entrainment).

Guided by the detailed examples of feedback in the Virgo and Perseus
clusters and the observed mismatch between the X-ray cooling rate and
the star formation rates in clusters, the first models of black hole
feedback in the context of galaxy clusters were presented in
\citet{binney:93}.

\subsubsection{The {\em Chandra} and {\em XMM-Newton} view}
\label{sec:cavities}

The role of AGN in regulating the cooling of gas in cool core clusters
was brought into clear focus with the launches of {\em Chandra} and
{\em XMM-Newton} in two ways:

{\em Chandra} observations revealed the presence of cavities just like
those found in the center of Perseus in virtually every cool core
cluster, providing the observational confirmation that the radio
galaxies present in these clusters {\em actively perturb} the gas.
Generically, the cavities appear to be surrounded by relatively cool
gas\footnote{This was surprising because one might naively expect the
gas most strongly affected by feedback to be hot.}.  Deep {\em
Chandra} observations sometimes reveal multiple cavities on different
scales, which has been interpreted as evidence for variability in the
AGN power\footnote{But see \citet{morsony:10} for arguments why the
presence of cavities is not a sufficient argument for AGN duty
cycles.}.

At the same time, {\em XMM-Newton's} high-resolution X-ray spectra of
the cluster centers revealed that radiative cooling must be impeded
below a threshold temperature of order 10 million degrees (about 1
keV), at a temperature where cooling should be efficient and rapid due
to the flattening of the cooling curve for thermal gas as atomic line
cooling becomes dominant \citep{peterson:03}.

This result is consistent with the observed lack of star formation in
central cluster galaxies, compared to the hundreds of solar masses of
star formation per year that would have been expected based on simple
cooling flow models.  It moves the discrepancy of cool gas missing at
{\em molecular} temperatures to cool gas missing at {\em X-ray}
temperatures below about a keV.  In either case, a heating agent is
needed, but in the ``revised'' cooling flow problem, the gas that must
be preferentially targeted is at about a keV and will thus occupy a
much larger volume fraction, which should make it easier to interact
with for any feedback mechanism.

Thus, after about a decade of study, a standard paradigm has emerged
from the high incidence of radio loud AGN in cool core clusters and
from the theoretical requirement of a heating agent that maintains the
temperature distributions in galaxy clusters: Radio galaxies provide
the energy needed to counterbalance cooling in the centers of
clusters.

From these observations, and from theoretical modeling
\citep{begelman:89,reynolds:01}, a simple understanding of radio
source evolution has been developed that underpins the radio galaxy
feedback paradigm. According to this picture, radio source evolution
separates into three stages \citep[e.g.][]{reynolds:01}:
\begin{enumerate}
\item{In the initial supersonic phase, jet plasma inflates cocoons
    that are strongly over-pressured relative to the environment.
    These cocoons must expand, and the rapid energy release implies
    that this expansion is supersonic in the frame of the environment.

    The expansion is similar to that of a wind-blown bubble described
    by \citet{castor:75}: The cavity radius roughly follows the
    self-similar scaling\footnote{This expression applies to bubbles
      smaller than a cluster pressure scale height. It is straight
      forward to extend it to stratified powerlaw atmospheres.}
    \begin{equation}
      R \propto \left(\frac{P_{\rm jet}t^3}{\rho_{\rm
            ICM}}\right)^{1/5}
      \label{eq:castor}
    \end{equation}
    which is functionally equivalent to the Sedov-Taylor solution if
    the blast energy is replaced by the injected energy over time,
    $P_{\rm jet}t$.  In this expression, $\rho_{\rm ICM}$ is the
    density of the environment.

    Note that this argument neglects the actual jet propagation
    completely and assumes the jet energy is randomized as the jet
    encounters the environment (in more powerful FR II sources) or
    through entrainment (which has been suggested as the dissipation
    agent in FR I source; \citealt{laing:02}).}
\item{As the source expands, the pressure inside the cocoon and in the
    shell eventually approaches the pressure of the environment and
    the expansion becomes sub-sonic.  The initially generated shock
    wave of the supersonic expansion will continue to coast outward,
    leaving behind a sub-sonically expanding cavity.  As the expansion
    velocity becomes sub-sonic, the confinement is dominated by the
    thermal pressure of the environment, so the solution changes to a
    pressure confined bubble, 
    \begin{equation}
      R \propto \left(\frac{P_{\rm jet}t}{p_{\rm e}}\right)^{1/3}
    \end{equation}}
\item{Finally, as the source expansion velocity drops below the
    buoyancy speed or once the pressure of the source drops below the
    dynamic pressure of motions in the environment, the
    cavities/cocoon will detach and float away from the black hole
    buoyantly or advectively.
    
    Once cold gas has refilled the central region of the cluster, any
    jet activity will start the cycle anew.}
\end{enumerate}

It is clear from eq.~(\ref{eq:castor}) that jet power and density are
the controlling variables in the initial evolution: Lower power jets
in denser environments will be more easily frustrated (with more
slowly expanding cocoons that become sub-sonic and unstable at smaller
sizes).  

\subsubsection{Estimating the kinetic power of a radio source}

Before the discovery of X-ray cavities, measurements of the {\em
  kinetic} power of radio galaxies (that is, the total power traveling
down the jets) were limited to estimates based on the observed
synchrotron emission (see discussion in \S\ref{sec:clusterfeedback}).
These were hampered by several factors:

\begin{itemize}
\item{Without knowledge of the field strength, an observed synchrotron
    flux, along with an estimate of the emitting volume could only
    provide a lower limit of the total energy in the radio plasma
    (essentially assuming equipartition between the energy in
    electrons and magnetic field).}
\item{The estimate of the volume depends strongly on the volume
    filling fraction, which is not measurable.}
\item{Synchrotron aging can cause electrons to cool and develop a
    sharp cutoff in the synchrotron spectrum.  Thus, significant
    amounts of energy in lower-energy electrons would be unaccounted
    for in the total power budget.}
\end{itemize}

One of the most important results from the discovery of X-ray cavities
in clusters is a robust, independent way to estimate the power of
cluster radio sources.  It is based on the fluid mechanics of inflated
bubbles: In order to inflate a cavity in the intra-cluster gas, the
jet must (a) displace the material in the environment into a shell
surrounding the cavity, which is of the order of $E_{pV} \sim pV$ (and
depends on the details of the inflation history of the bubble) and (b)
replace it with relativistic, magnetized gas.  At a minimum, the
amount of energy needed to do this is the work done on the cluster gas
and the internal energy of the radio plasma.  If the expansion of the
cavity is adiabatic, the total energy needed is
\begin{equation}
  E_{\rm cavity} = \frac{\gamma pV}{\left(\gamma - 1\right)} \sim 4 pV
\end{equation}
where $\gamma$ is the adiabatic index of the gas inside the cavity and
is typically assumed to be $\gamma = 4/3$, given the presence of
relativistic electrons and tangled magnetic fields.

Estimating the age of the cavity, and thus the jet power needed to
inflate it, is more difficult and introduces some uncertainty.  Direct
kinematic measurements of the expansion velocities of cavities are
impossible with current X-ray telescopes.  However, in many cases, the
observations suggest that the temperature of the shells surrounding
the cavities is {\em low}.  In this case, it is safe to assume that
the recent expansion of the cavity was sub-sonic.  Thus, the sound
crossing time of the bubble radius is a reasonable {\em lower} limit
on the cavity age:
\begin{equation}
  t_{\rm cavity} \geq \tau_{\rm sonic} = \frac{R_{\rm cavity}}{c_{\rm
      s}}
\end{equation}
where $R_{\rm cavity}$ is the radius of the cavity and $c_{\rm s}$ the
sound speed of the cluster gas.

Given that most cavities are found in cluster {\em centers}, a
reasonable {\em upper} limit on the age of the cavity is the buoyant
rise time $\tau_{\rm buoy}$ of the bubble in the gravitational
potential of the cluster (since an bubble older than $\tau_{\rm buoy}$
would have risen out of the cluster center):
\begin{equation}
  t_{\rm cavity} \leq \tau_{\rm buoy} \sim \frac{2R_{\rm cavity}}{v_{\rm
      buoy}} \sim \frac{2R_{\rm
      cavity}}{c_{\rm s}\sqrt{\frac{4}{3}\frac{d\ln{(P)}}{d\ln{R}}\frac{1}{C_{\rm
          W}}}} = 2\tau_{\rm sonic}\sqrt{\frac{3C_{\rm
        W}}{4}\frac{d\ln{(R)}}{d\ln{(P)}}} \sim 2\tau_{\rm sonic}
\end{equation}
where $C_{\rm W}$ is the drag coefficient of the rising cavity and
typically assumed to be of order $C_{\rm W} \sim 0.5$ and the exact
numerical value of the expression under the square root depends on the
pressure scale height of the cluster in bubble radii, but should be of
order unity for typical observed bubble radii\footnote{The buoyancy
  speed can never exceed the sound speed.}.  For {\em detached}
cavities, the appropriate age to use is the buoyancy time for the
projected distance instead of the bubble diameter.

Given that buoyancy sets in after the source becomes sub-sonic, this
loosely brackets the power inferred for the jets from the measurement
of cluster cavities to
\begin{equation}
  \frac{E_{\rm cavity}}{\tau_{\rm buoy} + \tau_{\rm sonic}} \sim
  \frac{E_{\rm cavity}}{3\tau_{\rm sonic}} \lesssim P_{\rm jet}
  \lesssim \frac{E_{\rm cavity}}{\tau_{\rm sonic}}
\end{equation}

Estimates of the central cluster density and temperature (and thus
pressure and sound speed) are readily obtained from X-ray images and
spectra.  The most difficult part is the estimate of the cavity
volume, since errors in the estimated cavity radius and viewing angle
uncertainties can compound to errors of up to an order of magnitude in
source power.  Nonetheless, this method has afforded us with a large
number of reliable estimates of jet powers to within a factor of a few
for dozens of radio galaxies in nearby clusters\footnote{It should be
  kept in mind that the inferred powers are averages over the cavity
  age, which can be between millions to hundreds of millions of years
  old.}.

While early, shallower {\em Chandra} exposures of cluster centers only
showed cool shells in the vicinity of the cavities, deep observations
of a number of important clusters later also showed the presence of
shocks surrounding at least some of the cavities
\citep[e.g.]{mcnamara:05,wise:07,forman:07}.
Given the generic picture of how radio sources evolve over time
described in \S\ref{sec:cavities}, the presence of {\em weak} shocks
should be expected (and had been predicted in \citealt{reynolds:01}).

In fact, an important corollary from eq.~(\ref{eq:castor}) is that the
initial strongly supersonic phase is short lived in typical cluster
environments: The expansion velocity of the shell is
\begin{equation}
  v_{\rm shell} = \frac{dR}{dt} \propto
  \left(\frac{P_{\rm jet}}{\rho_{\rm
        ICM}R^2}\right)^{1/3} \propto R^{-2/3} \propto t^{-2/5}
  \label{eq:shock}
\end{equation}
Given that cavities should remain in the cluster centers for about 2-3
sound crossing times before becoming buoyant, the fraction $f_{>M}$ of
time a given radio galaxy spends expanding super-sonically at or above
a given Mach number $M$, relative to the total dynamic lifetime before
buoyant removal, should only be of order
\begin{equation}
  f_{>M} \sim \frac{\tau_{>M}}{3\tau_{\rm sonic}} \sim \frac{1}{3}M^{-5/2}
  \label{eq:shocks}
\end{equation}
Thus, the observational lack of evidence for sources expanding at
large Mach numbers does not rule out that radio sources go through
this strong shock phase.  It is, however, short lived and only a small
mass fraction of AGN's environment passes through a strong shock.

The detection of a shock (which requires not just the detection of a
surface brightness jump, but also a temperature jump, which is most
easily identified through a harder X-ray color) offers a significantly
better diagnostic of jet power than the cavity method. Because the
shock strength is an indication of the expansion velocity, a measured
shock radius, brightness, and strength can be modeled using simple 1D
spherical shock models to give a reliable source power.  Typical Mach
numbers for shocks detected in clusters are between one and two,
consistent with this argument.

This rather simple parametric description of radio source evolution is
complemented by a growing body of numerical simulations of jet-driven
feedback.  Initial 2-dimensional simulations generally supported the
simple picture \citep{reynolds:01}.  However, a thorough understanding
requires full 3-dimensional simulations, and work on understanding the
details of feedback {\em ab initio} is still in the early stages.

A key problem posed by 2D and early 3D simulations is the apparent
contradiction of highly bipolar release of energy in the jet and the
need for mostly isotropic heating \citep{vernaleo:06}.  While outgoing
shocks assume a spheroidal shape relatively quickly, heating by shocks
is insufficient to solve the cooling flow problem (see
eq.~\ref{eq:shocks}).  In addition, jets evacuate cocoons around them.
Ongoing, unidirectional jet activity was found to propagate inside this
cocoon, dynamically and energetically isolated from the inner cluster
gas and thus unable to counteract cooling in the center.

The solution might lie in the interaction with the cluster and in the
internal dynamics of jets: \citet{heinz:06} found that the jets can
continue to efficiently couple to the inner cluster if their axes are
subjected to a moderate wobble and, crucially, the cluster itself is
dynamically evolved in the context of a cosmological simulation.
Simulations generally show that the complex X-ray appearance of
clusters, and even the appearance of multiple successive cavities, can
be generated by a single, ongoing episode of jet activity and a
dynamic cluster atmosphere \citep{morsony:10}.  This suggests that the
interaction between jets and clusters goes both ways: cluster dynamics
affects jet dynamics and vice versa.

\subsubsection{Cluster radio sources as a population}

\begin{figure}[t]
\begin{center}
\resizebox{!}{0.416\textwidth}{\includegraphics{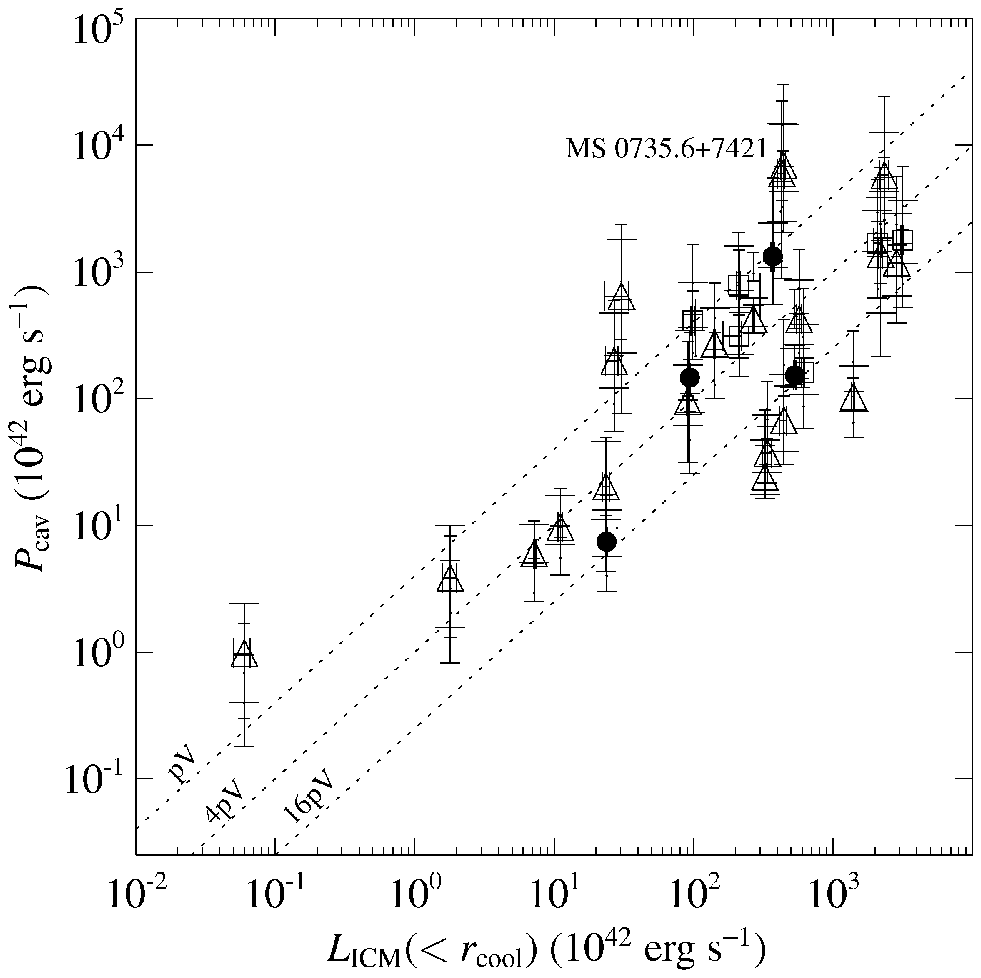}}
\hspace*{16pt}
\resizebox{!}{0.415\textwidth}{\includegraphics{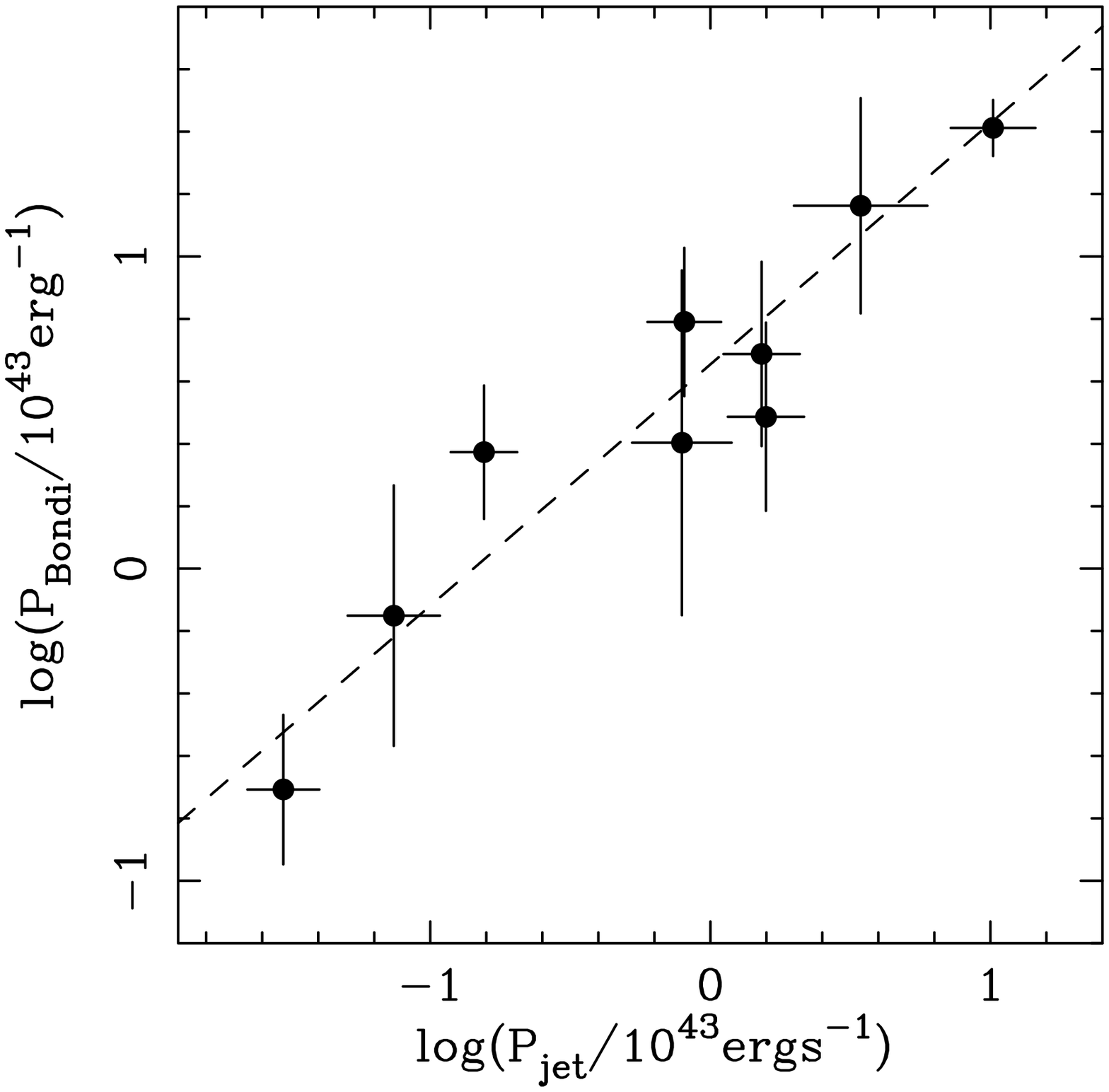}}
\end{center}
\caption{{\it Left}: jet power inferred from cluster cavities plotted
  against X-ray cooling luminosity from within the cluster cooling
  radius, adopted from \citet{rafferty:06}; the diagonal lines
  indicate the efficiency required for the jet power to offset
  cooling, with the line labeled as $4pV$ equivalent to 100\%
  efficiency, adopted from \citet{rafferty:06}; \label{fig:lr_lx} {\em
    right:} jet power plotted against Bondi accretion rate, adopted
  from \citet{allen:06}.\label{fig:pj_mdot}}
\end{figure}

A census of nearby clusters with cavities reveals a number of
important insights about the statistical and global properties of {\em
  central} cluster radio sources:
\begin{itemize}
\item{When plotting jet power vs.~X-ray luminosity, radio sources
    straddle the heating=cooling line: About half of the radio
    sources sampled have kinetic powers that are higher than the
    cooling luminosity of the cluster (see Fig.~\ref{fig:lr_lx}
    adapted from \citealt{rafferty:06}).}
\item{More importantly, the jet power appears to be correlated with
    the cooling luminosity of the cluster: More powerful radio sources
    are found in clusters with higher cooling rates. This is exactly
    the signature one would expect in an AGN feedback picture
    \citep{rafferty:06}.}
\item{The jet power measured from cavities appears to be related to
    the Bondi accretion rate of the cluster.  That is, in clusters
    with higher central densities and/or lower core temperatures, the
    jet power is higher \citep{allen:06}.  While this correlation does
    not imply that black holes actually accrete at the Bondi rate in
    clusters, the relatively close match suggests that black holes
    accrete directly from the cluster gas as it flows into the central
    galaxy (see Fig.~\ref{fig:pj_mdot}).}
\end{itemize}

Together, these results suggest that the balance between heating and
cooling in clusters is relatively tight.  In fact, the high
efficiencies suggested by these findings, in a global sense, as well
as from studies of individual, powerful black holes with relatively
low masses, have led to the suggestion that the black holes powering
the jets in cluster centers might even require the extraction of black
hole spin \citep{mcnamara:11}.

However, given the significant uncertainties in the jet power
estimates, and given that clusters are, in all likelihood, not steady
state systems (thus, cooling rates can temporarily exceed heating
rates, as long as they are balanced on average), a sufficient region
of parameter space is still allowed in which AGN can balance cooling
without requiring extreme efficiencies, accretion rates, or black hole
masses.

All of the detailed studies of radio galaxy feedback in clusters have
been limited to clusters at relatively low redshifts, because of the
need for high-fidelity X-ray images of cavities and shocks.  In
addition, high-redshift cluster samples are sparse.  Consequently,
constraints on cluster feedback at higher redshift are much harder to
obtain.  Radio surveys of high-redshift clusters do indicate an
increase in the cluster radio luminosity function \citep{branchesi:06}
hinting at an increase in feedback activity.

Finally, it is worth mentioning that, while the central massive
elliptical galaxies harbor by far the most massive black holes in
clusters, other cluster galaxies can be radio loud.  This situation
typically leads to the formation of a bent radio source.  The
exploration of the statistics of non-central cluster radio sources is
ongoing, and it has been suggested that this population of sources
could contribute to cluster feedback \citep{hart:09}.  However, it is
difficult to envision a scenario whereby a population of such sources
will dominate the heating rate in the average cluster.  It is also
unclear how they could respond to central gas cooling as would be
required for thermal regulation of the cluster gas.

\subsection{Feedback in groups}

Due to the lower temperatures and densities in the intra-group medium,
detecting evidence for group-wide feedback from X-ray observations is
significantly more difficult than in clusters.

\begin{figure}[t]
\begin{center}
\resizebox{!}{0.386\textwidth}{\includegraphics{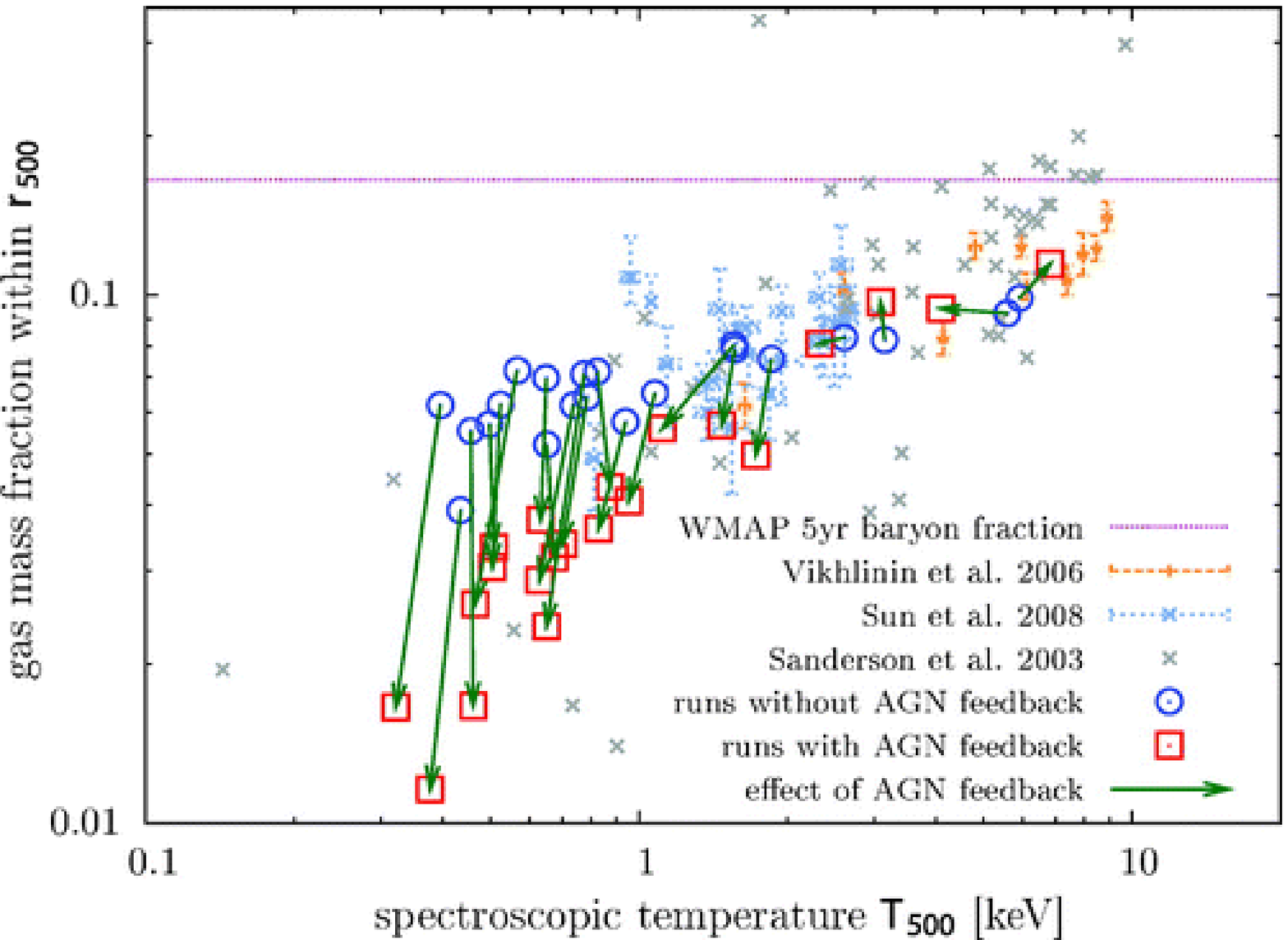}}
\resizebox{!}{0.386\textwidth}{\includegraphics{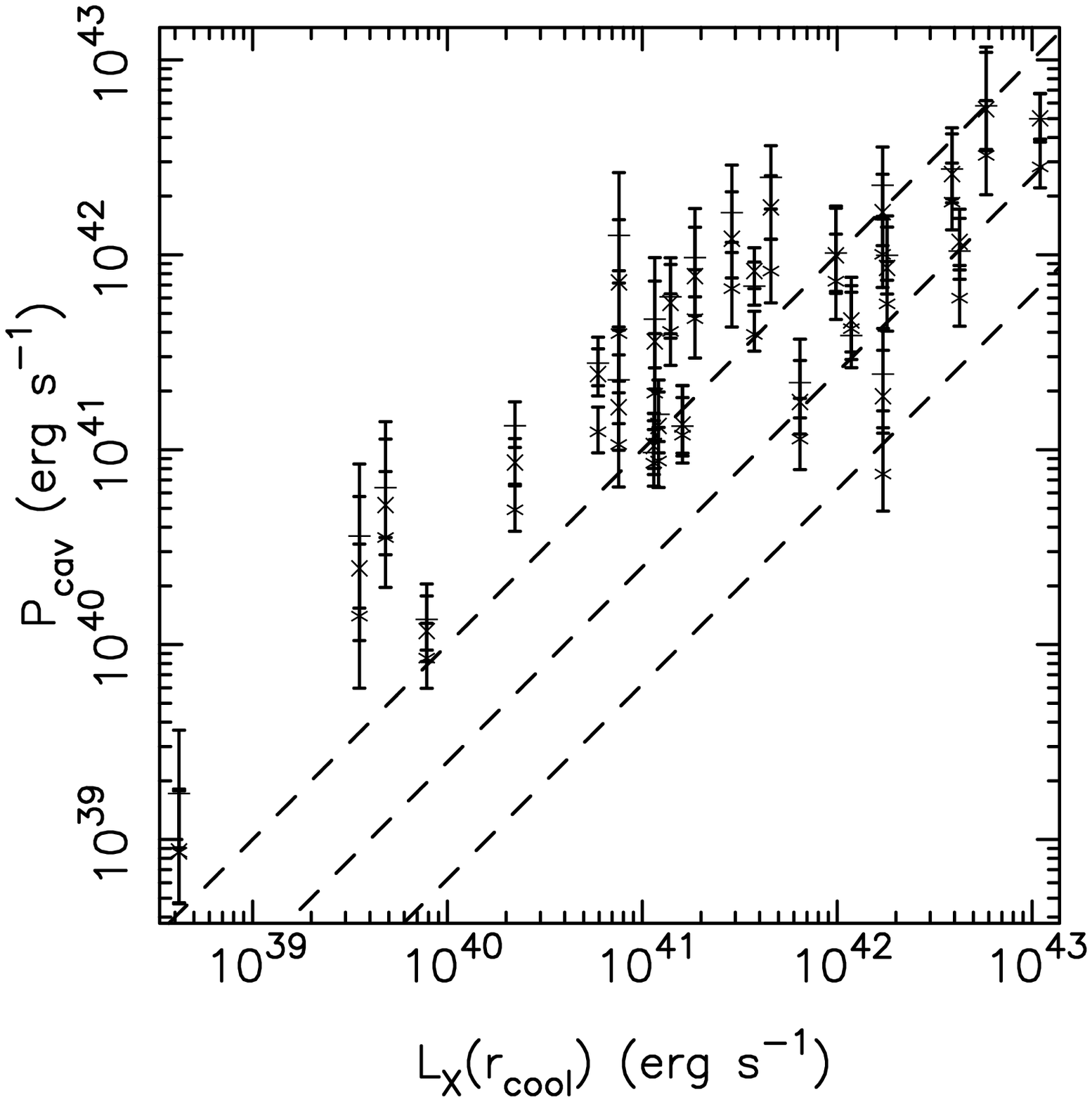}}
\end{center}
\caption{{\it Left}: Simulations show the increasingly important
  effect of AGN feedback on lower mass clusters and groups; plotted is
  the central gas mass fraction as a function of virial temperature
  (mass); adopted from \citet{puchwein:08} ; {\em right:} same plot as
  in the left panel of Fig.~\ref{fig:lr_lx}, but for atmospheres of
  elliptical galaxies instead of clusters; adopted from
  \citet{nulsen:07}.\label{fig:nulsen_puchwein}}
\end{figure}

Statistical X-ray studies of cluster and group samples show that,
generically, lower mass halos have excess central entropy when
compared to self-similar models of halo formation, indicating that an
additional source of non-gravitational heating must have injected
entropy preferentially into the low mass systems
\citep[e.g.][]{ponman:03}.  AGN have been suggested as a possible
heating mechanism \citep[e.g.][]{short:09}.  Figure
\ref{fig:nulsen_puchwein} shows how the inclusion of feedback in
detailed models of cluster and group atmospheres affects primarily low
mass systems and raises their central entropies.

And, in fact, surveys of groups suggest that such a scenario might
work: \citet{dong:10} find that a fraction of at least 25\% of the
groups in their survey contain clearly detectable cavities, with a
clear preference for cavities to be found in groups with {\em cooler}
cores (as is the case in clusters).  In their sample, the presence of
cavities does not appear to be correlated with the 1.4\,{\rm GHz}
radio flux.

On the other hand, a survey of radio properties of groups shows that
the central temperatures of groups with central radio sources are {\em
  elevated} compared to a radio quiet sample of groups, which was
suggested as a possible indication of ongoing radio galaxy feedback
\citep{croston:05}.

\subsection{Radio mode feedback in galaxies}

\begin{figure}[t]
\resizebox{!}{0.32\textwidth}{\includegraphics{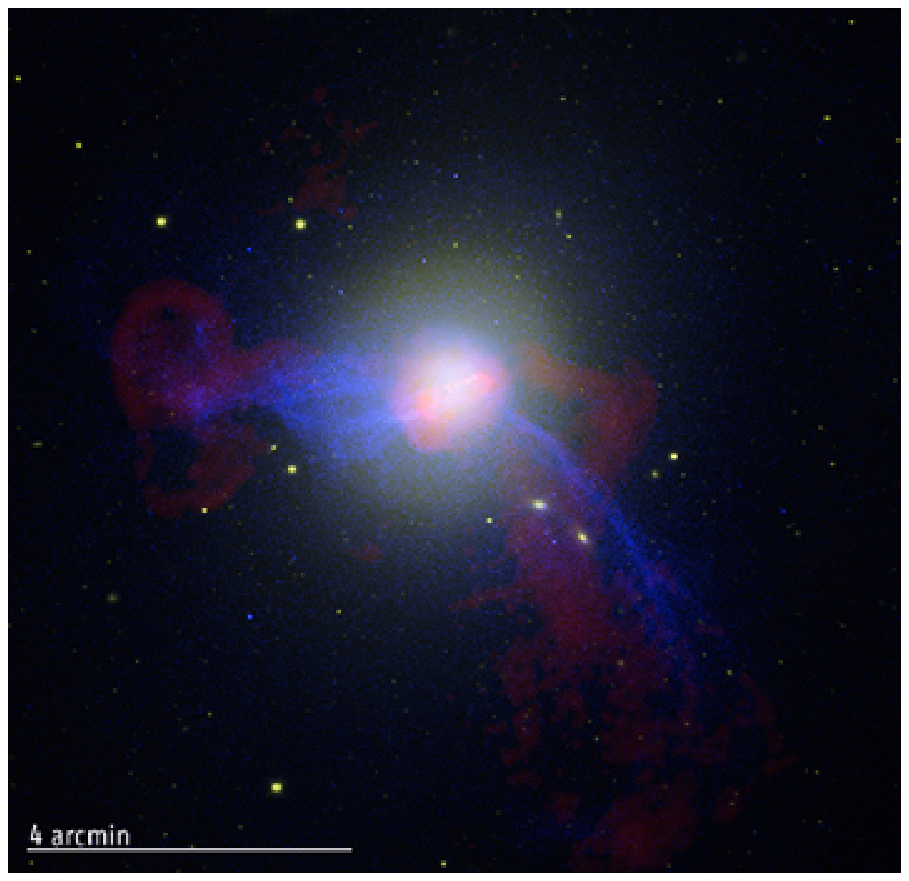}}
\resizebox{!}{0.32\textwidth}{\includegraphics{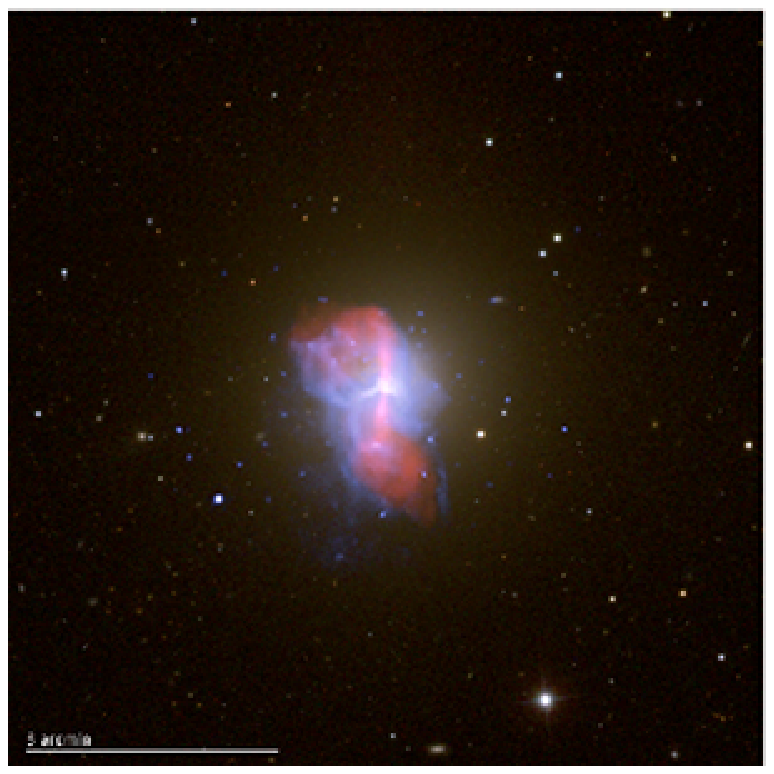}}
\resizebox{!}{0.32\textwidth}{\includegraphics{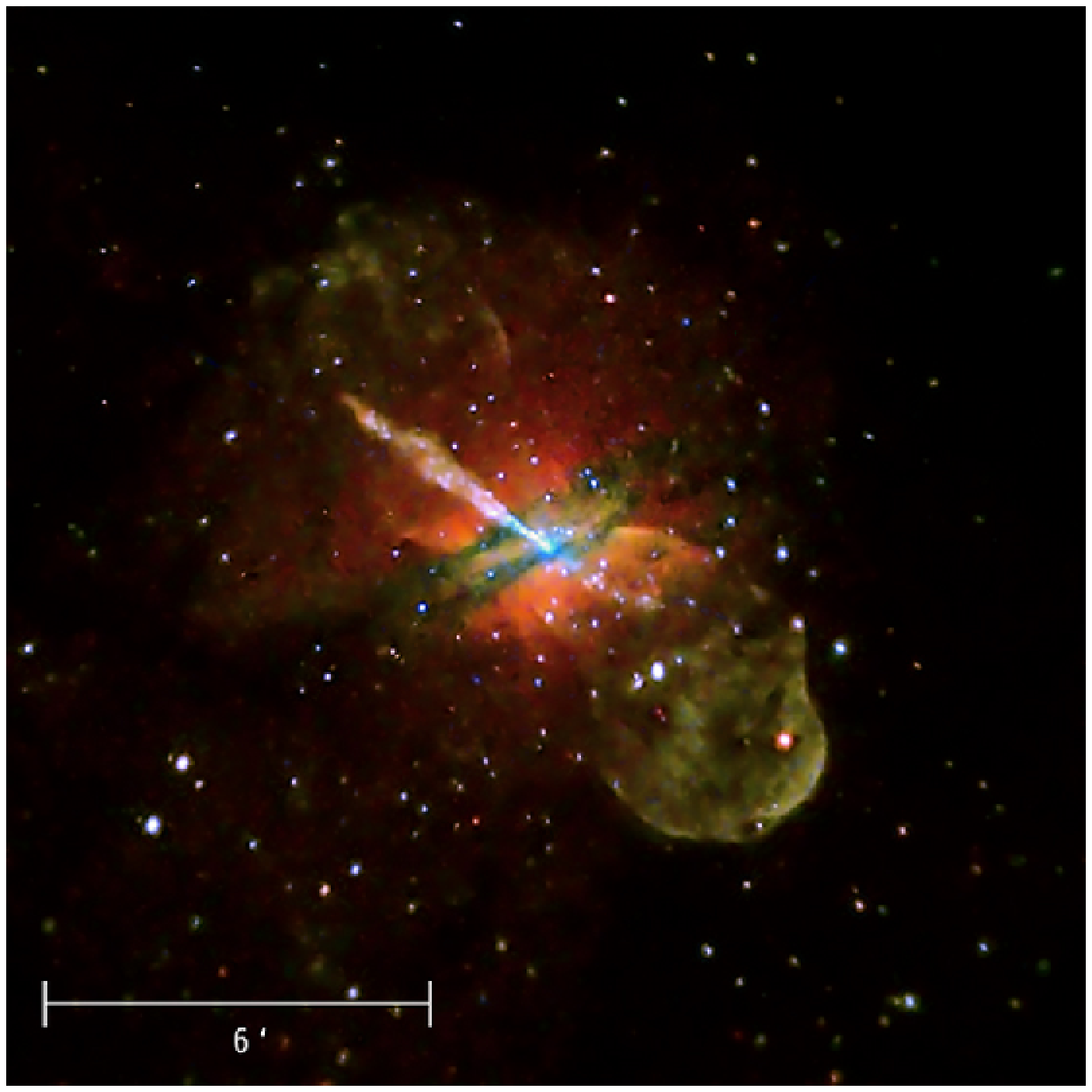}}
\caption{{\it Left}: multi-wavelength image of the central region of
  the Virgo cluster, showing the influence of the black hole on
  sub-galactic scales; {\em middle:} multi-wavelength image of the
  galaxy M84 in the Virgo cluster (blue: radio, red: X-ray, yellow:
  visible); {\em right:} multi-wavelength image of Centaurus A.}
\label{fig:galaxies}
\end{figure}

The best direct evidence for black holes affecting the surrounding gas
within galaxies comes again from combined radio and {\em Chandra}
observations of nearby objects: Fig.~\ref{fig:galaxies} shows images
of three nearby galaxies where jets clearly excavate cavities and
affect gas on sub-galactic scales: M87, M84, and NGC5128 (Centaurus
A).  In the latter case, the entrainment of gas into the jet, and the
formation of a strong shock driven by the expanding south-western
radio lobe is directly visible in X-rays.

These are clear local examples of relatively powerful, evolved radio
galaxies.  Numerous other examples have been studied individually.
Yet, perhaps the most convincing argument for the {\em importance} and
prevalence of jets in massive galaxies come from statistical studies.

In a study of X-ray cavities in and around nearby ellipticals,
\citet{nulsen:07} found that AGN power more than matched the cooling
luminosity of the gas in the hot galactic X-ray halo; compared to the
same study in clusters (shown in the left hand panel of
Fig.~\ref{fig:lr_lx}), the galaxies fall consistently above the
heating=cooling line \citep{mcnamara:07}.

\begin{figure}[t]
\begin{center}
\resizebox{!}{0.406\textwidth}{\includegraphics{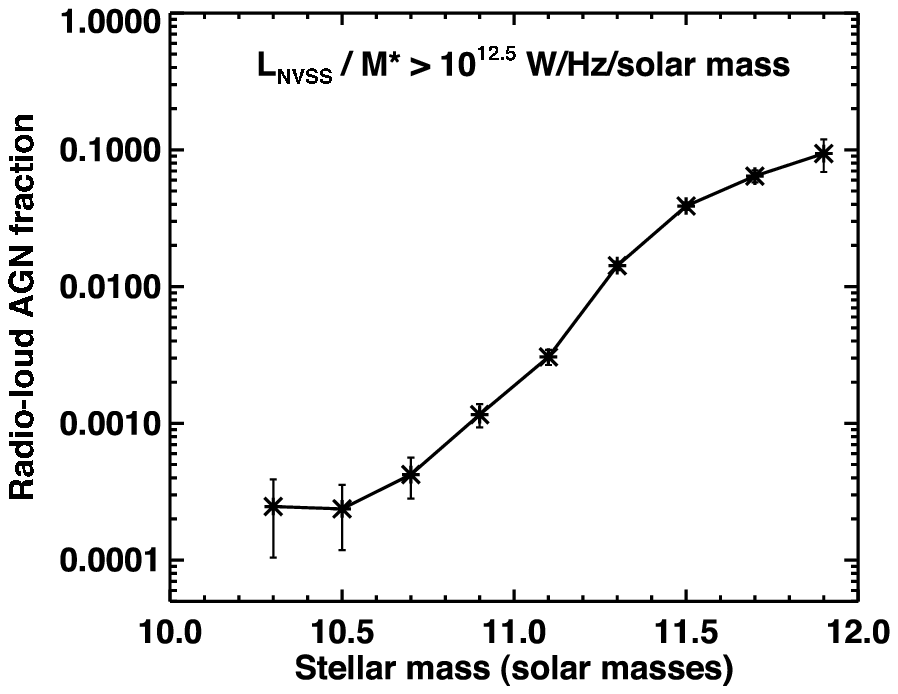}}
\resizebox{!}{0.406\textwidth}{\includegraphics[clip=true, trim=12.5cm
  0cm 0cm 0cm]{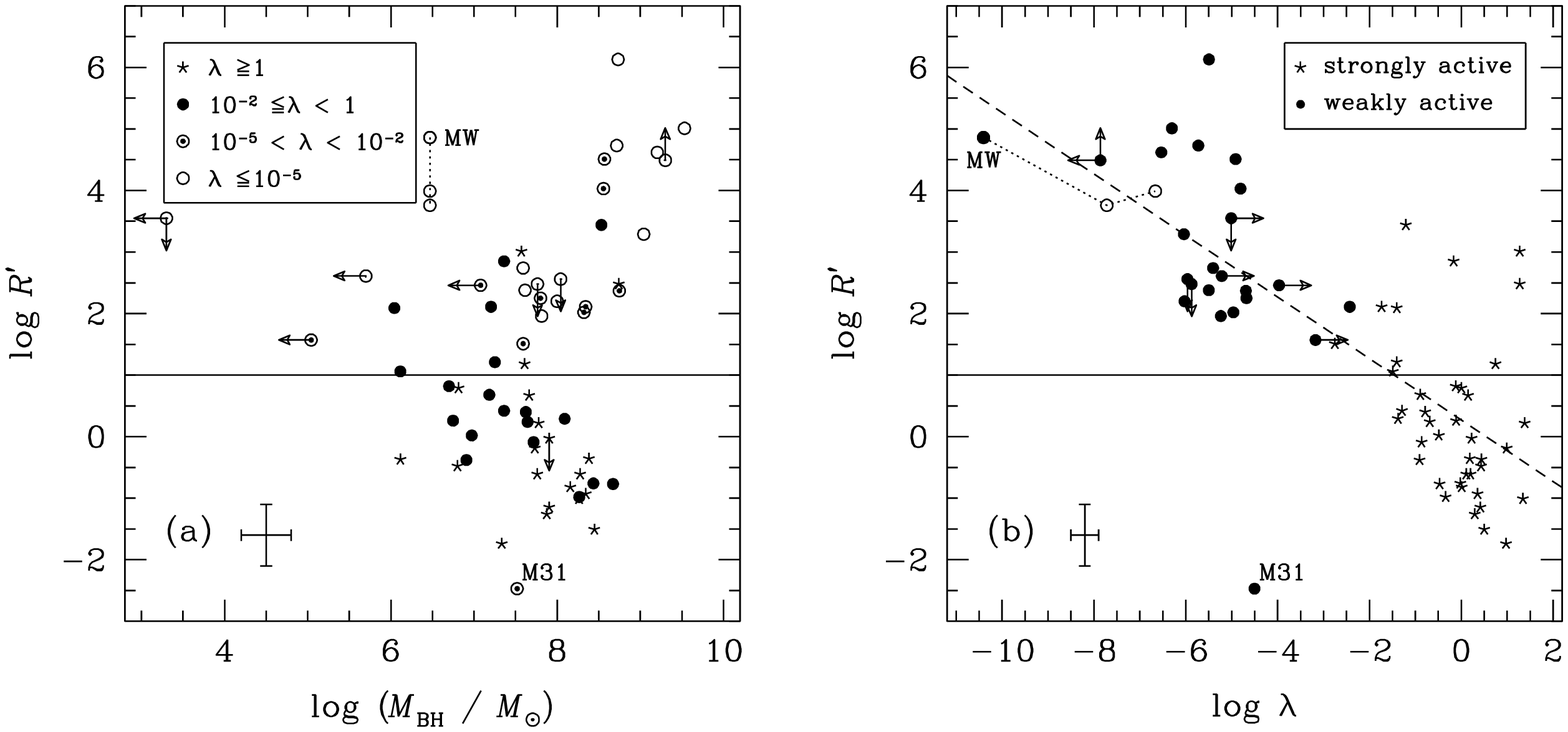}}
\end{center}
\caption{{\it Left}: Fraction of radio loud objects (defined as
  objects with a radio luminosity larger than $10^{12.5}\,{\rm
    W\,Hz^{-1}}$ per solar mass of stars of its host galaxy) plotted
  against stellar mass of the host, adopted from \citet{best:05}; {\em
    right:} Radio loudness $R'$ (here defined as 5GHz radio flux to
  2500$\AA$ flux) as a function of Eddington ratio $\lambda$ (defined
  as the ratio of bolometric AGN flux to Eddington flux), adopted from
  \citet{ho:02}.\label{fig:radioloud}}
\end{figure}

The implication that radio-mode feedback is important in massive
galaxies (those that have detectable X-ray halos) is complemented by
statistical studies of the radio source incidence in galaxies of
different type and mass: The left panel of Fig.~\ref{fig:radioloud}
shows a steady increase of the fraction of radio loud galaxies (that
is, galaxies above a fixed radio luminosity per stellar mass) with
stellar mass of the host galaxy \citep{best:05}.  The most massive
galaxies exhibit the highest radio fluxes {\em per stellar mass},
indicating that radio mode feedback is most active in today's most
massive galaxies, with 10\% or more of the most massive galaxies
hosting radio loud AGN.

Given that more massive galaxies harbor more massive black holes, one
might naively expect this result to suggest that radio loudness is a
fraction of black hole mass.  Early studies seemed to suggest this
\citep[e.g.][]{franceschini:98}.  However, deep radio surveys of a
wider class of black holes show that the radio loudness, defined
relative to the bolometric flux, actually {\em increases} for {\em
  decreasing} Eddington ratios \citep{ho:02}: The right panel of
Fig.~\ref{fig:radioloud} shows that radio emission increases in
relative brightness (compared to the bolometric luminosity) for lower
luminosity AGN.

This suggests that black holes become relatively {\em more} efficient
at liberating energy in the form of jets as their luminosity (and
presumably accretion rate) drops, such that {\em all} black holes at
sufficiently low accretion rate appear to be driving some form of
radio-loud outflow.

The idea that low luminosity black holes are universally radio loud
arose roughly in parallel also in the study of X-ray binary black
holes \citep{gallo:03}, where it is possible to track individual black
holes across outburst and decline into quiescence.  These observations
showed that a jet was always present at low luminosities, with
increasing {\em relative} radio flux at decreasing X-ray luminosities.

While a thorough theoretical understanding is still missing, it has
been shown that low-efficiency accretion leads to the formation of
geometrically thick flows, as opposed to the geometrically thin
accretion disks found in, for example, Seyfert galaxies and quasars.
In such a geometrically thick (quasi-spherical) flow, it might be much
easier to build up significant magnetic flux, even just from
stochastic turbulent dynamo processes in the disk, which could in turn
drive the jet (see chapter ``Active Galactic Nuclei'' by E. Perlman in
this book).

Anecdotally, the case of the M87 jet makes the perfect illustration of
this point: The total radiative output from the black hole is
unimpressive\footnote{Much of this radiation may actually be in the
  form of X-rays from the unresolved base of the jets itself} at
$L_{\rm M87} \sim 10^{41}\,{\rm ergs\,s^{-1}}$, about two orders of
magnitude below the luminosity corresponding to efficient accretion at
the Bondi rate of that particular black hole. Meanwhile, the estimated
jet power from this object is about two orders of magnitude larger
\citep{forman:07}, consistent with the accretion power one would
derive if the black hole were accreting at the Bondi accretion rate,
thus making M87 a jet-dominated low-luminosity black hole.

While a direct translation of radio loudness into feedback efficiency
is challenging, this result suggests that black hole feedback might be
more prevalent than just in the most powerful AGN in the most massive
galaxies.  In fact, the ubiquity of low-luminosity AGN, and their 
increased dominance at low redshift (see Fig.~\ref{fig:smolcic}) supports both
the idea of a radio maintenance mode of ongoing feedback in early type
galaxies \citep{croton:06}, as well as the picture of gentle
feedback (through buoyancy rather than shocks) not just in clusters
but also in galaxies.

Additionally, ongoing low-level jet activity seems much better suited
at targeting gas within the galaxy than explosive outbursts in the
form of powerful radio galaxies that would transport most of the
energy far beyond galactic scales, and the prominent examples of radio
mode feedback shown in Fig.~\ref{fig:galaxies} should be considered
extreme outliers from more typical radio-mode feedback.  This
statement can, in fact, be quantified based on the global AGN
evolutionary models outlined in chapter \ref{sec:mf_evol}, which we
shall discuss next.

%%%%%

\subsection{Quantifying the efficiency of the radio mode}

The observed omni-presence of radio cores\footnote{The ``core'' of a
  jet is the brightest innermost region of the jet, where the jet just
  becomes optically thin to synchrotron self absorption, i.e., the
  synchrotron photosphere of the jet.} in low luminosity AGN and the
observed increase in radio loudness of X-ray binaries at low
luminosities can be placed on a solid theoretical footing.  Jets
launch in the innermost regions of accretion flows around black holes,
and at low luminosities, these flows likely become mechanically (i.e.,
advectively) cooled.

Such flows can, to lowest order, be assumed to be scale invariant: a
low luminosity accretion flow around a 10 solar mass black hole,
accreting at a fixed, small fraction of the Eddington accretion rate,
will be a simple, scaled down version of the same flow around a
billion solar mass black hole (with the spatial and temporal scales
shrunk by the mass ratio).  It follows, then, that jet formation in
such a flow should be similarly scale invariant.

This assumption is sufficient to derive a very generic relation
between the radio luminosity emitted by such a scale invariant jet and
the total (kinetic and electromagnetic) power carried down the jet,
{\em independent} of the unknown details of how jets are launched and
collimated \citep{heinz:03}: The synchrotron radio luminosity
$L_{\nu}$ of a self-absorbed jet core depends on the jet power $P_{\rm
  jet}$ through
\begin{equation}
  L_{\rm radio} \propto P_{\rm jet}^{\frac{17+8\alpha}{12}}M^{-\alpha} \sim
  P^{\frac{17}{12}}\label{eq:scaling}
\end{equation}
where $M$ is the mass of the black hole and $\alpha \sim 0$ is the
observable, typically flat radio spectral index of the synchrotron
power-law emitted by the core of the jet.  This relation is a result
of the fact that the synchrotron photosphere (the location where the
jet core radiates most of its energy) moves further out as the size
scale and the pressure and field strength inside the jet increase
(corresponding to an increase in jet power).  As the size of the
photosphere increases, so does the emission.  The details of the
power-law relationship are an expression of the properties of
synchrotron emission.

For a given black hole, the jet power should depend on the accretion
rate as $P_{\rm jet} \propto \dot{M}$ (this assumption is implicit in
the assumed scale invariance).  On the other hand, the emission from
optically thin low luminosity accretion flows itself depends
non-linearly on the accretion rate, roughly as $L_{\rm acc} \propto
\dot{M}^2$, since two body processes like bremsstrahlung and inverse
Compton scattering dominate, which depend on the square of the
density.  Thus, at low accretion rates, $L_{\rm radio} \sim L_{\rm
  bol}^{\frac{17}{24}}$, which implies that black holes should become
more radio loud at lower luminosities
\citep{heinz:03,merloni:03,falcke:04}.  It also implies that more
massive black holes should be relatively more radio loud than less
massive ones, at the same {\em relative} accretion rate $\dot{M}/M$.

Equation \ref{eq:scaling} is a relation between the observable core
radio flux and the underlying jet power.  Once calibrated using a
sample of radio sources with known jet powers, it can be used to
estimate the jet power of other sources based on their radio
properties alone (with appropriate provisions to account,
statistically, for differences in Doppler boosting between different
sources).

\begin{figure}[t]
\begin{center}
\resizebox{!}{0.466\textwidth}{\includegraphics{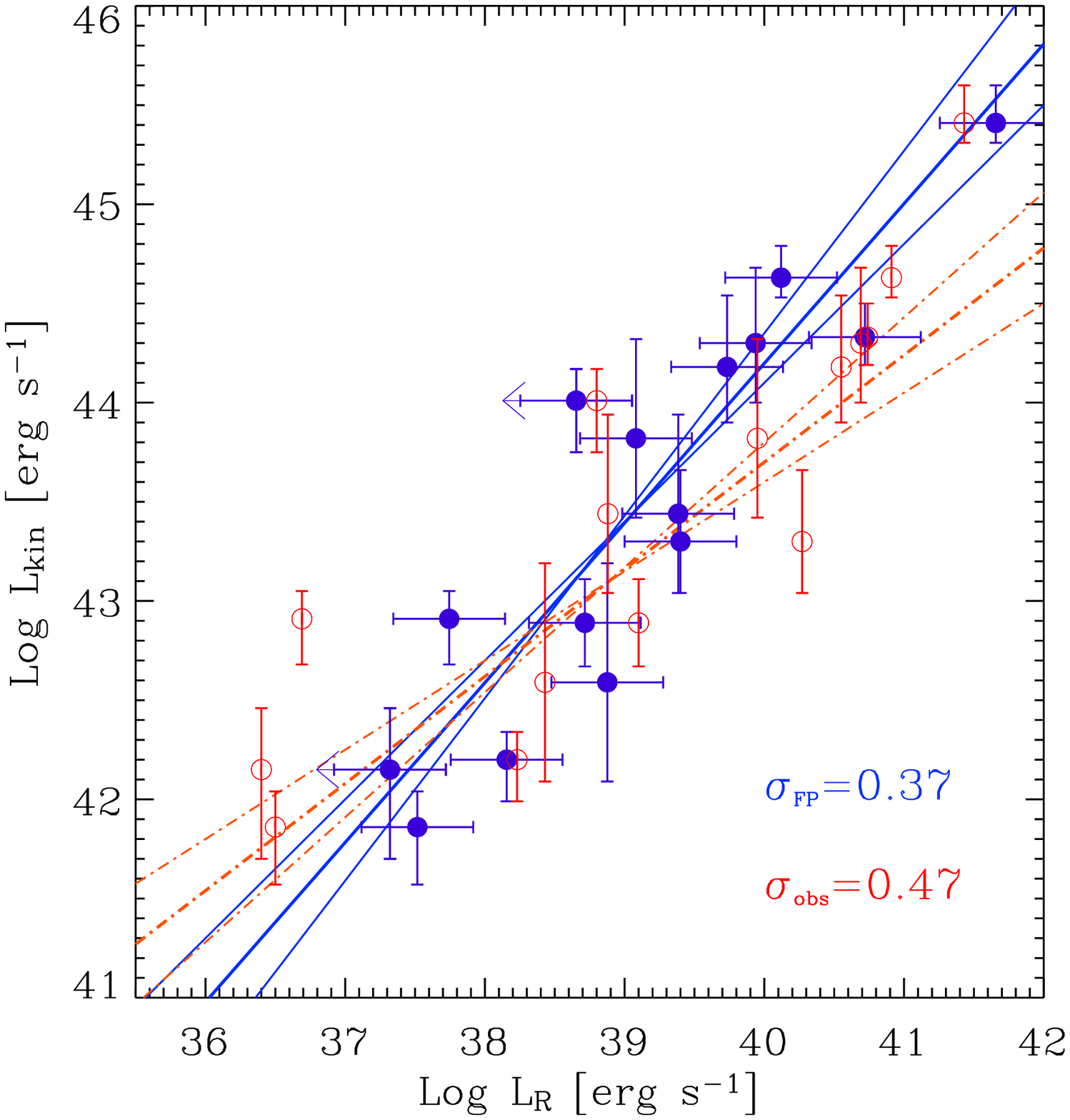}}
\resizebox{!}{0.526\textwidth}{\includegraphics{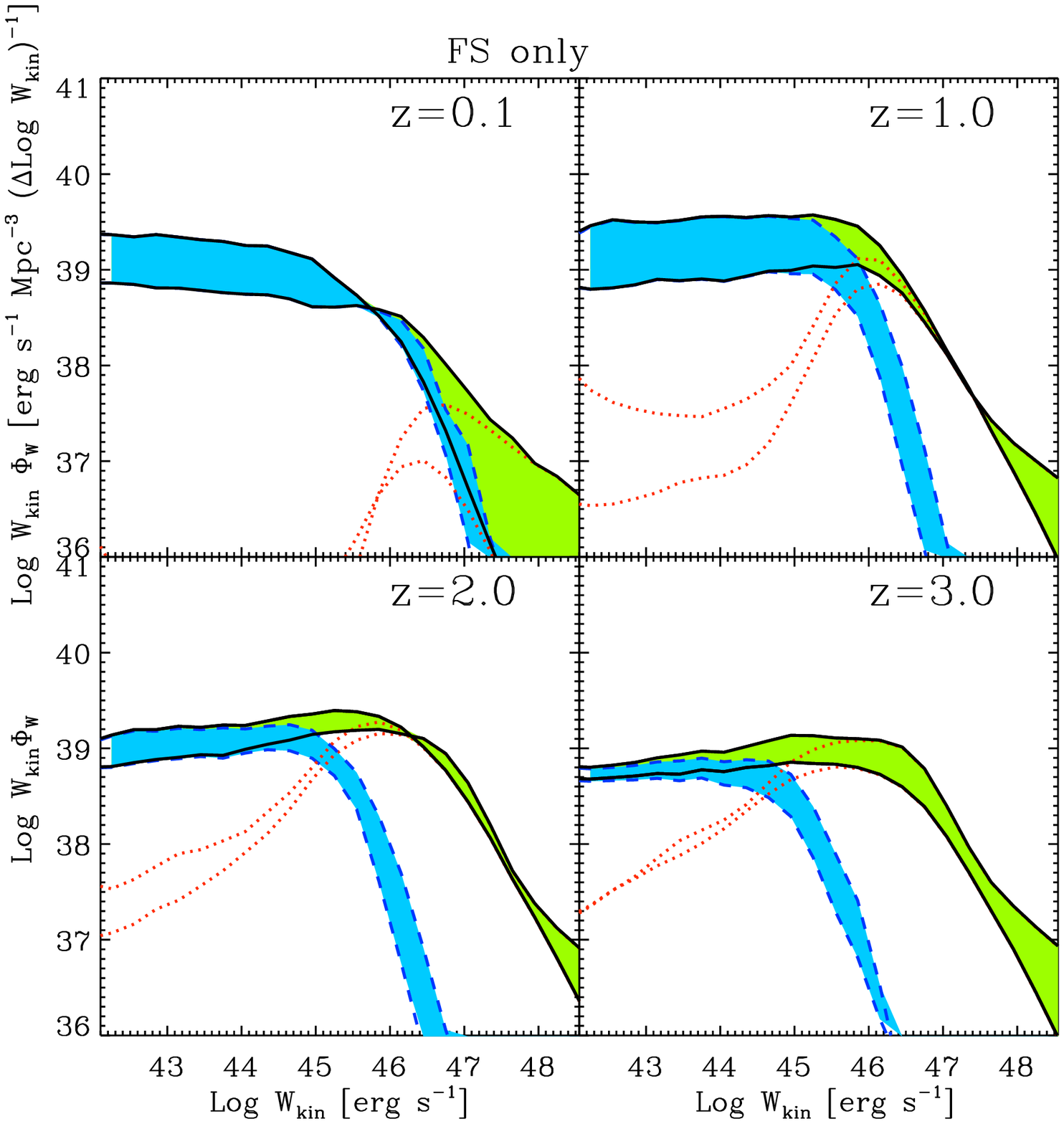}}
\end{center}
\caption{{\it Left}: Red open circles show jet power (measured from
  X-ray cavities) plotted against 5 GHz core radio luminosity (also
  shown in blue solid circles is a Doppler boosting corrected version
  of the same points) along the power-law fit given in
  eq.~(\ref{eq:power}); adopted from \citet{merloni:07}; {\em right:}
  Kinetic jet luminosity function derived from flat spectrum (FS)
  radio galaxy surveys, for different redshift bins (green
  curve shows all radio sources, blue shows radio sources in the radio
  mode, red shows radio sources in quasar mode); adopted from
  \citet{merloni:08}.\label{fig:scaling}}
\end{figure}

The cluster radio sources shown in Fig.~\ref{fig:scaling} provide such a
sample.  Plotting the core (unresolved) radio power against the jet
power inferred from cavity and shock analysis shows a clear non-linear
relation between the two variables \citep{merloni:07}.  Fitting this
relation provides the required constant of proportionality and is
consistent (within the uncertainties) with the power-law slope of
$17/12$ predicted by eq.~(\ref{eq:scaling})
\begin{equation}
  P_{\rm jet} = P_{0}\left(\frac{L_{\rm radio}}{L_{0}}\right)^{\zeta}
  \sim 1.6\times 10^{36}\,{\rm ergs\,s^{-1}} \left(\frac{L_{\rm
        radio}}{10^{30}\,{\rm ergs\,s^{-1}}}\right)^{0.81}
  \label{eq:power}
\end{equation}
with an uncertainty in the slope $\zeta$ of 0.11, where $L_{\rm
  radio}=\nu L_{\nu}$ is measured at $\nu=5\,GHz$.

Because this relation was derived for the {\rm cores} of jets, which
display the characteristic flat self-absorbed synchrotron spectrum,
care has to be taken when applying it to a sample of objects: only the
core emission should be taken into account, while extended emission
should be excluded.  As discussed in \S\ref{sec:radio_lf}, radio
luminosity functions are separated spectrally into flat and steep
sources, and we can use both samples to limit the contribution of flat
spectrum sources from both ends.

Given a radio luminosity function $\Phi_{\rm rad}$ and an appropriate
correction for relativistic boosting, eq.~\ref{eq:power} can be used
to derive the kinetic luminosity function of jets
\citep{heinz:07,merloni:08}:
\begin{equation}
  \Phi_{\rm kin}(P_{\rm jet}) = \Phi_{\rm
    rad}\left[L_{0}\left(\frac{P_{\rm
          jet}}{P_{0}}\right)^{\frac{1}{\zeta}}\right]
  \frac{1}{\zeta}\frac{L_{0}}{P_{0}}
  \left(\frac{P_{\rm jet}}{P_{0}}\right)^{\frac{1-\zeta}{\zeta}}
\end{equation}
The resulting kinetic luminosity functions for the flat spectrum radio
luminosity functions\footnote{Comparison to the steep spectrum
  luminosity function shows that the error in $\Phi_{P}$ from the
  sources missed under the steep spectrum luminosity function is at
  most a factor of two} from \citet{dunlop:90} and \citet{dezotti:05}
are plotted in the right panel of Fig.~\ref{fig:scaling}.

Since the figure plots $P\cdot\Phi_{\rm P}$, the curves show {\em
  directly} the total contribution of AGN at a given jet power to the
total feedback power at a given redshift.  At the low luminosity end,
these curves are roughly flat, implying that low luminosity source
contributed a significant fraction of the total power.  These are the
low-luminosity AGN presumably responsible for radio mode feedback, and
they dominate the total jet power output at low redshift.

Integrating the luminosity function over $P_{\rm jet}$ gives the local
jet power density $\rho_{\rm Pjet}$, which, at redshift zero, is of
the order of $\langle \rho_{\rm Pjet}\rangle \sim 6\times
10^{39}\,{\rm ergs\,s^{-1}\,Mpc^{-3}}$, which is comparable to the
local power density from supernovae, but will be significantly above
the supernova power in early type galaxies (which harbor massive black
holes prone to accrete in the radio mode but no young stars and thus
no type 2 supernovae).

Finally, integrating $\Phi_{\rm kin}$ over redshift gives the total
kinetic energy density $u_{\rm Pjet}$ released by jets over the
history of the universe, $u_{\rm Pjet} \sim 3\times 10^{57}\,{\rm
  ergs\,Mpc^{-3}}$.  By comparing this to the local black hole mass
density $\rho_{\rm BH}$ we can derive the average conversion
efficiency $\eta_{\rm jet}$ of accreted black hole mass to jet power:
\begin{equation}
  \eta \equiv \frac{u_{\rm Pjet}}{\rho_{\rm BH}c^2} \approx 0.2\% -0.5\%
\end{equation}
In other words, about half a percent of the accreted black hole rest
mass energy gets converted to jets, {\em averaged} over the growth
history of the black hole.

Since most black hole mass was accreted during the quasar epoch, when
black holes were mostly radio quiet, about 90\% of the mass of a given
black hole was accreted at zero efficiency (assuming that only 10\% of
quasars are radio loud).  Thus, the average jet production efficiency
during radio loud accretion must be at least a factor of 10 higher,
about 2\%-5\%, comparable to the {\em radiative} efficiency of
quasars.  These are exactly the kinds of efficiencies needed for radio
mode feedback to work.

%%%%%

\subsection{Quasar mode feedback}
\label{sec:qso_feedback}

\begin{figure}[t]
\begin{center}
\resizebox{!}{0.466\textwidth}{\includegraphics{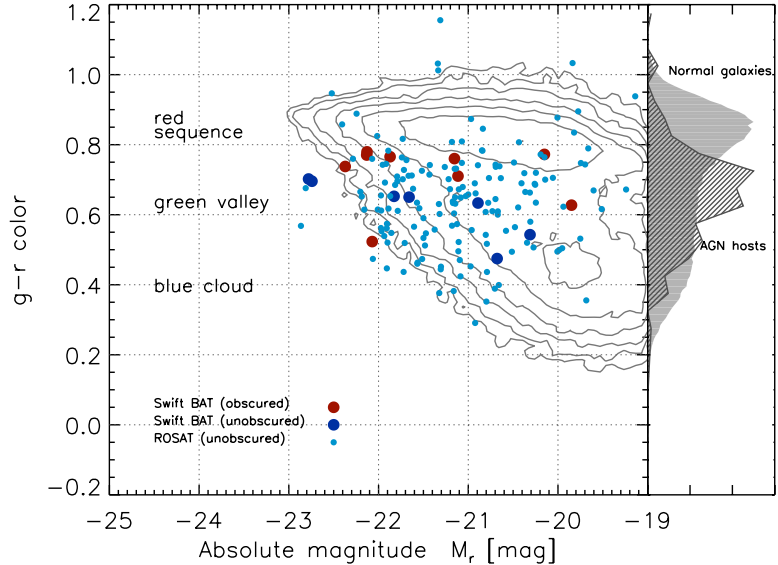}}
\end{center}
\caption{Contours of galaxy counts as a function of R-band magnitude
  and g-r color, showing the existence of two separate populations -
  the red sequence on top and the blue cloud on the
  bottom. Over-plotted are the locations of X-ray selected AGN (hard
  and soft), with a clear preference for a location in the ``green
  valley'' in between both populations; adopted from
  \citet{schawinski:09}.}
\label{fig:qsofeedback}
\end{figure}

Arguments for black hole feedback on galactic scales stem primarily
from three facts:
\begin{itemize}
\item{The deviation of the bright end of the galaxy luminosity
    function from self-similar predictions \citep{springel:05}, that
    is, a dearth of bright galaxies.}
\item{The bimodality of the galaxy distribution in color-magnitude
    space \citep{strateva:01}, with early type galaxies forming the
    ``red sequence'' and late type galaxies forming the ``blue
    cloud''.  Between these two populations lies, naturally, the
    so-called ``green valley'', which is relatively sparsely
    populated.  This bimodality is shown in
    Fig.~\ref{fig:qsofeedback}. }
\item{The tight relation between stellar bulge mass and black hole
    mass, which suggests a common formation scenario. Given that
    massive black holes grew mostly as quasars during the epoch of
    star formation, this suggests a relationship between both.  In
    fact, the argument for quasar mode feedback was first made in part
    to motivate black hole -- galaxy scaling relations
    \citep[e.g.][]{silk:98,wyithe:03}.}
\end{itemize}
Rapidly growing black holes are attractive as agents of feedback on
ongoing star formation because they have similar growth histories,
they can be found in the centers of galactic bulges (the stellar
populations their feedback is supposed to influence), they can release
large amounts of energy isotropically, and they are likely to be
fueled rapidly in response to galaxy mergers, which also trigger star
formation.

Numerical simulations of black hole feedback in individual galaxy
mergers have produced impressive visualizations of how rapid,
isotropic energy injection by a growing black hole can heat and
disperse the cool, star forming gas, in essence explosively
terminating star formation and black hole growth \citep{dimatteo:05}.
In part as a result of these successes, quasar mode feedback is now
routinely incorporated into cosmological simulations of structure
formation and semi analytic models of galaxy formation
\citep{croton:06,bower:06}.

In these simulations the prescription of how black holes accrete is
simplified to variations of the Bondi accretion rate, necessitated by
the unresolvably vast dynamic range of the problem.  Energy is
injected isotropically, which is an appropriate zero order choice
given our lack of knowledge about the actual channel through which the
energy is delivered.  What the simulations tell us is that efficient
black hole feedback {\em can} regulate star formation and black hole
growth.  But because black hole feedback and supernova feedback are
operationally very similar, and because the presumed AGN feedback
mechanism is generic, it is difficult to extract more detailed
information about the AGNs themselves from the models.

In addition, the causality of the interaction of black holes with the
star forming gas surrounding them is not yet fully established.  It is
also plausible that star formation itself provides the feedback
through supernovae, and that competitive accretion starves both black
holes and stars, leading to a passive link between black holes and
stars.

Identifying currently ongoing episodes of feedback has proven to be
difficult, in part because of the large degrees of visible- and soft
X-ray extinction towards star forming regions and because of the small
angular scales involved.  Proving the {\em causal} connection between
AGN activity and terminated star formation is even more difficult.

Generally, one would assume that galaxies caught in the act of
feedback should just start their transition from the blue cloud to the
red sequence, as the population of recently formed early stars fades
without any replenishment.  The relative under-density of galaxies in
the green valley suggests that this transition is a relatively rapid
process (one would expect it to occur roughly on A-star life times).

Stellar population modeling has successfully been used to identify
such post-starburst galaxies.  And indeed, sources have been found
among this class that show clear evidence for very fast outflows in
excess of $1000\,{\rm km\,s^{-1}}$ \citep{tremonti:07} that might be
the smoking gun.  Estimating the mass in the outflow has proven to be
difficult, and we have to await deep imaging that can directly resolve
the outflow to quantify the energetic impact of the AGN on the
galactic gas.

In addition, surveys of (hard) X-ray selected AGN find these sources
to preferentially lie in the green valley \citep[from the all-sky {\it
  Swift-BAT} survey; ][]{schawinski:09}, as
can be seen in Fig.~\ref{fig:qsofeedback}.  Since AGN accretion time scales
can be expected to be shorter than the transition time across the
green valley, this observation suggests that the AGN activity comes
{\em after} star formation has been terminated.  Since this is true
also for hard X-ray selected AGN, this conclusion should not be
affected by obscuration unless AGN in the act of feedback are so
heavily obscured that even {\it Swift} cannot detect them.

Thus, the question of what is at the heart of the putative quasar mode
feedback is left unanswered.  Generally, AGN can release energy via
three channels: Through radiation, through jets, and through
un-collimated outflows (i.e., ``winds'').

The most obvious source of feedback energy in efficiently accreting
black holes is, of course, the radiation itself.  Since most bright
AGN are obscured, we can infer that a high fraction of the initially
emitted light is reabsorbed by the surrounding gas.  If some of the
energy is deposited on sufficiently large scales (rather than into gas
bound to the black hole), it can in principle supply the energy for
feedback.  Models of radiative feedback \citep{ciotti:01,sazonov:05}
generally rely on Compton heating.  The efficiency of radiative
feedback requires  about 10\% of the radiation to be absorbed on
scales outside of the Bondi radius but within the star forming region
of the host.  Whether radiative transfer will always conspire to
provide such an arrangement is an open question.

Efficiently accreting AGN (quasars and Seyferts) are also known to
generate massive winds: Optical absorption line studies show outflows
at velocities of thousands of kilometers per second.  The most
dramatic demonstration comes from the class of broad absorption line
quasars (BAL QSOs), which show high column densities of absorption in
visible and X-rays (indicating high mass fluxes) and wind velocities
up to $50,000\,{\rm km\,s^{-1}}$.

If column density measurements made in the X-ray band trace the same
gas that produces the large outflow velocities in optical absorption
lines, the power and mass contained in these winds would be of the
same order as the total radiative power of these objects
\citep{furlanetto:01}. Driving such a wind would presumably require
some form of mechanical input (for example, from magneto-centrifugal
launching as described in chapter ``Active Galactic Nuclei'' by
E. Perlman in this same volume) in
addition to radiative driving.

The ubiquity of outflow signatures in efficiently accreting AGN,
coupled with the large wind efficiencies inferred from the more
extreme cases, have made AGN winds the primary mechanism invoked in
feedback models \citep[see e.g.][]{king:05}.  Given the uncertainties
in column density of the high-velocity gas, a direct imaging detection
of outflow signatures (like the cavities in the case of AGN feedback
in clusters) would provide more certainty that winds can affect the
surrounding gas on the scales needed for feedback to operate.

The uncertainty about whether winds are powerful enough to drive
feedback raises the interesting question whether episodes of {\em
  powerful jet} activity in quasars can lead to feedback on galactic
scales, and whether they can be observed.  In the simple framework of
radio source dynamics laid out in \S\ref{sec:cavities}, an episode of
jet activity will inflate a supersonically expanding cocoon, the size
scale of which depends on the jet power and the density of the
environment.

In the dense environments of star forming regions, one might thus
expect sources to go through a compressed evolution, with slowed or
even stalled expansion as sources run into dense gas.  In such a
scenario, the initial expansion might produce strong shocks (given the
cold gas they encounter) but at much reduced shock temperatures given
the slower expansion.  Is it possible that powerful radio sources in
dense environments can heat the gas sufficiently to provide the quasar
mode feedback postulated by semi-analytic models?

Given that about 10\% of all powerful quasars are radio loud, and
given that the required {\em average} feedback efficiency for the
quasar mode can be an order of magnitude lower still, jet powered
feedback may actually contribute significantly to the quasar mode as
well.  In fact, a class of sources that might represent these powerful
radio quasars in the act of feedback exists in the so-called
Compact-Steep-Spectrum (CSS) and the Gigahertz-Peaked-Spectrum (GPS)
sources \citep{odea:98}.

These are small scale radio sources that show clear signs of strong
absorption to the radio spectrum (indicating a high local pressure and
thus ISM density) and otherwise appear similar to classical radio
sources but on smaller scales.  The cause for their compactness has
been debated since their discovery: They might be young sources, in
the very early stages of supersonic expansion, or frustrated older
sources, caught in very dense environments.  In either case, this
would be a population of sources directly heating the dense gas in the
centers of galaxies, where quasar mode feedback is observed.

Recent evidence does suggest that these sources are indeed young, and
that we are looking at infant powerful radio sources
\citep[e.g.][]{holt:08,kunert_bajraszewska:10}.  The high rate of
incidence, compared to bona fide quasars, suggests that they are a
short-lived phenomenon, which would make them effective short-cycle
thermostats in a feedback scenario.

The detection of compact radio sources in high-redshift star forming
environments seems to support the role of jets in quasar mode
feedback: A number of high-power compact radio sources have been found
in actively star forming regions with powerful outflows
\citep{nesvadba:07,nesvadba:11} and in dense, high-z cooling flow
environments \citep{siemiginowska:10}.  Because we know jet feedback
works in the context of clusters and likely in the ``maintenance''
mode of feedback, and because we know that CSS and GPS sources are (a)
frequent and (b) powerful, they present an attractive alternative to
the wind-driven QSO mode of feedback.

Simulations of jets in dense, multi-phase environments, as might be
expected in star forming galaxies, already show significant promise in
solving the question of how bipolar, highly collimated jets in even
very powerful radio sources could efficiently heat the gas in galaxies
\citep{wagner:11}.

\section{Cosmogony}
\label{sec:cosmogony}

We have seen in section~\ref{sec:global_growth} that the total mass
density estimated in relic supermassive black holes at $z\sim 0$ is
consistent with the total mass accreted by growing black holes during
(obscured and un-obscured) AGN phases for a radiative efficiency of
the accretion process ($0.06<\epsilon_{\rm rad}<0.2$, depending on the
bolometric corrections and local mass density exact estimate), well in
line with the prediction of classical relativistic accretion disc
models \citep{novikov:73}.

In fact, the validation of the \citet{soltan:82} argument implies that
the last few e-folds of a SMBH's mass are mainly grown via
(radiatively efficient) classical accretion discs, rather than through
mergers or radiatively inefficient accretion.  If this is true,
however, the very process of cosmological black hole growth through
accretion quickly erases the initial condition, namely, the primordial
mass function of seed black holes, making it almost impossible to
deduce the physical properties of early black hole formation from
observations probing redshifts smaller than that corresponding to the
most efficient growth ($z\approx 2-3$).  That is, unless a specific
range of BH masses is identified which is less affected by the complex
process of AGN activation during structure formation.

Indeed, some have argued that small mass black holes in isolated,
small mass galaxies could have maintained a 'memory' of the seeding
mechanism (in their location with respect to the scaling relations
defined for more massive systems, for example), being less affected by
the multiple generations of hierarchical mergers in the emerging
cosmic web (see e.g.~\citealt{volonteri:10}).  Very few observational
constraints are available for this class of objects, however.

On the other hand, the observation of luminous quasars at $z\simeq 6$
\citep[e.g.][]{fan:01} has shown that it is possible to probe directly
the earlier epochs of massive black hole assembly, and thus to try to
directly constrain the various physical processes responsible for
planting the seeds that grow into the giant monsters in the nuclei of
galaxies.

\subsection{The first black holes: observational constraints and
  theoretical ideas}
\label{sec:first}

\begin{figure*}
\begin{center}
\resizebox{!}{0.42\textwidth}{\includegraphics{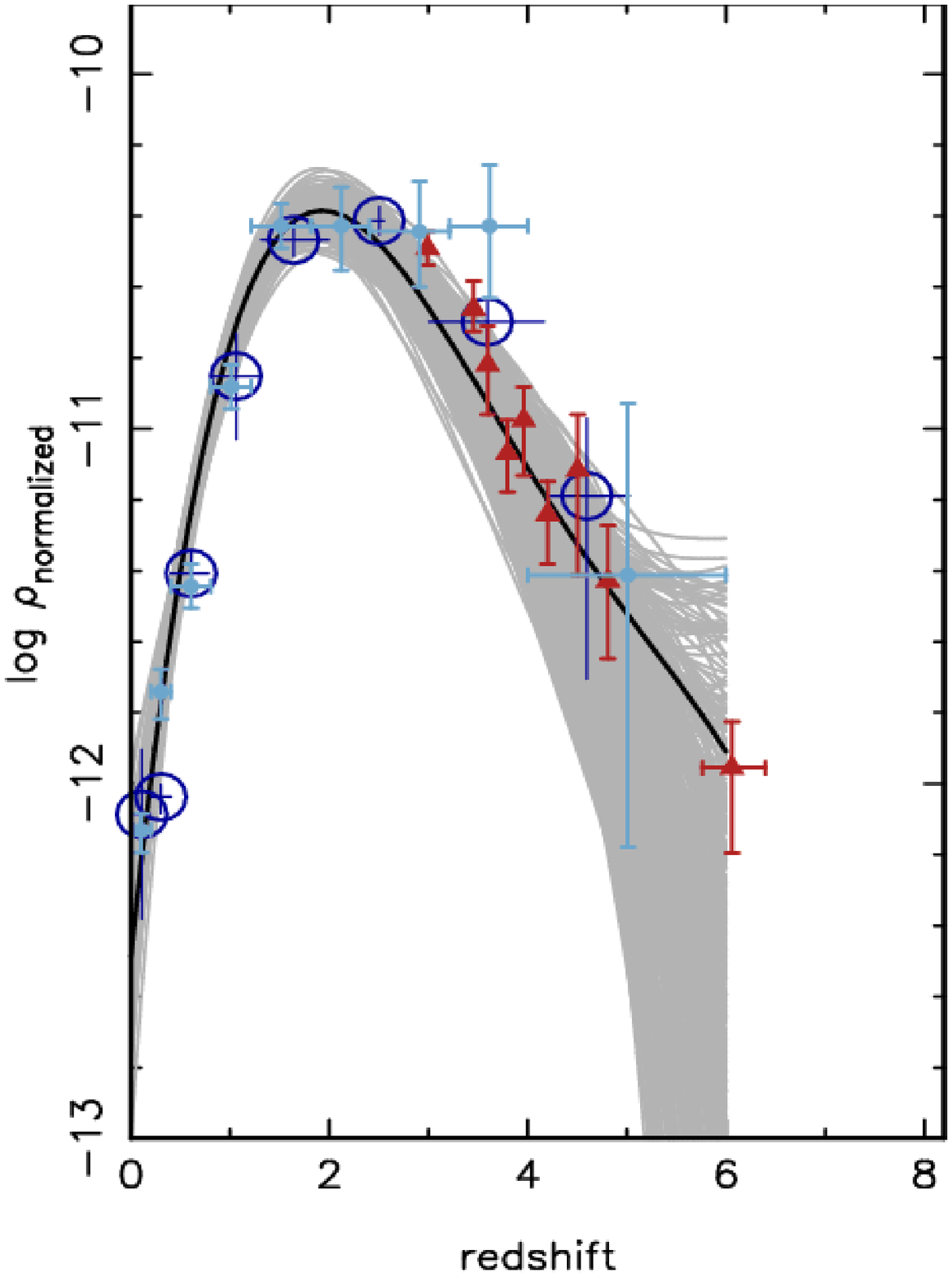}}
\resizebox{!}{0.42\textwidth}{\includegraphics{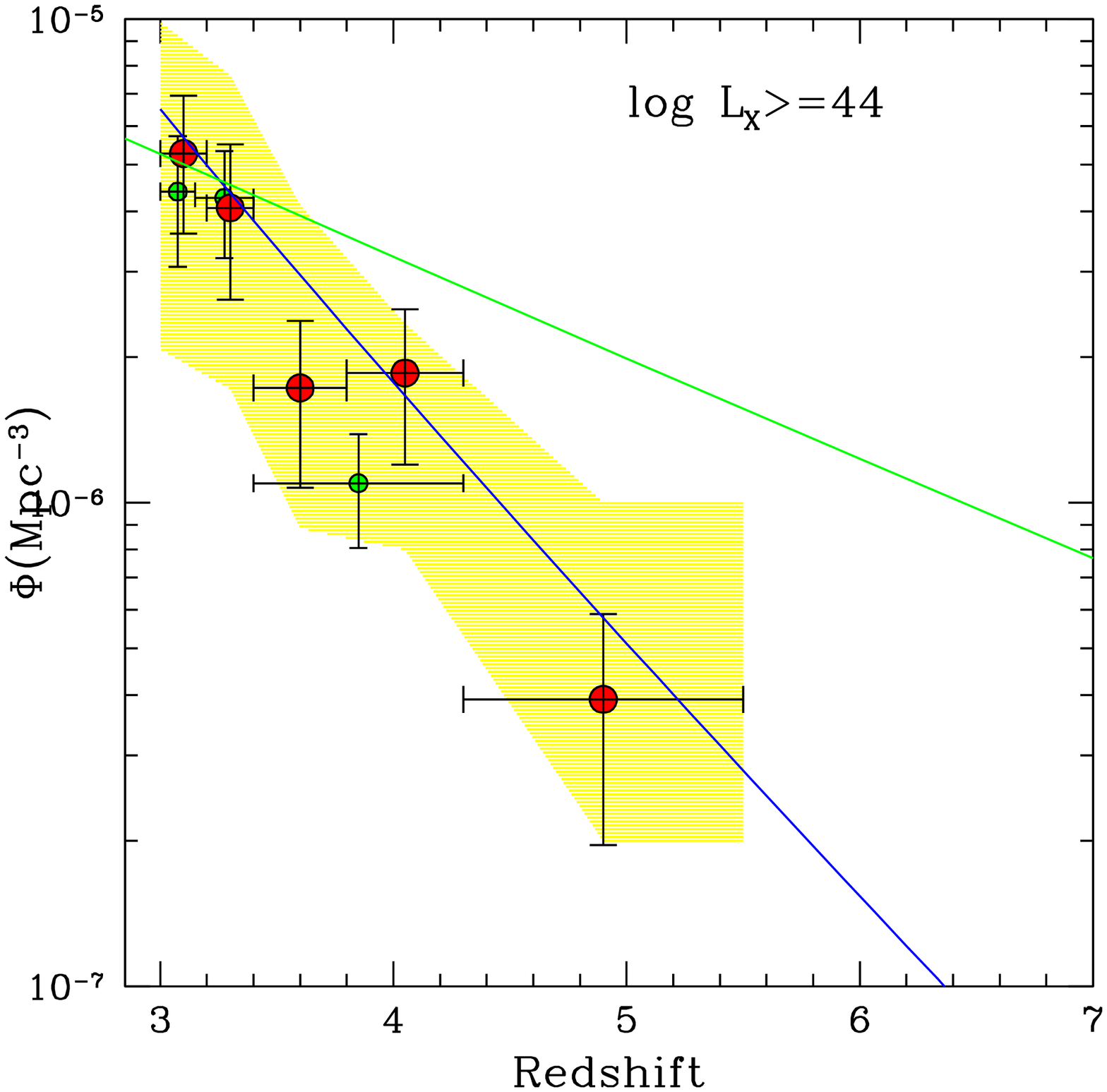}}
\end{center}
\caption{Observational constraints on the decline of QSO number
  densities at high redshift. {\it Left}: Relative space density of
  QSOs ($\rho$) as a function of redshift. The shaded area and black
  line represent the current QSO space-density determination from the
  Parkes quarter-Jansky flat-spectrum sample of radio selected
  QSOs. Light blue filled and dark blue open circles show the soft
  X-ray data from ROSAT, Chandra and XMM-Newton surveys. Space density
  behavior of optically-selected QSOs is given by the set of dark red
  triangles.  The X-ray and optical QSO data were scaled vertically to
  match the current determination of space density at redshifts 2 to
  2.5. From \citet{wall:05}. {\it Right}: The comoving space density
  at bright 2-10 keV X-ray luminosity from the Chandra-COSMOS survey,
  computed taking into account the effect of obscuration. The blue
  curve corresponds to the X–ray selected AGN space density computed
  for the same luminosity limit from models of the CXRB.  The yellow
  shaded area represents the maximum and minimum space density. The
  green symbols correspond to the data of XMM-COSMOS. From
  \citet{civano:11}.}
\label{fig:high_z_radio_x}
\end{figure*}

The observed luminosity functions of AGN suggest a rapid decline of
the total luminosity density above $z\approx 3$ (see
Figure~\ref{fig:lf_bol_param}).  At face value, the constraints on the
very high redshift evolution of the population at $z>5$ come primarily
from bright optical quasars detected in very large area surveys, but
recent indications from large area and moderately deep radio
\citep{wall:05} and X-ray \citep{civano:11} surveys do provide a
consistent picture for the evolution of the most luminous AGN over all
observational wavebands (Figure~\ref{fig:high_z_radio_x}).

The observed rapid decline of QSO number density towards high
redshift translates into a rapid decrease in the number of
AGN-generated ionizing photons towards the end of the re-ionization
epoch. Accurate determination of the QSO rest-frame UV luminosity
function are thus crucial to assess the role of growing black hole
might have played in re-ionizing the universe. Current estimates
suggest that galaxies do dominate the comoving emissivity
of ionizing photons escaping in the inter-galactic medium at $z>4$
\citep[][and references therein]{haardt:12}.

Despite their rarity, very high-redshift QSOs can provide interesting
constraints on the early evolution (and even formation mechanisms) of
nuclear supermassive black holes.  The high metal enrichment observed
in high-z AGNs (see section~\ref{sec:sed_vs_z}), even in those close
to reionization ($z \sim 6$), indicate that the host galaxies of these
AGNs must have undergone a powerful and rapid burst of star
formation. And indeed, vigorous star formation is observed in such
high-z quasars, as inferred from the detection of prominent PAH (Polycyclic Aromatic Hydrocarbons)
features, strong far-IR emission and from the detection of the
[CII]158μm line \citep[see][]{maiolino:09}, redshifted to millimeter
wavelengths. Most likely, we are witnessing the coeval, rapid
formation of massive bulges along with their supermassive central
black holes.

These luminous quasars detected at $z>6$, when the universe was less
than 1 Gyr old have estimated BH masses (from the ``virial method'')
in excess of $\approx 10^9 M_{\odot}$, and it is by no means a trivial
task to grow such massive holes in the relatively short time
available.  Assuming continuous growth at an Eddington ratio of
$\lambda=L/L_{\rm Edd}$:
\begin{equation}
  \frac{dM_{\rm BH}}{dt}=(1-\epsilon_{\rm rad}) \lambda \times L_{\rm Edd}
  /(\epsilon_{\rm rad} c^2)
\end{equation}
we have, for the final BH mass as a function of the initial mass
\begin{equation}
\label{eq:eddington_growth}
M_{\rm BH,f}(t)=M_i \exp{(1-\epsilon_{\rm rad})/\epsilon_{\rm rad}
  \times (t/\tau_{\rm salp})} \,,
\end{equation}
where we have defined the typical e-folding time (the so-called
``Salpeter time'') as
\begin{equation}
  \tau_{\rm salp}=\frac{\lambda c \sigma_{\rm T}}{4\pi G m_{\rm p}} =
  0.45 \left(\frac{1}{\lambda}\right) {\rm Gyr}
\end{equation}

\begin{figure}
\begin{center}
  \resizebox{!}{0.426\textwidth}{\includegraphics{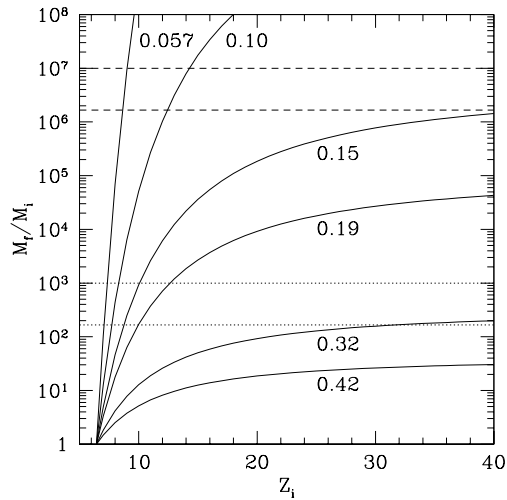}}
\end{center}
\caption{Ratio of final-to-initial black hole mass at $z=6$ calculated
  from eq.~(\ref{eq:eddington_growth}) assuming $\lambda=1$
  (i.e. continuous Eddington-limited accretion) as a function of the
  formation redshift of the seed black hole.  Each solid curve is
  labeled with the corresponding value of the radiative efficiency
  $\epsilon_{\rm rad}$.  Dotted horizontal lines show the ratio
  implied by the observed $z\sim 6$ QSOs for massive seeds, while the
  dashed ones that for stellar mass (Pop III; see
  section~\ref{sec:first} for details).  From \citet{shapiro:05}.}
   \label{fig:shapiro}
\end{figure}

Depending on the redshift of formation of the seed of mass $M_i$, and
on the average radiative efficiency of the accretion process, only a
limited range of final BH masses can be reached at $z=6$, as shown in
Figure~\ref{fig:shapiro} \citep{shapiro:05}.  If the BH seed masses
are in the range expected from Pop III remnants, of the order of a few
hundred solar masses, then highly radiative efficient ($\epsilon_{\rm
  rad} > 0.1$) accretion is excluded, as it would not allow enough
mass to be accumulated into the black hole rapidly enough.

Such a scenario has a number of major difficulties. First of all, if
the accretion is indeed continuous and proceeds at high Eddington
rates, the accretion flow should form a geometrically thin, optically
thick disk. If this is the case, even a black hole with zero initial
dimensionless angular momentum $a_i=0$ will be rapidly spun-up by the
angular momentum captured with the accreting gas.  \citet{bardeen:70}
has shown that in this case the final spin obeys:
\begin{equation}
  a_f=\frac{r_{\rm ms,i}^{1/2}}{3}\frac{M_i}{M_f}\left[4-\sqrt{3r_{\rm
        ms,i}\left(\frac{M_i}{M_f}\right)^2-2}\right] 
\end{equation}
where $r_{\rm ms,i}$ is the radius of the marginally stable orbit of
the initial black hole (itself a monotonic function of spin, with $r_{\rm
  ms}(a=0)=6$ and $r_{\rm ms}(a=1)=1$ in gravitational units).  Thus,
an initially non-spinning BH is spun up by accretion of gas in a
classical geometrically thin and optically thick accretion disc as
soon as\footnote{Note that the above calculation assumes that there is
  no torque at the inner boundary of the accretion disc
  \citep{novikov:73}. Magnetic linkage between the disc, the plunging
  region and the event horizon can modify the above picture, reducing
  the maximal spin a BH can reach \citep{krolik:05}.  Nonetheless,
  most numerical models of geometrically thin magnetized discs are
  still consistent with a rapid spin of the BH.}  $M_f/M_i=\sqrt{6}$.
Since an accretion disc around a maximally rotating hole will radiate
with an efficiency $\epsilon_{\rm rad}=0.42$, we are left with the
impossibility of growing black holes larger than a few thousands of
solar masses via prolonged coherent accretion onto a stellar mass seed
(see Figure~\ref{fig:shapiro}).

A few solutions have been proposed to the above problem.  

\begin{enumerate}
\item{First, accretion might not be coherent, but rather stochastic,
    such that the accreting gas comes into the gravitational sphere of
    influence of the black hole in parcels with randomly oriented
    angular momenta.  If the mass of each parcel is small enough
    (roughly speaking, if $\Delta M \ll \sqrt{6} M_i$), then the BH
    spin vector performs a random walk, but as it is easier to spin a
    black hole down than up, the net effect is a relatively low
    average spin (and correspondingly lower radiative efficiency) of
    the final hole \citep{king:06}.}
\item{In the primordial cosmological setup where proto-galaxies form,
    the first black holes grow in a very gas rich environment, and
    there is no reason to believe that the gas could not flow towards
    the central black holes at vastly super-Eddington rates.  It is
    not clear, however, whether the hole can swallow matter that fast.
    On the one hand, quasi-spherical inflows can be established, where
    the radial velocity of the accreting gas is so high that the
    photons produced inside the disc by the viscous torques cannot
    escape \citep{frank:02}.  In those cases, although the
    emerging luminosity is at most logarithmically in excess of the
    Eddington limit, the accretion rate onto the hole can be orders of
    magnitude larger.  On the other hand, the accretion flow can start
    blowing out matter at the (large) radius where the locally
    produced energy exceeds the local Eddington limit
    \citep{shakura:73}.  A powerful wind ensues, which may prevent the
    mass accretion rate onto the black hole from exceeding the
    Eddington limit by more than a factor of a few.}
\item{Black hole seeds in the early universe can be more massive than
    the remnants of Pop III stars.  These would have formed by direct
    collapse of primordial massive stars \citep[see
    e.g.][]{volonteri:10}.  For example, some theoretical models have
    argued that the gas chemistry in the most massive, hottest
    primordial DM halos can prevent fragmentation of the cooling gas
    and lead to the formation of very massive stars.  Their cores will
    rapidly collapse, leaving a black hole at the center of a
    quasi-static, radiation pressure-supported supermassive star.  The
    resulting object, called a ‘quasistar’, resembles a red giant with
    a luminosity comparable to a Seyfert nucleus.  The black hole
    grows inside it until the cooling photosphere can no-longer
    sustain its own radiation pressure and the envelope disperses,
    leaving behind the naked seed black hole of typically
    $10^4$--$10^5 M_{\odot}$ \citep{begelman:10}.}
\item{Finally, a large number of major mergers could help relax the
    demands on the efficiency and stability of gas accretion on the
    first black holes by enhancing the final-to-initial mass ratio by
    a factor of the order of the number of equal-mass (or major)
    mergers along the main tree of the hierarchy.

    Cosmological numerical simulations \citep{li:07} provide a
    possible route to the formation of a $10^9M_{\odot}$ at $z\sim 6$
    by starting with Pop III stellar mass seeds at $z\sim 30$ which
    experience an early phase of continuous, Eddington-limited
    accretion (subject, however, to the limitations discussed above),
    before entering the merger tree at $z\sim 14$, where a large
    number of major merger events is able to accumulate the final
    mass, even in the presence of AGN feedback.}
\end{enumerate}

\begin{figure}
\centering
  \psfig{figure=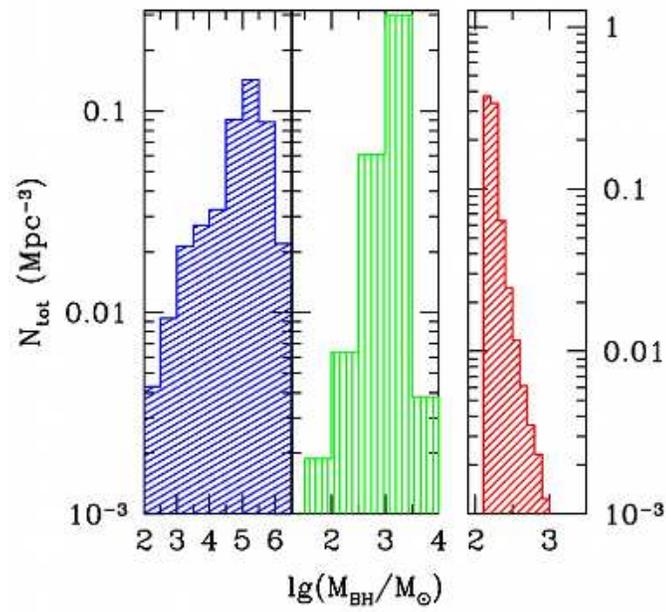,height=9cm}
  \caption{Mass function of seed massive Black Holes for three
    different formation scenarios: direct collapse (left), runaway
    stellar mergers in high-redshift stellar clusters (center) and
    Population III remnants (right).  Note the different y-axis scale
    for the Pop III case.  From \citep{volonteri:10}.}
   \label{fig:seeds}
\end{figure}

Analytic and semi-analytic models of the early assembly of massive
black holes that include (with varying degrees of sophistication) the
many competing processes (mergers, gravitational wave recoil and
nuclear black hole ejection, dynamical friction on wandering BHs,
pristine gas accretion, etc.) make clear predictions for the early
seed mass distributions, the initial conditions for SMBH growth that
we would like to probe.  Figure~\ref{fig:seeds} shows one such
prediction, comparing the outcomes of three different formation
scenarios: direct collapse, runaway stellar mergers in high-redshift
stellar clusters, and Population III remnants.  A more complete census
of the AGN population at $z\sim 6$ or even a few detections of $z \sim
10$ AGN with the next generation of large astronomical facilities
could provide direct means to distinguish among these simple formation
scenarios, allowing us to glimpse into the obscured epoch when the
first nuclear black holes formed.

\section*{Acknowledgments}
We would like to thank warmely our Editor William C. Keele for the kind
invitation to write this chapter in the new edition of the ``Planets,
Stars and Stellar System'' book, and for the patience and
endurance he demonstrated in putting up with our delays. Obviously, 
this work could not have been possible without the many
contributions from a vast number of colleagues and collaborators. In
particular, we whish to thank Viola Allevato, 
Roberto Assef, Silvia Bonoli, Niel Brandt, Marcella Brusa, Johannes
Buchner, Nico Cappelluti, Scott Croom, Francesco De Gasperin, Andy
Fabian, Guenther Hasinger, Phil Hopkins, Brian McNamara, Eric Perlman, 
Gordon Richards, Marta Volonteri.
AM's work is partially supported by the DFG Cluster of Excellence
``Origin and Structure of the Universe''.

\section*{Cross References: Chapters in the same book}
\noindent ``Active Galactic Nuclei'' by E. Perlman\\
\noindent ``Clusters of Galaxies'' by R. Bower\\
\noindent ``Galaxies in the cosmological context'' by G. De Lucia\\
\noindent ``Large Scale Structure of the universe'' by A. Coil\\

\bibliographystyle{apj} \bibliography{feedback}
\end{document}